%% file: RateDistortion_0609.tex
\documentclass[12pt,journal,onecolumn]{IEEEtran}
\usepackage{amsmath,amssymb, dsfont}
\usepackage{epsfig,latexsym,amssymb,amsmath,graphics}
\usepackage{graphicx}
\usepackage[linesnumbered,ruled]{algorithm2e}
\usepackage{cite}
\usepackage{url}

\input{pream_bm}

\renewcommand{\baselinestretch}{1}

\def\IR{\mathbb R}

\renewcommand{\baselinestretch}{1.6}

\newtheorem{theorem}{Theorem}

\newcommand{\calN}{{\cal N}}

\newcommand{\X}{{\cal X}}
\newcommand{\Y}{{\cal Y}}
\newcommand{\C}{{\cal C}}
\newcommand{\U}{{\cal U}}

\newcommand{\V}{{\cal V}}

\newcommand{\R}{{\cal R}}

\newcommand{\B}{{\cal B}}

\newcommand{\A}{{\cal A}}
\newcommand{\Q}{{\cal Q}}

\title{On Finite Block-Length Quantization Distortion}

\author{Chen~Gong,\thanks{Dept. of Electrical Engineering, Columbia Univ., New York,
NY 10027.} \and Xiaodong Wang$^{*}$ }
\date{}
\begin{document}
\maketitle{}

\renewcommand{\baselinestretch}{1.3}

\begin{abstract}
We investigate the upper and lower bounds on the quantization distortions
for independent and identically distributed sources in the finite block-length regime.
Based on the convex optimization framework of the rate-distortion theory,
we derive a lower bound on the quantization distortion under finite block-length,
which is shown to be greater than the asymptotic distortion given by the rate-distortion theory.
We also derive two upper bounds on the quantization distortion based on random quantization codebooks,
which can achieve any distortion above the asymptotic one.
Moreover, we apply the new upper and lower bounds to two types of sources,
the discrete binary symmetric source and the continuous Gaussian source.
For the binary symmetric source, we obtain the closed-form expressions of the upper and lower bounds.
For the Gaussian source, we propose a computational tractable method to numerically compute the upper and lower bounds,
for both bounded and unbounded quantization codebooks.
Numerical results show that the gap between the upper and lower bounds is small for reasonable block length and hence the bounds are tight.
\end{abstract}
\newpage
\renewcommand{\baselinestretch}{1.6}

\thispagestyle{empty}

\section{Introduction}
The rate distortion theory provides an achievable asymptotic lower bound on the distortion of the lossy quantization,
as the block length of the source sequence approaches infinity.
This bound is described by a single-symbol probability transition function
between the original alphabet and the reconstruction alphabet,
which produces the minimum asymptotic distortion provided that the mutual information is below the quantization rate.
For sufficiently large block length, any distortion above the asymptotic bound can be achieved;
and any distortion below the asymptotic bound cannot be achieved for any block length~\cite{ElementsInfoTheory}.
Existing works on the rate distortion focuses on either the rate-distortion functions
of various types of sources~\cite{ExtremeProperty,EstimateRD,DSCRD,RDFeedforward,RDMemory,RDQuotient},
or the analysis of various exponents
as the quantization block length approaches infinity~\cite{ListExponent,DiscreteSources,TrellisCoding,TradeoffExponents,RDNoisySource}.

On the other hand, in the non-asymptotic regime, up till now there has been no analysis on the lower bounds
for the vector quantization in the finite block-length regime, even for i.i.d. sources.
Note that, since the rate-distortion theory provides an asymptotic achievable lower bound,
any meaningful lower bound in the finite block-length regime should be larger
than this asymptotic lower bound.
Also, to the best of our knowledge, the only upper bound for random quantization codebook is provided in~\cite{DiscreteSources}.
This upper bound can be improved.

In this paper, we derive new upper and lower bounds on the optimal quantization error in the finite block length regime.
More specifically, based on the convex optimization formulation of the rate-distortion problem,
we derive a lower bound for the quantization distortion,
which is shown to be larger than the asymptotic distortion and thus is non-trivial.
Using this lower bound, we analyze the duality between the quantization distortion
and the error probability of the equivalent channel characterized by  the optimal probability  transition function.
We further specialize the lower bounds to cases where the codewords are bounded and where the source is symmetric, respectively.
We then provide two improved upper bounds on the vector quantization distortion assuming random quantization codebooks.
It is shown that any quantization distortion above the asymptotic lower bound can be achieved by the proposed upper bounds.

Furthermore, we apply new upper and lower bounds to two types of sources,
the discrete binary symmetric source and the continuous Gaussian source.
For the binary symmetric source, we provide closed-form expressions for the upper and lower bounds.
For the Gaussian source, we further analyze the upper and lower bounds for codebooks with bounded and unbounded codewords, respectively,
and provide an efficient method to compute these bounds.

The remainder of this paper is organized as follows.
In Section~\ref{sec.Notation}, we introduce the problem formulation.
In Sections~\ref{sec.LowerBound} and~\ref{sec.AchievableD}, we derive lower and upper bounds
for the optimal quantization distortion, respectively.
In Sections~\ref{sec.BinarySymSource} and~\ref{sec.GaussianSource}, we specialize  the new bounds to the binary symmetric source
and identically and the Gaussian source, respectively.
Finally, Section~\ref{sec.Conclusions} contains the concluding remarks.
All proofs are relegated to the Appendix.

\section{Background and Problem Statement} \label{sec.Notation}
Consider a memoryless source with an original symbol alphabet $\X$, which is to be reconstructed
using a reconstruction symbol alphabet $\Y$.
For each pair $(x , y)\in \X \times \Y$, let $d(x,y)$ denote the distortion of representing symbol $x$ using symbol $y$.
Note that $d(x,y) \geq 0$ for all $(x , y)\in \X \times \Y$.
Assume that the original symbol $x \in \X$ has a probability density function (pdf) $p(x)$.
Let $q(y|x)$ denote the conditional pdf of the reconstruction symbol $y \in \Y$ given the
original symbol $x \in \X$.
Define $D(q)$ as the expected distortion of the conditional pdf $q(y|x)$, given by
\be \label{equ.DistortionQ}
D(q) \dff \mathbb{E}_{(x,y) \sim p(x)q(y|x)} \Big(d(x,y)\Big) = \int \int p(x) q(y|x) d(x,y) dx dy.
\ee
We adopt conventional notations $H(\cdot)$, $H(\cdot|\cdot)$, and $I(\cdot;\cdot)$ to denote the
entropy, conditional entropy, and mutual information, respectively.
We define $I(q) \dff I(X;Y)$ as the mutual information between $X$ and $Y$ under the conditional pdf $q(y|x)$.

Consider a length-$n$ source symbol block $\bx \dff (x_1,x_2,...,x_n) \in \X^n$ that is to be reconstructed
using a length-$n$ reconstruction symbol block $\by \dff (y_1,y_2,...,y_n) \in \Y^n$.
Define the following distortion metric between $\bx$ and $\by$,
\be \label{equ.DistortionS}
d(\bx , \by) \dff \frac{\sum^n_{i = 1}d(x_i , y_i)}{n}.
\ee
Then the expected distortion with respect to the reconstructing pdf $q(\by|\bx)$ is
\be \label{equ.DistortionS2}
D(q) \dff \mathbb{E}_{(\bx,\by) \sim p(\bx)q(\by|\bx)}\Big(d(\bx,\by)\Big) = \int\int p(\bx) q(\by|\bx) d(\bx,\by) d\by d\bx.
\ee

We consider reconstructing $\bx \in \X^n$ using a size-$Q$
quantization codebook $\Y_Q \dff \{\by_1,\by_2,...,\by_Q\} \subseteq \Y^n$.
Let $q_Q(\by_j|\bx)$ be the probability of quantizing $\bx$ to $\by_j$.
Define the distortion with respect to the quantization function $q_Q(\by_j|\bx)$ as
\be \label{equ.DistortionS4}
D^n(q_Q) = \int p(\bx) \sum^Q_{j = 1} q_Q(\by_j|\bx) d(\bx,\by_j) d\bx.
\ee

It is easily seen that the distortion $D^n(q_Q)$ is minimized when $q_Q(\by_{j^*}|\bx) = 1$, where
\be \label{equ.DistortionS5}
j^* = \arg \min_{1 \leq k \leq Q} d(\bx , \by_k),
\ee
and $q_Q(\by_j|\bx) = 0$ for $j \neq j^*$. If there are multiple codewords satisfying (\ref{equ.DistortionS5}),
then $j^*$ is the smallest index $j$ among such codewords.
Define $\R_j$  as the quantization region of $\by_j$, given by
\be \label{equ.DistortionS6}
\R_j \dff \{\bx: q_Q(\by_j|\bx) = 1\}, \ \mbox{or} \ \ q_Q(\by_j|\bx) = \mathbf{1}_{\{\bx \in \R_j\}}.
\ee
Note that the subsets $\R_j$, $1 \leq j \leq Q$, are non-overlapping and $\cup^Q_{j = 1} \R_j = \X^n$.

We denote $D^n(\Y_Q) \dff D^n(q_Q)$ as the quantization distortion of the codebook $\Y_Q$,
and summarize the above arguments as follows.

\begin{theorem} \label{thm.Thm1}
Consider the non-overlapping subsets $\R_j$, $1 \leq j \leq Q$, of $\X^n$ given by (\ref{equ.DistortionS6}).
Then, for any quantization probability function $\tilde{q}_Q$,
\be \label{equ.DistortionS7}
D^n(\tilde{q}_Q) \geq D^n(\Y_Q) = \sum^Q_{j = 1} \int_{\R_j} p(\bx) d(\bx , \by_j) d\bx.
\ee
\end{theorem}

$\hfill \Box$

The rate-distortion theory provides an asymptotic lower bound to the quantization distortion $D^n(q_Q)$ in (\ref{equ.DistortionS4})
as the block length $n$ approaches infinity.
Let the size of the quantization codebook $\Y_Q$ be $Q = 2^{nR}$.
According to the rate-distortion theory, any distortion $D$ can be achieved if $D > D(R)$,
where
\be \label{equ.RateDistortion}
D(R) \dff \min_{q(y|x), I(q) \leq R} D(q),
\ee
for sufficiently large block length $n$; and any distortion $D < D(R)$ is not achievable for any block length $n$.
Note that in (\ref{equ.RateDistortion}), $q(y|x)$ is the single-alphabet conditional pdf.

The rate distortion theory provides a lower bound to the quantization error as the block length $n$ approaches infinity.
Such an asymptotic bound typically cannot accurately approximate the distortion for finite block length.
In this work, we aim to obtain upper and lower bounds for the quantization distortion $D^n(q_Q)$ in (\ref{equ.DistortionS4})
for finite $n$.

In the remainder of this paper, let $\hat q(y|x)$ be the optimal solution to $q(y|x)$
which minimizes the distortion $D(q)$ for infinite quantization block length,
i.e., the optimal solution to the rate-distortion problem (\ref{equ.RateDistortion}).
We denote $D^* \dff D(\hat q)$ as the resulting minimum distortion from the rate-distortion theory.
For length-$n$ symbol blocks $\bx = (x_1,x_2,...,x_n)$ and $\by = (y_1,y_2,...,y_n)$,
we define the following product pdf of the optimal quantization function $\hat q(y|x)$,
\be \label{equ.ProbabilityOpt}
\hat q(\by|\bx) \dff \prod^n_{i = 1} \hat q(y_i|x_i).
\ee

\section{A Lower Bound on Quantization Distortion} \label{sec.LowerBound}
In this section, we derive a lower bound on the quantization distortion $D^n(q_Q)$ in (\ref{equ.DistortionS4}).
We start from the convex optimization formulation of the rate distortion problem in (\ref{equ.RateDistortion}).

\subsection{Convex Optimization Formulation of the Rate Distortion Problem}
The rate-distortion problem in (\ref{equ.RateDistortion}) can be expressed as
\be \label{equ.CvxOptimization}
&&\min_{q(y|x)} \int \int p(x)q(y|x) d(x,y) dx dy, \nonumber \\
\mbox{s.t.} && I(X;Y) = \int \int p(x)q(y|x) \log\frac{q(y|x)}{\int_{\X} p(x')q(y|x') dx'} dx dy \leq R, \nonumber \\
&& \int q(y|x) dy = 1, \ x \in \X.
\ee
Note that the mutual information $I(X;Y)$ is a convex function in terms of $q(y|x)$ for a fixed $p(x)$~\cite{ElementsInfoTheory}.
Hence the above is a convex optimization problem.
The Lagrangian form of (\ref{equ.CvxOptimization}) can be written as
\be \label{equ.CvxOptimization2}
J(q , \lambda , v) &=& \int \int p(x)q(y|x) d(x , y) dx dy + \int v(x) \Big(\int q(y|x) dy - 1\Big)dx \nonumber \\
&& \ \ + \lambda \Big(\int\int p(x)q(y|x) \log\frac{q(y|x)}{\int p(x')q(y|x') dx'}dxdy - R\Big).
\ee

We then have
\be \label{equ.Different1}
\frac{\partial J}{\partial q(y|x)} = p(x)d(x , y) + \lambda p(x) \log\frac{q(y|x)}{q(y)} + v(x),
\ee
where $q(y) = \int_{\X} p(x)q(y|x)dx$.
Let $\hat q(y) = \int_{\X} p(x)\hat q(y|x)dx$, where $\hat q(y|x)$ is an optimal solution to (\ref{equ.RateDistortion}).
Then, by the KKT condition, we have
\be \label{equ.Different2}
\frac{\partial J}{\partial q(y|x)}\Big|_{q(y|x) = \hat q(y|x)} = 0 \ \mbox{for} \ \hat q(y|x) > 0, \ \mbox{and} \
\frac{\partial J}{\partial q(y|x)}\Big|_{q(y|x) = \hat q(y|x)} \geq 0 \ \mbox{for} \ \hat q(y|x) = 0,
\ee
which can be compactly written as
\be \label{equ.Different3}
\hat q(y|x)\frac{\partial J}{\partial q(y|x)}\Big|_{q(y|x) = \hat q(y|x)} = 0.
\ee

The Slater condition \cite{ConvexOptimization} for the convex optimization problem says that the dual gap is zero if there is a feasible conditional pdf $q(y|x)$
such that $I(q) < R$.
This can be easily verified since we can set $q(y|x) = 1$ for $y = y_0$ and $q(y|x) = 0$ otherwise for all $x$, for some fixed $y_0$,
such that $I(q) = I(X;Y) = 0$.
Therefore, for the optimal solution to (\ref{equ.CvxOptimization}) the duality gap is zero.
Thus, we have the following result.

\begin{theorem} \label{thm.Thm2}
Let $\hat q(y|x)$ be an optimal conditional pdf, $\hat q(y)$ be the corresponding marginal pdf of $y$,
and $\hat \lambda$ and $\hat v(x)$ be the optimal dual variables.
Let $\hat q(x|y) = \frac{p(x)\hat q(y|x)}{\int_{x}p(x)q(y|x)dx}$.
Then, we have the following conditions to characterize the optimal solution to (\ref{equ.CvxOptimization}):
\be
p(x)d(x , y) + \hat \lambda p(x) \log\frac{\hat q(y|x)}{\hat q(y)} + \hat v(x) &=& 0, \ \mbox{for} \ \hat q(y|x) > 0; \label{equ.Different41} \\
p(x)d(x , y) + \hat \lambda p(x) \log\frac{\hat q(y|x)}{\hat q(y)} + \hat v(x) &\geq& 0, \ \mbox{for} \ \hat q(y|x) = 0; \label{equ.Different42} \\
p(x)\hat q(y|x)d(x,y) + \hat \lambda p(x) \hat q(y|x) \log\frac{\hat q(y|x)}{\hat q(y)} + \hat q(y|x)\hat v(x) &=& 0, \ x \in \X, y \in \Y; \label{equ.Different43} \\
\hat \lambda \Big(\int\int p(x)\hat q(y|x) \log\frac{\hat q(y|x)}{\hat q(y)}dxdy - R\Big) &=& 0; \label{equ.Different44} \\
\hat v(x) \Big(\int \hat q(y|x) dy - 1\Big) &=& 0, \ x \in \X. \label{equ.Different45}
\ee
\end{theorem}

$\hfill \Box$

\subsection{A Lower Bound on Distortion} \label{subsec.DistortionLN}
The following result gives a lower bound on the distortion $D^n(q)$ for block size $n$ and a given reconstruction conditional pdf $q(\by|\bx)$.

\begin{theorem} \label{thm.Thm30}
Let $\hat q(y|x)$ be an optimal conditional pdf, and $\hat q(y)$ be the corresponding marginal pdf of $y$,
and $\hat \lambda$ and $\hat v(x)$ be the optimal dual variables for the rate-distortion problem (\ref{equ.CvxOptimization}).
Then, for any reconstruction conditional pdf $q(\by|\bx)$, we have
\be \label{equ.DistortionOpt0}
D^n(q) - D^* \geq \frac{\hat \lambda}{n} \Big(n I(\hat q) - \int\int p(\bx) q(\by|\bx) \log\frac{\hat q(\bx|\by)}{p(\bx)} d\by d\bx \Big);
\ee
and the equality holds if $\hat q(y|x) > 0$ for all $x \in \X$ and $y \in \Y$.

\begin{proof}
Since the source is memoryless, we have that $p(\bx) = \prod^n_{i = 1}p(x_i)$, and thus
\be \label{equ.DistortionOpt2}
p(\bx) d(\bx,\by) = \prod^n_{j = 1}p(x_j) \frac{\sum^n_{i = 1}d(x_i,y_i)}{n}
= \frac{1}{n} \sum^n_{i = 1}\prod_{j \neq i}p(x_j)p(x_i)d(x_i,y_i).
\ee
From Theorem~\ref{thm.Thm2}, it follows that
\be \label{equ.DistortionOpt3}
p(x_i)d(x_i , y_i) \geq - \hat \lambda p(x_i) \log\frac{\hat q(y_i|x_i)}{\hat q(y_i)} - \hat v(x_i).
\ee
Substituting (\ref{equ.DistortionOpt3}) into (\ref{equ.DistortionOpt2}), we have
\be \label{equ.DistortionOpt4}
p(\bx) d(\bx,\by) &\geq& - \frac{1}{n} \sum^n_{i = 1}\prod_{j \neq i}p(x_j) \hat \lambda p(x_i) \log\frac{\hat q(y_i|x_i)}{\hat q(y_i)}
- \frac{1}{n} \sum^n_{i = 1}\prod_{j \neq i}p(x_j) \hat v(x_i) \nonumber \\
&=& - \frac{\hat \lambda}{n} p(\bx) \log\frac{\hat q(\by|\bx)}{\hat q(\by)}
- \frac{1}{n} \sum^n_{i = 1}\prod_{j \neq i}p(x_j) \hat v(x_i).
\ee
Therefore, we have,
\be \label{equ.DistortionOpt5}
D^n(q) &=&  \int\int p(\bx) q(\by|\bx) d(\bx,\by) d\bx d\by \nonumber \\
&\geq& - \frac{\hat \lambda}{n} \int\int p(\bx) q(\by|\bx)
\log\frac{\hat q(\by|\bx)}{\hat q(\by)} d\bx d\by 
- \frac{1}{n} \int\int q(\by|\bx) \sum^n_{i = 1}\prod_{j \neq i}p(x_j) \hat v(x_i) d\bx d\by \nonumber \\
&=& - \frac{\hat \lambda}{n} \int\int p(\bx) q(\by|\bx) \log\frac{\hat q(\bx|\by)}{p(\bx)} d\bx d\by
- \frac{1}{n} \int \sum^n_{i = 1}\prod_{j \neq i}p(x_j) \hat v(x_i) d\bx.
\ee

On the other hand, using a similar argument as above and noting that (\ref{equ.Different3}) holds for the optimal solution $\hat q(y|x)$,
we have that
\be \label{equ.DistortionOpt8}
D^n(\hat q) &=& - \frac{\hat \lambda}{n} \int\int p(\bx) \hat q(\by|\bx)
\log\frac{\hat q(\by|\bx)}{\hat q(\by)} d\bx d\by
- \frac{1}{n} \int \sum^n_{i = 1} \prod_{j \neq i} p(x_j) \hat v(x_i) d\bx \nonumber \\
&=& - \hat \lambda I(\hat q)
- \frac{1}{n} \int \sum^n_{i = 1} \prod_{j \neq i} p(x_j) \hat v(x_i) d\bx.
\ee

Recall that $D^* = D^n(\hat q)$.
Finally, using (\ref{equ.DistortionOpt5}) and (\ref{equ.DistortionOpt8}), we obtain:
\be \label{equ.DistortionOpt10}
D^n(q) - D^* = D^n(q) - D^n(\hat q) \geq \frac{\hat \lambda}{n} \Big(n I(\hat q) -
\int \int p(\bx) q(\by|\bx) \log\frac{\hat q(\bx|\by)}{\hat q(\bx)} d\bx d\by\Big).
\ee
Moreover, if $\hat q(y|x) > 0$ for any $x \in \X$ and $y \in \Y$, then the ``$\geq$'' becomes ``$=$'' in
(\ref{equ.DistortionOpt3}) - (\ref{equ.DistortionOpt5}), and thus the equality holds in (\ref{equ.DistortionOpt10}).
\end{proof}
\end{theorem}

Now consider the case of quantization using a size-$Q$ codebook $\Y_Q = \{\by_1,\by_2,...,\by_Q\}$.
For the vector quantizer $q_Q$ given by (\ref{equ.DistortionS6}),
based on Theorem~\ref{thm.Thm30}, we have the following lower bound on the quantization distortion.

\emph{Corollary 3}:
Assume that $I(\hat q) = R$ and $Q = 2^{nR}$.
The distortion $D^n(q_Q)$ of a vector quantizer with the quantization regions ${\cal R}_j, j=1, ..., Q$, satisfies
\be
D^n(q_Q) - D^* &\geq& \frac{\hat \lambda}{n}
\Big(nR - \sum^Q_{j = 1} \int_{\R_j} p(\bx) \log\frac{\hat q(\bx|\by_j)}{p(\bx)} d\bx\Big) \label{equ.DistortionOpt13} \\
&\geq& 0; \label{equ.DistortionOpt131}
\ee
and the equality in (\ref{equ.DistortionOpt13}) holds if $\hat q(y|x) > 0$ for any $x \in \X$ and $y \in \Y$.

\begin{proof}
(\ref{equ.DistortionOpt13}) follows from Theorem~\ref{thm.Thm30} where the reconstruction pdf is given by $q_Q(\by|\bx) = \mathbf{1}_{\bx \in \R_j}$.
The tightness of this lower bound follows from the same argument as that for Theorem~\ref{thm.Thm30}.
Next we prove (\ref{equ.DistortionOpt131}).
Note that
\be \label{equ.DistortionOpt132}
&& nR - \sum^Q_{j = 1} \int_{\R_j} p(\bx) \log\frac{\hat q(\bx|\by_j)}{p(\bx)} d\bx \nonumber \\
&=& \sum^Q_{j = 1} \int_{\R_j} p(\bx) \log\frac{p(\bx)}{\hat q(\bx|\by_j)/Q} d\bx \nonumber \\
&=& \int_{\X^n} p(\bx) \log\frac{p(\bx)}{\frac{1}{Q}\sum^Q_{j=1} \hat q(\bx|\by_j)\cdot \mathbf{1}_{\bx \in \R_j}} d\bx.
\ee
We let $T \dff \int_{\bx} \frac{1}{Q}\sum^Q_{j=1} \hat q(\bx|\by_j)\cdot \mathbf{1}_{\bx \in \R_j} d\bx$, and define the following pdf
\be \label{equ.DistortionOpt133}
t(\bx) = \frac{\hat q(\bx|\by_j)\cdot \mathbf{1}_{\bx \in \R_j}}{T}.
\ee
Note that we have
\be \label{equ.DistortionOpt134}
T &=& \int_{\bx} \frac{1}{Q}\sum^Q_{j=1} \hat q(\bx|\by_j)\cdot \mathbf{1}_{\bx \in \R_j} d\bx \nonumber \\
&=& \frac{1}{Q}\sum^Q_{j=1} \int_{\bx} \hat q(\bx|\by_j)\cdot \mathbf{1}_{\bx \in \R_j} d\bx \nonumber \\
&\leq& \frac{1}{Q}\sum^Q_{j=1} 1 = 1.
\ee
Then, we have
\be \label{equ.DistortionOpt135}
(\ref{equ.DistortionOpt132}) &=& \int_{\X^n} p(\bx) \log\frac{p(\bx)}{T \cdot t(\bx)} d\bx  \nonumber \\
&=& \log \frac{1}{T} + \int_{\X^n} p(\bx) \log\frac{p(\bx)}{t(\bx)} d\bx = \log \frac{1}{T} + D(p||t) \geq 0,
\ee
where $D(p||t)$ is the Kullback-Leibler (KL) distance and thus $D(p||T) \geq 0$.
\end{proof}

\subsubsection*{Non-singular Optimal Reconstruction pdf $\hat q(y|x)$}
The optimal solution $\hat q(y|x)$ to the rate distortion problem is non-singular if and only if $I(\hat q) = R$
and $\hat q(y|x) > 0$ for all $x \in \X$ and $y \in \Y$.
Define the following residue term
\be \label{equ.DistortionOpt15}
\Delta D^n(q_Q)
\dff nR - \sum^Q_{j = 1} \int_{\R_j} p(\bx) \log\frac{\hat q(\bx|\by_j)}{p(\bx)} d\bx.
\ee
Then by \emph{Corollary 3}, for non-singular optimal reconstruction pdf we have
\be \label{equ.DistortionOpt150}
D^n(q_Q) - D^* = \frac{\hat \lambda}{n}\Delta D^n(q_Q).
\ee

\emph{Remark 1:}
For a binary symmetric source, the optimal reconstruction pdf is given by
$\hat q(y|x) = q_0 \mathbf{1}_{y \neq x} + (1 - q_0) \mathbf{1}_{y = x}$,
where $q_0$ is determined by the relation $H(q_0) \dff - q_0 \log_2 q_0 - (1 - q_0) \log_2 (1 - q_0) = R$.
For a Gaussian source, the optimal reconstruction pdf is determined by the conditional distribution $\hat q(x|y) \sim {\cal N}(y,\sigma^2)$,
where $\sigma^2$ is such that $I(\hat q) = R$.
Hence for both cases, the optimal reconstruction pdfs are non-singular.

\emph{Remark 2:} For sources with non-singular reconstruction pdfs, the regions $\{\R_j\}^Q_{j = 1}$
that minimize the distortion $D^n(q_Q)$ in (\ref{equ.DistortionS4})
and those that minimize $\Delta D^n(q_Q)$ in (\ref{equ.DistortionOpt15}) are equivalent.
To see this, note that due to (\ref{equ.Different41}), we have
\be \label{equ.DistortionOpt18}
&&d(x_i , y_i) + \hat \lambda \log\frac{\hat q(x_i|y_i)}{p(x_i)} + \frac{\hat v(x_i)}{p(x_i)} \nonumber \\
&=& d(x_i , y_i) + \hat \lambda \log\frac{\hat q(y_i|x_i)}{q(y_i)} + \frac{\hat v(x_i)}{p(x_i)} = 0, \ \ \mbox{for} \ i=1, ..., n.
\ee
Summing up the above equations over $i$, we have that for $\bx \in \X^n$ and $\by \in \Y^n$,
\be \label{equ.DistortionOpt18}
n d(\bx,\by) + \hat \lambda \log\frac{\hat q(\bx|\by)}{p(\bx)} = - \sum^n_{i = 1} \frac{\hat v(x_i)}{p(x_i)}.
\ee
It then follows from (\ref{equ.DistortionOpt18}) that the $\by_j \in \Y_Q$ that minimizes $d(\bx,\by_j)$ is the one
that maximizes $\log\frac{\hat q(\bx|\by_j)}{p(\bx)}$, and thus minimizes $\Delta D^n(q_Q)$.

\subsection{Two Looser Lower Bounds on Quantization Distortion} \label{subsec.Duality}
We next give two looser lower bounds on the quantization distortion based on \emph{Corollary 3},
which are derived using the following simple inequality
\be \label{equ.ExponentialEQ}
G - 1 + e^{-G} \geq 0, \ \forall G \in \mathbb{R}.
\ee

\subsubsection{A Lower Bound from the Source-Channel Duality}
Consider the equivalent channel corresponding to the rate-distortion source model,
where the input to the channel is a codeword from $\Y_Q = \{\by_j\}_{1 \leq j \leq Q}$,
and the memoryless channel is characterized by $\hat q(x|y)$.

Assume that the probability of transmitting each codeword is $1 / Q$.
Then the decoding rule is given by $\hat\by = \argmax_{\by_k} \hat q(\by_k|\bx) = \argmax_{\by_k} \hat q(\bx|\by_k)$.
For non-singular reconstruction pdf, according to (\ref{equ.DistortionOpt15}),
the decoding region for $\by_j$, $1 \leq j \leq Q$, is exactly $\R_j$.
Then, the decoding error probability is given by
\be \label{equ.Duality1}
p_e = 1 - \frac{\sum^Q_{j = 1}\int_{\R_j} \hat q(\bx|\by_j) d\bx}{Q}.
\ee

Define
\be \label{equ.Duality11}
G(\bx) \dff \ln \frac{p(\bx)}{\frac{1}{Q} \max_{1 \leq j \leq Q} \hat q(\bx|\by_j)}.
\ee
Since $\R_j$ is the optimal quantization region for $\by_j$, according to (\ref{equ.DistortionOpt15})
we have $\hat q(\bx|\by_j) = \max_{1 \leq k \leq Q} \hat q(\bx|\by_k)$ for $\bx \in \R_j$, and thus
\be \label{equ.Duality12}
\Delta D^n(q_Q)
&=& nR + \sum^Q_{j = 1} \int_{\R_j} p(\bx) \log\frac{p(\bx)}{\hat q(\bx|\by_j)} d\bx \nonumber \\
&=& \sum^Q_{j = 1} \int_{\R_j} p(\bx) \log\frac{p(\bx)}{\frac{1}{Q} \max_{1 \leq k \leq Q} \hat q(\bx|\by_k)} d\bx = \frac{1}{\ln 2}\int_{\X^n} p(\bx)G(\bx)d\bx \nonumber \\
&\geq& \frac{1}{\ln 2}\int_{\X^n} p(\bx)\Big(1 - e^{-G(\bx)}\Big)d\bx
= \frac{1}{\ln 2}\Big(1 - \int_{\X^n} \frac{1}{Q} \max_{1 \leq j \leq Q} \hat q(\bx|\by_j)d\bx\Big) \label{equ.Duality13} \\
&=& \frac{1}{\ln 2} \Big(1 - \frac{\sum^Q_{j = 1}\int_{\R_j} \hat q(\bx|\by_j) d\bx}{Q}\Big) = \frac{p_e}{\ln 2},
\ee
where (\ref{equ.Duality13}) follows from (\ref{equ.ExponentialEQ}).

\begin{theorem} \label{thm.Thm4}
If the optimal reconstruction pdf is non-singular, then the quantization residue $\Delta D^n(q_Q)$ in (\ref{equ.DistortionOpt15})
and the decoding error probability $p_e$ in (\ref{equ.Duality1}) are related as follows
\be \label{equ.Duality2}
\Delta D^n(q_Q) \geq \frac{p_e}{\ln 2}.
\ee
\end{theorem}
$\hfill \Box$

\subsubsection{A Further Lower Bound}
From (\ref{equ.Duality13}) we have that $\frac{1}{\ln 2}\int_{\X^n} p(\bx)\Big(1 - e^{-G(\bx)}\Big)d\bx \geq 0$.
Then, using (\ref{equ.Duality13}) we have
\be \label{equ.Computation3}
\Delta D^n(q_Q) &\geq& \frac{1}{\ln 2}\int_{\X^n} p(\bx) G(\bx) d\bx - \frac{1}{\ln 2}\int_{\X^n} p(\bx) \Big(1 - e^{-G(\bx)}\Big) d\bx \nonumber \\
&=& \frac{1}{\ln 2}\int_{\X^n} p(\bx) \Big(G(\bx) - 1 + e^{-G(\bx)}\Big) d\bx \nonumber \\
&=& \frac{1}{\ln 2} \mathbb{E}\Big(G(\bx) - 1 + e^{-G(\bx)}\Big) \nonumber \\
&=& \frac{1}{\ln 2} \int^{+\infty}_{0} \mathbb{P}\Big(G(\bx) - 1 + e^{-G(\bx)} \geq \lambda\Big) d\lambda,
\ee
where (\ref{equ.Computation3}) follows from the fact that for a non-negative random variable $Z$, $E(Z) = \int^{+\infty}_{0} \mathbb{P}\Big(Z \geq \lambda\Big)d\lambda$.
Based on the above arguments, we provide the following lower bound.

\begin{theorem} \label{thm.Thm40}
If the optimal reconstruction pdf $\hat q(y|x)$ is non-singular, then a lower bound on $\Delta D^n(q_Q)$ is given by (\ref{equ.Computation3}),
which is nonnegative.
\end{theorem}

Theorem~\ref{thm.Thm40} can be used to derive a lower bound on $\Delta D^n(q_Q)$ for a quantization codebook $\Y_Q$,
that is constrained to be in a subset of $\X^n$.
This will be illustrated for computing the lower bound on the quantization distortion for Gaussian sources in Section~\ref{sec.GaussianSource}.

When the source alphabet  is discrete and finite, then $G(\bx) - 1 + e^{-G(\bx)}$ also takes finite number of values,
denoted as $\lambda_k$, $k = 1, 2, ..., M$ such that $\lambda_1 < \lambda_2 < ... < \lambda_M$. Then the lower bound in (\ref{equ.Computation3}) becomes
\be \label{equ.DistortionC9}
\Delta D^n(q_Q) \geq \sum_{j=1}^{M-1} (\lambda_{j+1} - \lambda_{j})\mathbb{P}\Big(G(\bx) - 1 + e^{-G(\bx)} > \lambda_{j+1}\Big),
\ee
which can be used to derive a lower bound for the quantization distortion when the quantization codebook $\Y_\Q \subset \X^n$.

An example for the application of (\ref{equ.Computation3}) is the lower bound for the quantization distortion for Gaussian source with bounded codewords,
which is given in Section~\ref{sec.GaussianSource}.

\subsection{More Properties of $\Delta D^n(q_Q)$} \label{subsec.SymmetricCase}
We now provide another lower bound on $\Delta D^n(q_Q)$ for symmetric reconstruction alphabet.
Given the codebook $\Y_Q = \{\by_1, \by_2, ..., \by_Q\}$ and $\gamma \in \mathbb{R}^+$,
we define the following three regions
\be \label{equ.Symmetric4}
\A_j(\gamma) \dff \Big\{\bx: \frac{\hat q(\bx|\by_j)}{p(\bx)} > \gamma \Big\}, \
\bar \A_j(\gamma) \dff \Big\{\bx: \frac{\hat q(\bx|\by_j)}{p(\bx)} = \gamma \Big\}, \
\A^c_j(\gamma) \dff \Big\{\bx: \frac{\hat q(\bx|\by_j)}{p(\bx)} < \gamma \Big\}.
\ee
We then have the following result for the second term on the right-hand side of (\ref{equ.DistortionOpt15}), which leads to
a lower bound on $\Delta D^n(q_Q)$.

\begin{theorem} \label{thm.Thm51}
Given the codebook $\Y_Q$ and the regions $\A_j(\gamma)$,
$\bar \A_j(\gamma)$, and $\A^c_j(\gamma)$ given by (\ref{equ.Symmetric4}), we have
\be \label{equ.Symmetric5}
&&\sum^Q_{j = 1} \int_{\R_j} p(\bx) \log\frac{\hat q(\bx|\by_j)}{p(\bx)} d\bx \nonumber \\
&\leq& \sum^Q_{j = 1} \int_{\A_j(\gamma)} p(\bx) \log\frac{\hat q(\bx|\by_j)}{p(\bx)} d\bx
+ \alpha \sum^Q_{j = 1} \int_{\bar \A_j(\gamma)} p(\bx) \log\frac{\hat q(\bx|\by_j)}{p(\bx)} d\bx,
\ee
where $\gamma > 0$ is such that
\be \label{equ.Symmetric60}
\gamma = \sup_{\beta}\{ \sum^Q_{j = 1} \int_{\A_j(\beta)} p(\bx) d\bx \leq 1 \};
\ee
and $0 \leq \alpha \leq 1$ is such that
\be \label{equ.Symmetric6}
\sum^Q_{j = 1} \int_{\A_j(\gamma)} p(\bx) d\bx + \alpha \sum^Q_{j = 1} \int_{\bar \A_j(\gamma)} p(\bx) d\bx = 1.
\ee

\begin{proof}
Denote $\tilde \R_j = \A_j(\gamma) \setminus \R_j$, $\bar \R_j = \R_j \cap \bar \A_j(\gamma)$, and $\R^c_j = \R_j \cap \A^c_j(\gamma)$.
We have
\be \label{equ.Symmetric6}
0 &=& \sum^Q_{j = 1} \int_{\A_j(\gamma)} p(\bx) d\bx + \sum^Q_{j = 1} \int_{\bar \A_j(\gamma)} \alpha p(\bx) d\bx
- \sum^Q_{j = 1} \int_{\R_j} p(\bx) d\bx \nonumber \\
&=& \sum^Q_{j = 1} \int_{\tilde \R_j} p(\bx) d\bx
+ \sum^Q_{j = 1} \Big(\int_{\bar \A(\by_j , \gamma)} \alpha p(\bx) d\bx - \int_{\bar \R_j} p(\bx)d\bx\Big) \nonumber \\
&& \ \ \ - \sum^Q_{j = 1} \int_{\R^c_j} p(\bx) d\bx.
\ee

Then, using (\ref{equ.Symmetric6}), we can write
\be \label{equ.Symmetric7}
&&\sum^Q_{j = 1} \int_{\A_j(\gamma)} p(\bx) \log\frac{\hat q(\bx|\by_j)}{p(\bx)} d\bx
+ \alpha \sum^Q_{j = 1} \int_{\bar \A_j(\gamma)} p(\bx) \log\frac{\hat q(\bx|\by_j)}{p(\bx)} d\bx \nonumber \\
&& \ \ \ - \sum^Q_{j = 1} \int_{\R_j} p(\bx) \log\frac{\hat q(\bx|\by_j)}{p(\bx)} d\bx \nonumber \\
&=& \sum^Q_{j = 1} \int_{\tilde \R_j} p(\bx) \log\frac{\hat q(\bx|\by_j)}{p(\bx)} d\bx
+ \sum^Q_{j = 1} \Big(\int_{\bar \A_j(\gamma)} \alpha p(\bx) \log\frac{\hat q(\bx|\by_j)}{p(\bx)} d\bx
- \int_{\bar \R_j} p(\bx) \log\frac{\hat q(\bx|\by_j)}{p(\bx)} d\bx\Big)\nonumber \\
&& \ \ \ - \sum^Q_{j = 1} \int_{\R^c_j} p(\bx) \log\frac{\hat q(\bx|\by_j)}{p(\bx)} d\bx \nonumber \\
&\geq& \log\gamma \sum^Q_{j = 1} \int_{\tilde \R_j} p(\bx) d\bx
+ \log\gamma \sum^Q_{j = 1} \Big(\int_{\bar \A_j(\gamma)} \alpha p(\bx)d\bx
- \int_{\bar \R_j} p(\bx)d\bx\Big)\nonumber \\
&& \ \ \ - \log\gamma \sum^Q_{j = 1} \int_{\R^c_j} p(\bx)d\bx = 0.
\ee
\end{proof}
\end{theorem}

Theorem~\ref{thm.Thm51} leads to a lower bound on $\Delta D^n(q_Q)$
that depends on the codewords $\{\by_j\}_{1 \leq j \leq Q}$.
In the following we consider sources with a symmetric property, under which the above lower bound can be
simplified and no longer depends on the codewords.

\subsubsection*{Symmetric Reconstruction Alphabet $\Y^n$}
According to the form of $\Delta D^n(q_Q)$ in (\ref{equ.DistortionOpt15}), for any $\by_j \in \Y^n$, we define the following function
\be \label{equ.Symmetric3}
\theta(\by_j , \epsilon) \dff \left\{
\begin{array}{lll}
  &&\max_{\B_j \subset \X^n} \int_{\bx \in \B_j} p(\bx) \log\frac{\hat q(\bx|\by_j)}{p(\bx)} d\bx, \\
  &&\mbox{s.t.} \ \int_{\B_j} p(\bx) d\bx = \epsilon.
\end{array}\right.
\ee
Intuitively, the optimal solution $\B_j$ defines a region $\B_j$ of $\bx$ with a probability mass $\epsilon$,
which contains the largest values of $\log\frac{\hat q(\bx|\by_j)}{p(\bx)}$.
The reconstruction alphabet is called \emph{symmetric} if $\theta(\by_j , \epsilon)$ does not depend on $\by_j$.
An example of the \emph{symmetric} reconstruction alphabet is the binary symmetric sources, where the
$p(\bx) = 2^{-n}$ and $\hat q(\bx|\by_j)$ is decreasing with the Hamming distance between $\bx$ and $\by_j$.
In that case, the optimal solution $\B_j$ for any $\by_j \in \{0,1\}^n$ is a ball within some Hamming distance around $\by_j$.

We have the following result that characterizes
the solution to the optimization problem (\ref{equ.Symmetric3}).
The proof is similar to that of Theorem~\ref{thm.Thm51} and thus is omitted here.
The basic idea is that, given $p(\bx)$, the set $\B_j$ that maximizes $\int_{\B_j} p(\bx) \log\frac{\hat q(\bx|\by_j)}{p(\bx)} d\bx$
is the set of $\bx$ consisting of the largest values of $\log\frac{\hat q(\bx|\by_j)}{p(\bx)}$, specified by the set $\A_j(\gamma)$,
where $\gamma$ is determined by the constraint $\int_{\B_j} p(\bx) d\bx= \epsilon$.

\begin{theorem} \label{thm.Thm52}
The solution to (\ref{equ.Symmetric3}) is given by
\be \label{equ.Symmetric31}
\theta(\by_j , \epsilon) = \int_{\A_j(\gamma)} p(\bx) \log\frac{\hat q(\bx|\by_j)}{p(\bx)} d\bx
+ \alpha \int_{\bar \A_j(\gamma)} p(\bx) \log\frac{\hat q(\bx|\by_j)}{p(\bx)} d\bx,
\ee
where the parameters $\gamma > 0$ and $0 \leq \alpha < 1$ are determined by the following,
\be
\gamma &=& \sup\{\beta|\int_{\A_j(\beta)} p(\bx) d\bx \leq \epsilon\}, \label{equ.Symmetric32} \nonumber \\
\epsilon &=& \int_{\A_j(\gamma)} p(\bx) d\bx + \alpha \int_{\bar \A_j(\gamma)} p(\bx) d\bx, \label{equ.Symmetric321}
\ee
where $\A_j(\gamma)$ and $\bar \A_j(\gamma)$ are given by (\ref{equ.Symmetric4}).
\end{theorem}
$\hfill \Box$

The next result is on the concavity of $\theta(\by_j, \epsilon)$ and its proof is given in the Appendix.

\begin{theorem} \label{thm.Thm53}
Given $\by_j$, the function $\theta(\by_j, \epsilon)$ is concave in terms of $\epsilon$, i.e.,
for any $\epsilon_1,\epsilon_2 > 0$, and $0 < \beta < 1$, we have that
\be \label{equ.Symmetric8}
\theta(\by_j , \epsilon_\beta) \geq \beta\theta(\by_j , \epsilon_1) + (1 - \beta)\theta(\by_j , \epsilon_2),
\ \ \epsilon_\beta = \beta \epsilon_1 + (1 - \beta)\epsilon_2.
\ee

\begin{proof}
According to Theorem~\ref{thm.Thm53}, we can write, for $\kappa = 1,2$ and $\beta$,
\be \label{equ.Symmetric9}
\theta(\by , \epsilon_\kappa) = \int_{\A_j(\gamma_\kappa)} p(\bx) \log\frac{\hat q(\bx|\by_j)}{p(\bx)} d\bx
+ \alpha_\kappa \int_{\bar \A_j(\gamma_\kappa)} p(\bx) \log\frac{\hat q(\bx|\by_j)}{p(\bx)} d\bx,
\ee
where the parameters $\gamma_\kappa$ and $\alpha_\kappa$ satisfy
\be \label{equ.Symmetric10}
\gamma_\kappa &=& \sup\{\beta|\int_{\A_j(\beta)} p(\bx) d\bx \leq \epsilon_\kappa\}\nonumber \\
\epsilon_\kappa &=& \int_{\A_j(\gamma_\kappa)} p(\bx) d\bx + \alpha_\kappa \int_{\bar \A_j(\gamma_\kappa)} p(\bx) d\bx.
\ee
Note that from (\ref{equ.Symmetric8}) we have
\be \label{equ.Symmetric12}
\beta(\epsilon_\beta - \epsilon_1) - (1 - \beta)(\epsilon_2 - \epsilon_\beta) = 0.
\ee

According to (\ref{equ.Symmetric9}) and (\ref{equ.Symmetric12}), we have
\be \label{equ.Symmetric13}
&&\beta\Big(\theta(\by_j , \epsilon_\beta) - \theta(\by_j , \epsilon_1)\Big)
- (1 - \beta)\Big(\theta(\by_j , \epsilon_2) - \theta(\by_j , \epsilon_\beta)\Big) \nonumber \\
&\geq& \Big(\beta(\epsilon_\beta - \epsilon_1) - (1 - \beta)(\epsilon_2 - \epsilon_\beta)\Big) \log\gamma_\beta = 0.
\ee
Therefore, by (\ref{equ.Symmetric13}),
the convexity of $\theta(\by_j, \epsilon)$ is proved.
\end{proof}
\end{theorem}

For a symmetric reconstruction alphabet, we can write $\theta(\by , \epsilon) = \theta(\epsilon)$ for any $\by \in \Y^n$.
We have the following lower bound on $\Delta D^n(q_Q)$ for symmetric reconstruction alphabets.

\begin{theorem} \label{thm.Thm54}
If $\theta(\by , \epsilon) = \theta(\epsilon)$ for any $\by \in \Y^n$, then for any quantization function $q_Q$,
we have
\be \label{equ.Symmetric15}
\sum^Q_{j = 1} \int_{\R_j} p(\bx) \log\frac{\hat q(\bx|\by_j)}{p(\bx)} d\bx \leq Q \cdot \theta\Big(\frac{1}{Q}\Big).
\ee
Thus, we have the following lower bound for $D^n(q_Q)$,
\be
D^n(q_Q) &\geq& nR - Q \cdot \theta\Big(\frac{1}{Q}\Big) \label{equ.Symmetric151} \\
&\geq& 0.\label{equ.Symmetric151}
\ee
\begin{proof}
Denote $\zeta_j = \int_{\R_j}p(\bx) d\bx$ and then $\sum^Q_{j = 1}\zeta_j = 1$.
By the concavity of $\theta(\by, \epsilon)$ we have
\be \label{equ.Symmetric16}
\sum^Q_{j = 1} \int_{\R_j} p(\bx) \log\frac{\hat q(\bx|\by_j)}{p(\bx)} d\bx \leq \sum^Q_{j = 1}\theta(\zeta_j)
\leq Q \cdot \theta\Big(\frac{\sum^Q_{j = 1}\zeta_j}{Q}\Big) = Q \cdot \theta\Big(\frac{1}{Q}\Big).
\ee
Next we prove (\ref{equ.Symmetric151}). For any $\by \in \Y^n$, let $\B$ be its optimal region for the optimization problem (\ref{equ.Symmetric3}) for $\epsilon = \frac{1}{Q}$.
We let $T = \int_{\by \in \B} \hat q(\bx|\by) d\bx \leq 1$ and $t(\bx) = \frac{\hat q(\bx|\by)}{T}$.
Note that $\int_\B p(\bx) d\bx = \frac{1}{Q}$.
Then we have
\be \label{equ.Symmetric17}
nR - Q \cdot \theta\Big(\frac{1}{Q}\Big)
&=& nR + Q \int_{\B} p(\bx) \log\frac{p(\bx)}{\hat q(\bx|\by)} d\bx \nonumber \\
&=& \int_{\B} Q p(\bx) \log\frac{Q p(\bx)}{\hat q(\bx|\by)} d\bx \nonumber \\
&=& \int_{\B} Q p(\bx) \log\frac{Q p(\bx)}{t(\bx)} d\bx + \log\frac{1}{T}.
\ee
Note that both $Q\cdot p(\bx)$ and $t(\bx)$ are pdfs over $\B$.
Then, the first term of (\ref{equ.Symmetric17}) is the KL distance $D(Q p(\bx)||t(\bx))$ and the second terms is nonnegative.
Thus we have (\ref{equ.Symmetric151}).
\end{proof}
\end{theorem}

\section{Upper Bounds on Quantization Distortion} \label{sec.AchievableD}
\subsection{Existing Achievable Upper Bounds} \label{subsec.ExistingUB}
\subsubsection{Bounded Sources}
Assume that the source is bounded, i.e., $|x| < X \ \forall x \in {\cal X}.$
Consider a codebook $\Y_Q = \{\by_1,\by_2,...,\by_Q\}$ where $Q = 2^{nR}$.
In \cite{DiscreteSources} an upper bound on the quantization distortion is given considering a reference rate $R_0 < R$.
More specifically, let $\hat q_0(y|x)$ be the optimal solution to the rate-distortion problem (\ref{equ.RateDistortion}) with rate $R_0$,
and $D_0^* = D(R_0)$ be the corresponding distortion.
Assume that $\hat q_0(y|x)$ is non-singular. Let $\hat q_0(\by)$ be the corresponding marginal pdf of $\by$.
Consider random codebooks $\Y_Q$ of size $Q = 2^{nR}$, where $p(\Y_Q) = \prod^Q_{j = 1} \hat q_0(\by_j)$.
The expected quantization distortion is given by $\bar D^n_Q \dff \sum_{\Y_Q} p(\Y_Q) D(\Y_Q)$ over random codebooks $\Y_Q$,
The following result is found in~\cite{DiscreteSources} for \emph{bounded alphabets}.

\begin{theorem} \label{thm.Thm6}
Assume that the source is bounded and let $d_m = \max_{(x, y) \in \X \times \Y} d(x, y)$. Then, for any $0 < \epsilon < R - R_0$,
the distortion $\bar D^n_Q$ satisfies $\bar D^n_Q \leq D_0^* + d_m 2^{- (R - R_0 - \epsilon)n}$.
\end{theorem}
$\hfill \Box$

Hence there exists a quantization codebook $\Y_Q$ for which the distortion $D^n(\Y_Q) \leq D_0^* + d_m 2^{- (R - R_0 - \epsilon)n}$.
However, this upper bound is valid only for bounded sources and therefore not applicable to, e.g., Gaussian sources.
Since the above upper bound is based on a reference rate $R_0$, we call it the reference rate upper bound.

\subsubsection{Unbounded Sources}
For unbounded sources, it is shown in~\cite{GeneralRD} that if there exists $y_b \in \Y$, such that
\be \label{equ.UpperB32}
\int p(x) d(x , y_b) dx = \hat d < + \infty,
\ee
then for the distortion $D^*$ with respect to rate $R$, we have $\bar D^n_Q - D^* < \epsilon$ for sufficiently large block length $n$.
Note that (\ref{equ.UpperB32}) is a mild condition that is satisfied by, e.g., the Gaussian source with e.g., $y_b = 0$.
Recall that $\hat q(\by)$ is the pdf of $\by$ with respect to the optimal conditional pdf $\hat q(\by|\bx)$ for rate $R$.
In particular, consider a random codebook $\Y_Q = \{\by_1, \by_2, ..., \by_Q\}$,
where $\by_1$ is fixed to be $\by_1 \dff (y_b,y_b,...,y_b)$ and other codewords $\by_2, \by_3, ..., \by_Q$ are distributed according to
$p(\by_2, ..., \by_Q) = \prod^Q_{j = 2} \hat q(\by_j)$.
For any $\delta > 0$, we define the follow region $\B^n_\delta = \{\bx: d(\bx , \by_1) < \hat d + \delta\}$ and $\bar \B^n_\delta = \X^n \setminus \B^n_\delta$.

Denote the codebook other than the codewords $\by_1$ as $\Y_{Q - 1} \dff \{\by_2, ..., \by_Q\}$,
such that $p(\Y_{Q - 1}) = \prod^Q_{j = 2} \hat q(\by_j)$.
The average distortion is then,
\be \label{equ.UpperB321}
\bar D^n_Q = \int p(\bx) \int p(\Y_{Q-1}) d(\bx , \Y_{Q}) d\Y_{Q-1} d\bx.
\ee
The following result found in~\cite{GeneralRD} provides an upper bound on $\bar D_Q$.

\begin{theorem} \label{thm.Thm600}
We have
\be \label{equ.UpperB321}
\bar D^n_Q \leq  \int_{\B^n_\delta} p(\bx) \int p(\Y_{Q-1}) d(\bx , \Y_{Q}) d\Y_{Q - 1} d\bx
+ \int_{\bar \B^n_\delta} p(\bx) d(\bx , \by_b) d\bx.
\ee
Moreover, the second term can be made arbitrarily small for sufficiently large $n$.
\end{theorem}

Since the second term in (\ref{equ.UpperB321}) approaches zero for large $n$,
in the following we focus on the first term, denoted as
\be \label{equ.UpperB327}
\tilde D^n_Q = \int_{\B^n_\delta} p(\bx) \int p(\Y_{Q-1}) d(\bx , \Y_{Q}) d\Y_{Q - 1} d\bx.
\ee

We provide two upper bounds based on ordered statistics and reference rate, respectively.
The upper bound based on the reference rate is an improved version of that given in~\cite{DiscreteSources}.

\subsection{An Upper Bound Based on Ordered Statistics} \label{subsec.RefinedUBO}
Denote
\be \label{equ.Achieve22}
h_{Q - 1}(\bx) \dff \int p(\Y_{Q-1}) d(\bx , \Y_{Q}) d\Y_{Q - 1},
\ee
such that $\tilde D^n_Q = \int_{\B^n_\delta}p(\bx)h_{Q - 1}(\bx)d\bx$.
Next we give an upper bound on $h_{Q - 1}(\bx)$ based ordered statistics.

Since the quantization codewords $\by_j$, $2 \leq j \leq Q$, are chosen independently,
the distortions $d(\bx , \by_j)$, $2 \leq j \leq Q$, are $Q - 1$ independent random variables
with the cumulative distributive function
\be \label{equ.Achieve232}
\mathbb{P}\Big(d(\bx , \by_j) \leq d\Big) = \int_{\by: d(\bx , \by) \leq d} \hat q(\by) d\by \dff F(\bx , d).
\ee
Denote $\bar F(\bx , d) \dff 1 - F(\bx , d)$.
Then we have
\be \label{equ.Achieve233}
d(\bx , \Y_Q) = \min\Big\{d(\bx , \by_1), \min_{2 \leq j \leq Q} d(\bx , \by_j)\Big\}.
\ee
Based on this, we have the following expression for $h_{Q - 1}(\bx)$.

\begin{theorem} \label{thm.Thm601}
We have
\be \label{equ.Achieve234}
h_{Q - 1}(\bx) = \int^{d(\bx , \by_b)}_{0} \bar F^{Q - 1}(\bx , t) dt.
\ee

\begin{proof}
Since $d(\bx,\by) \geq 0$, we have
\be \label{equ.Achieve235}
h_{Q - 1}(\bx) = \mathbb{E}\Big(d(\bx , \Y_Q)\Big) = \int^{+\infty}_{0} \mathbb{P}\Big(d(\bx , \Y_Q) \geq t\Big) dt.
\ee
Moreover,
\be \label{equ.Achieve236}
\mathbb{P}\Big(d(\bx , \Y_Q) \geq t\Big)
&=& \mathbb{P}\Big( \min_{2 \leq j \leq Q} d(\bx , \by_j) \geq t\Big) \cdot \mathbf{1}_{d(\bx , \by_b) \geq t} \nonumber \\
&=& \prod^Q_{j = 2}\mathbb{P}\Big(d(\bx , \by_j) \geq t\Big) \cdot \mathbf{1}_{d(\bx , \by_b) \geq t} \nonumber \\
&=& \bar F^{Q - 1}(\bx , t) \cdot \mathbf{1}_{d(\bx , \by_b) \geq t}.
\ee
Substituting (\ref{equ.Achieve236}) into (\ref{equ.Achieve235}) we have
\be \label{equ.Achieve237}
h_{Q - 1}(\bx) &=& \int^{+\infty}_{0} \bar F^{Q - 1}(\bx , t) \cdot \mathbf{1}_{d(\bx , \by_b) \geq t} dt \nonumber \\
&=& \int^{d(\bx , \by_b)}_{0} \bar F^{Q - 1}(\bx , t) dt.
\ee
\end{proof}
\end{theorem}

In order to bound $h_{Q - 1}(\bx)$ in (\ref{equ.Achieve237}) in a more efficient manner, we
divide the interval $[0, d(\bx, \by_b)]$ into two parts $[0, t_\bx]$ and $[t_\bx, d(\bx, \by_b)]$, such that
\be \label{equ.Achieve2370}
\bar F^{Q - 1}(\bx , t) &\leq& \bar F(\bx , t) \cdot \epsilon, \ \ \mbox{for} \ t_\bx \leq t \leq d(\bx, \by_b),
\ee
for some small $\epsilon > 0$. Then we can write
\be \label{equ.Achieve23700}
h_{Q - 1}(\bx) \leq \int^{t_\bx}_{0} \bar F(\bx , t) dt + \epsilon\int^{d(\bx , \by_b)}_{t_\bx} \bar F(\bx , t) dt.
\ee

To this end we need to find a threshold $t_\bx$ such that (\ref{equ.Achieve2370}) is satisfied. Define
\be \label{equ.Achieve238}
t_\bx \dff \inf\{t: \bar F^{Q - 2}(\bx , t) \leq \epsilon\}.
\ee
We give the following Theorem~\ref{thm.Thm602} which provides a formal upper bound for $h_{Q - 1}(\bx)$.

\begin{theorem} \label{thm.Thm602}
Given the threshold $t_\bx$ specified in (\ref{equ.Achieve238}), we have the following upper bound for $h_{Q - 1}(\bx)$:
\begin{itemize}
  \item If $t_\bx \geq d(\bx , \by_b)$, then a trivial upper bound is given by $h_{Q - 1}(\bx) \leq d(\bx , \by_b)$;
  \item and if $t_\bx < d(\bx , \by_b)$, then
\be \label{equ.Achieve2310}
h_{Q - 1}(\bx) \leq (1 - \epsilon) t_{\bx} + \epsilon \int \hat q(\by) d(\bx , \by) d\by.
\ee
\end{itemize}
Furthermore, we have
\be \label{equ.Achieve2312}
\tilde D^n_Q \leq \int_{\B^n_\delta} p(\bx) \min\{d(\bx,\by_b), t_\bx\} d\bx
+ \epsilon \int \int p(\bx)\hat q(\by) d(\bx , \by) d\bx d\by.
\ee

\begin{proof}
If $t_\bx \geq d(\bx , \by_b)$, we have
\be \label{equ.Achieve2312}
h_{Q - 1}(\bx) = \int^{d(\bx , \by_b)}_{0} \bar F^{Q - 1}(\bx , t) dt \leq \int^{d(\bx , \by_b)}_{0} 1 \; dt = d(\bx , \by_b).
\ee

For $t_\bx < d(\bx , \by_b)$, from (\ref{equ.Achieve234}) we have the following,
\be \label{equ.Achieve270}
h_{Q - 1}(\bx) = \int^{t_{\bx}}_{0} \bar F^{Q - 1}(\bx , t) dt + \int^{d(\bx , \by_b)}_{t_{\bx}} \bar F^{Q - 1}(\bx , t) dt.
\ee
Since $\bar F(\bx , t) \leq 1$ for $0 \leq t \leq t_{\bx}$ and $\bar F^{Q - 2}(\bx , t) \leq \epsilon$ for $t \geq t_\bx$, then
\be \label{equ.Achieve271}
h_{Q - 1}(\bx) &\leq& \int^{t_{\bx}}_{0} \bar F(\bx , t) dt +  \epsilon\int^{+\infty}_{t_{\bx}} \bar F(\bx , t) dt \nonumber \\
&=& (1 - \epsilon) \int^{t_{\bx}}_{0} \bar F(\bx , t) dt + \epsilon\int^{+\infty}_{0} \bar F(\bx , t) dt \nonumber \\
&\leq& (1 - \epsilon) t_{\bx} + \epsilon \int \hat q(\by) d(\bx , \by) d\by.
\ee

Combining (\ref{equ.Achieve2312}) and (\ref{equ.Achieve271}), we obtain
\be \label{equ.Achieve272}
h_{Q - 1}(\bx) \leq  \min\{d(\bx,\by_b), t_\bx\} + \epsilon \int \hat q(\by) d(\bx , \by) d\by.
\ee
Finally $\tilde D_Q$ in (\ref{equ.UpperB321}) can be upper bounded as,
\be \label{equ.Achieve273}
\tilde D^n_Q &\leq& \int_{\B^n_\delta} p(\bx) h_{Q - 1}(\bx) d\bx \nonumber \\
&\leq& \int_{\B^n_\delta} p(\bx) \min\{d(\bx,\by_b), t_\bx\} d\bx
+ \epsilon \int_{\B^n_\delta} \int p(\bx) \hat q(\by) d(\bx , \by) d\by d\bx \nonumber \\
&\leq& \int_{\B^n_\delta} p(\bx) \min\{d(\bx,\by_b), t_\bx\} d\bx
+ \epsilon \int \int p(\bx) \hat q(\by) d(\bx , \by) d\by d\bx.
\ee
\end{proof}
\end{theorem}

Next we show that the upper bound given by (\ref{equ.Achieve2312}) can be arbitrarily close to the rate-distortion bound.
Assume that the reconstruction pdf $\hat q(y|x)$ and quantization codebook size $Q$
satisfy the following
\be \label{equ.Achieve274}
I(\hat q) = R, \ \mbox{and} \ \ Q = 2^{nR};
\ee
and that the following condition is satisfied,
\be \label{equ.Achieve275}
\int \int p(\bx) \hat q(\by) d(\bx , \by) d\by d\bx \dff \tilde D_0 < \infty.
\ee
We have the following result.

\begin{theorem} \label{thm.Thm603}
For any quantization rate $R_0 < R$ with the quantization distortion $D(R_0)$ from the rate distortion bound.
For any $\zeta > 0$, for sufficiently large quantization block length $n$, we have
\be \label{equ.Achieve276}
\tilde D^n_Q &<& D(R_0) + \zeta.
\ee
\end{theorem}
$\hfill \Box$

\subsection{An Improved Upper Bound based on Reference Rate} \label{subsec.RefinedUB}
In this section we provided an improved upper bound based on the reference rate, which can be proved to be tighter than
the upper bound based on the reference rate given in~\cite{DiscreteSources}.
We further analyze the term $\tilde D^n_Q$ given by (\ref{equ.UpperB327}).
We follow the main idea of~\cite{DiscreteSources}, which adds another codeword $\by_0$ into the current codebook yielding the optimal conditional distribution
$\hat q_0(\by_0|\bx)$ ($\by_0 \neq \by_b$) for another quantization rate $R_0 < R$,
and all other codewords yielding the independent distribution $\hat q_0(\by) = \int_{\bx} \hat q_0(\by|\bx) p(\bx) d\bx$.
For codebooks $\Y_Q$ and $\Y_{Q-1}$, we define the indicators $\Phi(\bx , \by_0 , \Y_Q) \dff \mathbf{1}_{d(\bx , \by_0) < d(\bx , \Y_Q)}$,
and $\Phi(\bx , \by_0 , \Y_{Q - 1}) \dff \mathbf{1}_{d(\bx , \by_0) < d(\bx , \Y_{Q-1})}$.
We have the following upper on the distortion gap $\tilde D^n_Q - D(R_0)$.

\begin{theorem} \label{thm.Thm60}
We have
\be \label{equ.UpperB33}
\tilde D^n_Q - D(R_0) \leq \int_{\B^n_\delta} p(\bx) h(\bx) d\bx,
\ee
where
\be \label{equ.UpperB34}
h(\bx) \dff \int \int \hat q_0(\by_0|\bx) p(\Y_{Q-1})
\Big(d(\bx , \tilde \Y_{Q}) -  d(\bx,\by_0)\Big) \Phi(\bx , \by_0 , \tilde \Y_{Q}) d\Y_{Q-1} d\by_0.
\ee

\begin{proof}
According to the definition of $\tilde D^n_Q$ and $D(R_0)$, we have the following
\be \label{equ.UpperB6}
&&\tilde D^n_Q - D(R_0) \nonumber \\
&=& \int_{\B^n_\delta} p(\bx) \int \hat p(\Y_{Q-1}) d(\bx , \Y_Q) d\bx d\Y_{Q-1}
- \int \int p(\bx) \hat q_0(\by_0|\bx) d(\bx,\by_0) d\bx d\by_0 \nonumber \\
&\leq& \int_{\B^n_\delta} p(\bx) \int \hat p(\Y_{Q-1}) d(\bx , \Y_{Q}) d\bx d\Y_{Q-1}
- \int_{\B^n_\delta} \int p(\bx) \hat q_0(\by_0|\bx) d(\bx,\by_0) d\bx d\by_0 \nonumber \\
&=& \int_{\B^n_\delta} p(\bx) d\bx \int \int \Big(\hat p(\Y_{Q-1}) d(\bx , \Y_{Q})
- \hat q_0(\by_0|\bx) d(\bx,\by_0)\Big) d\Y_{Q-1} d\by_0 \nonumber \\
&=& \int_{\B^n_\delta} p(\bx) d\bx \int \int \hat p(\Y_{Q-1}) \hat q_0(\by_0|\bx) \Big( d(\bx, \Y_{Q}) - d(\bx,\by_0)\Big)
d\Y_{Q-1} d\by_0.
\ee
Then, since
\be \label{equ.UpperB10}
\Big(d(\bx , \Y_Q) -  d(\bx,\by_0)\Big) \leq \Big(d(\bx , \Y_Q) -  d(\bx,\by_0)\Big) \Phi(\bx , \by_0 , \Y_Q),
\ee
from (\ref{equ.UpperB6}) we have the following
\be \label{equ.UpperB9}
&&\tilde D^n_Q - D(R_0) \nonumber \\
&\leq& \int_{\B^n_\delta} p(\bx) \int \int \hat q_0(\by_0|\bx) p(\Y_{Q-1})
\Big(d(\bx , \Y_{Q}) -  d(\bx,\by_0)\Big) \Phi(\bx , \by_0 , \Y_{Q}) d\bx d\by_0 d\Y_{Q-1} \nonumber \\
&=& \int_{\B^n_\delta} p(\bx) h(\bx) d\bx.
\ee
\end{proof}
\end{theorem}

We are interested in bounding the term $h(\bx)$.
We first define a dual of $h(\bx)$, and then analyze the dual using ordered statistics.
Finally we bound $h(\bx)$ based on the dual.

\subsubsection{Dual of $h(\bx)$}
define
\be \label{equ.UpperB7}
l(\bx , \by_0) \dff \int p(\Y_{Q-1}) \Big(d(\bx , \Y_{Q}) -  d(\bx,\by_0)\Big) \Phi(\bx , \by_0 , \Y_{Q}) d\Y_{Q-1},
\ee
such that
\be \label{equ.UpperB71}
h(\bx) = \int \hat q(\by_0|\bx) l(\bx , \by_0) d\by_0.
\ee
Next, we define a dual of $h(\bx)$ as follows
\be \label{equ.UpperB8}
\tilde h(\bx) \dff \int \hat q(\by_0) l(\bx , \by_0) d\by_0.
\ee

\subsubsection{Analysis of $\tilde h(\bx)$ Based on Ordered Statistics}
Given $\bx$, we consider the random variable $d(\bx, \by)$ with $\by \sim \hat q(\by)$.
Denote $d^{1}_\bx = (\bx , \by_0)$ and $d^{k}_\bx = d(\bx , \by_k)$ for $2 \leq k \leq Q$, then $d^k_\bx$ are i.i.d. random variables.
Suppose that they are ranked as $d^{(1)}_\bx \leq d^{(2)}_\bx \leq ... \leq d^{(Q)}_\bx$.
The following result upper bounds $\tilde h(\bx)$ using the ordered statistics of $\{d^{(k)}_\bx\}^Q_{k = 1}$.

\begin{theorem} \label{thm.Thm7}
We have
\be \label{equ.UpperB11}
\tilde h(\bx) \leq \frac{\mathbb{E}\Big(d^{(2)}_\bx\Big) - \mathbb{E}\Big(d^{(1)}_\bx\Big)}{Q}.
\ee

\begin{proof}
Note that we have the following
\be \label{equ.UpperB121}
\Big(d(\bx , \Y_{Q}) -  d(\bx,\by_0)\Big) \Phi(\bx , \by_0 , \Y_{Q})
&\leq& \Big(d(\bx , \Y_{Q - 1}) -  d(\bx,\by_0)\Big) \Phi(\bx , \by_0 , \Y_{Q}) \nonumber \\
&\leq& \Big(d(\bx , \Y_{Q - 1}) -  d(\bx,\by_0)\Big) \Phi(\bx , \by_0 , \Y_{Q - 1}),
\ee
and thus we have the following
\be \label{equ.UpperB122}
\tilde h(\bx) \leq \int \int \hat q_0(\by_0) p(\Y_{Q-1})
\Big(d(\bx , \Y_{Q-1}) -  d(\bx,\by_0)\Big) \Phi(\bx , \by_0 , \Y_{Q-1}) d\by_0 d\Y_{Q-1}.
\ee

In the following we rewrite the right side of (\ref{equ.UpperB122}) using ordered statistics.
Let $\bar d^k_\bx \dff \min_{l \neq k} d^l_\bx$ for $1 \leq k \leq Q$. According to (\ref{equ.UpperB8}), we have that
\be \label{equ.UpperB123}
\tilde h(\bx) &=& \int \hat q_0(\by_0) \prod^Q_{j = 2} \hat q_0(\by_j)
\Big(d(\bx , \Y_{Q - 1}) -  d(\bx,\by_0)\Big) \Phi(\bx , \by_0 , \Y_{Q - 1}) d\by_0d\by_2...d\by_Q \nonumber \\
&=& \mathbb{E}\Big(\Big(d(\bx , \Y_{Q-1}) -  d(\bx,\by_0)\Big) \cdot \mathbf{1}_{d(\bx , \Y_{Q-1}) > d(\bx,\by_0)}\Big) \nonumber \\
&=& \mathbb{E}\Big((\bar d^1_\bx - d^1_\bx) \cdot \mathbf{1}_{\bar d^1_\bx > d^1_\bx}\Big).
\ee
On the other hand, since all distances $\rho^k_\bx$, $1 \leq k \leq Q$, are independent and identically distributed,
then we have that for all $2 \leq l \leq Q$,
\be \label{equ.UpperB13}
\mathbb{E}\Big((\bar d^l_\bx - d^l_\bx) \cdot \mathbf{1}_{\bar d^l_\bx > d^l_\bx}\Big) =
\mathbb{E}\Big((\bar d^1_\bx - d^1_\bx) \cdot \mathbf{1}_{\bar d^1_\bx > d^1_\bx}\Big).
\ee
Thus according to (\ref{equ.UpperB122}), we have the following
\be \label{equ.UpperB14}
\tilde h(\bx) \leq \frac{\sum^Q_{j = 1}\mathbb{E}\Big((\bar d^j_\bx - d^j_\bx) \cdot \mathbf{1}_{\bar d^j_\bx > d^j_\bx}\Big)}{Q}
= \frac{\mathbb{E}\Big(\sum^Q_{j = 1}(\bar d^j_\bx - d^j_\bx) \cdot \mathbf{1}_{\bar d^j_\bx > d^j_\bx}\Big)}{Q}.
\ee

Note that we have the following equation,
\be \label{equ.UpperB15}
\sum^Q_{j = 1}(\bar d^j_\bx - d^j_\bx) \cdot \mathbf{1}_{\bar d^j_\bx > d^j_\bx} = d^{(2)}_\bx - d^{(1)}_\bx.
\ee
Therefore, according to (\ref{equ.UpperB14}) and (\ref{equ.UpperB15}), we have
\be \label{equ.UpperB16}
\tilde h(\bx) \leq \frac{\mathbb{E}\Big(d^{(2)}_\bx - d^{(1)}_\bx\Big)}{Q}
= \frac{\mathbb{E}\Big(d^{(2)}_\bx\Big) - \mathbb{E}\Big(d^{(1)}_\bx\Big)}{Q}.
\ee
\end{proof}
\end{theorem}

Recall that we have already defined $\bar F(\bx , t) \dff \mathbb{P}\Big(d(\bx , \by) \geq t\Big)$.
The following result provides an analytical expression for
$\mathbb{E}\Big(d^{(2)}_\bx\Big) - \mathbb{E}\Big(d^{(1)}_\bx\Big)$.

\begin{theorem} \label{thm.Thm8}
Based on the above definition of $\bar F(t)$, we have that
\be \label{equ.UpperB180}
\mathbb{E}\Big(d^{(2)}_\bx\Big) - \mathbb{E}\Big(d^{(1)}_\bx\Big) = \int^{+\infty}_{0} Q \bar F^{Q-1}(\bx , t)\Big(1 - \bar F(\bx , t)\Big)dt.
\ee
Furthermore, for any arbitrarily small $\eta > 0$, we have that for sufficient large $Q$,
\be \label{equ.UpperB181}
\tilde h(\bx) < \frac{\int \hat q(\by) d(\bx, \by)d\by}{Q(e - \eta)},
\ee
where $e$ is the natural base.

\begin{proof}
From the property of ordered statistics, we have the following
\be \label{equ.UpperB18}
\mathbb{P}\Big(d^{(1)}_\bx \geq t\Big) &=& \prod^Q_{k = 1} \mathbb{P}\Big(d^{k}_\bx \geq t\Big) = \bar F^{Q}(\bx , t), \nonumber \\
\mbox{and} \ \ \mathbb{P}\Big(d^{(2)}_\bx \geq t\Big) &=& \prod^Q_{k = 1} \mathbb{P}\Big(d^{k}_\bx \geq t\Big)
+ \sum^Q_{k = 1} \mathbb{P}\Big(d^{k}_\bx < t\Big) \prod_{l \neq k} \mathbb{P}\Big(d^{l}_\bx \geq t\Big) \nonumber \\
&=& \bar F^{Q}(t) + Q \bar F^{Q - 1}(t)\Big(1 - \bar F(t)\Big).
\ee
Then, we have that
\be \label{equ.UpperB21}
\mathbb{E}\Big(d^{(2)}_\bx\Big) - \mathbb{E}\Big(d^{(1)}_\bx\Big)
&=& \int^{+\infty}_{0} \mathbb{P}\Big(d^{(2)}_\bx \geq t\Big) dt - \int^{+\infty}_{0} \mathbb{P}\Big(d^{(1)}_\bx \geq t\Big) dt \nonumber \\
&=& Q \int^{+\infty}_{0} \bar F^{Q - 1}(\bx , t)\Big(1 - \bar F(\bx , t)\Big) dt.
\ee

Next we consider $\bar F^{Q - 1}(\bx , t)\Big(1 - \bar F(\bx , t)\Big)$. Define $w(x) \dff x^{Q - 2}(1 - x)$.
Then $w(x)$ is maximized when $x = \frac{Q - 2}{Q - 1}$, and thus
\be \label{equ.UpperB22}
w(x) \leq w\Big(\frac{Q - 2}{Q - 1}\Big) = \frac{1}{Q - 1}\Big(\frac{Q - 2}{Q - 1}\Big)^{Q - 2}, \ \ \mbox{for} \ 0 \leq x \leq 1.
\ee
Then, we have that
\be \label{equ.UpperB23}
\mathbb{E}\Big(d^{(2)}_\bx\Big) - \mathbb{E}\Big(d^{(1)}_\bx\Big) &=& Q \int^{+\infty}_{0} \bar F(\bx , t) w\Big( \bar F(\bx , t) \Big) dt
\leq \frac{Q}{Q - 1}\Big(\frac{Q - 2}{Q - 1}\Big)^{Q - 2}  \int^{+\infty}_{0} \bar F(\bx , t) dt \nonumber \\
&=& \frac{Q}{Q - 1}\Big(\frac{Q - 2}{Q - 1}\Big)^{Q - 2} \mathbb{E}\Big(d(\bx , \by)\Big).
\ee
Since $\lim_{Q \rightarrow +\infty} \frac{Q}{Q - 1} \Big(\frac{Q - 1}{Q}\Big)^{Q} = \frac{1}{e}$,
for any $\eta > 0$, we have that for sufficient large $Q$,
\be \label{equ.UpperB23}
\tilde h(\bx) \leq \frac{\mathbb{E}\Big(d^{(2)}_\bx\Big) - \mathbb{E}\Big(d^{(1)}_\bx\Big)}{Q}
< \frac{\mathbb{E}\Big(d(\bx , \by)\Big)}{Q(e - \eta)}.
\ee
\end{proof}
\end{theorem}

According to Theorem~\ref{thm.Thm8}, we define
\be \label{equ.UpperB182}
B(\bx) \dff \frac{\int \hat q(\by) d(\bx, \by)d\by}{Q(e - \eta)},
\ee
as an upper bound for $\tilde h(\bx)$. Then, we provide an upper bound on $h(\bx)$.

\subsubsection{Refined Upper Bound for $h(\bx)$} \label{subsubsec.RefinedUB}
We first prove that $l(\bx , \by)$ is bounded for all $\bx \in \B^n_\delta$ and $\by \in \Y$.

\begin{theorem} \label{thm.Thm100}
For all $\bx \in \B^n_\delta$ and $\by \in \Y^n$, we have that $l(\bx , \by) \leq \hat d + \delta$.

\begin{proof}
Note that for all $\bx \in \B^n_\epsilon$ and $\by \in \Y^n$, we have
\be \label{equ.RefinedUB01}
\Big(d(\bx , \Y_Q) -  d(\bx,\by_0)\Big) \Phi(\bx , \by_0 , \Y_Q)
&\leq& d(\bx , \Y_Q)\Phi(\bx , \by_0 , \Y_Q) \leq d(\bx , \Y_Q) \leq d(\bx , \by_b) \nonumber \\
&\leq& \hat d + \delta.
\ee
Therefore we have
\be \label{equ.RefinedUB02}
l(\bx , \by_0) &=& \int p(\Y_Q) \Big(d(\bx , \Y_Q) -  d(\bx,\by_0)\Big) \Phi(\bx , \by_0 , \Y_Q) d\Y_{Q-1} \nonumber \\
&\leq& \int p(\Y_{Q-1}) \Big(\hat d + \epsilon \Big) d\Y_{Q-1} = \hat d + \epsilon.
\ee
\end{proof}
\end{theorem}

Note that from (\ref{equ.UpperB71}) and (\ref{equ.UpperB8}) we have that $h(\bx) = \int \hat q_0(\by_0|\bx)l(\bx , \by_0) d\by_0$,
where $\tilde h(\bx) = \int \hat q_0(\by_0)l(\bx , \by_0) d\by_0 \leq B(\bx)$ and $l(\bx , \by_0) \leq \hat d + \epsilon$.
To drive an upper bound on $h(\bx)$, we free $l(\bx , \by_0)$ as variables that can be optimized to maximize $\tilde h(\bx)$,
and formulate the following optimization problem
\be \label{equ.RefinedUB2}
h^U(\bx) =
\left\{
\begin{array}{ll}
  \max_{k(\bx , \by)} &\int \hat q(\by|\bx)k(\bx , \by) d\by \\
  \mbox{s.t.} & \int \hat q(\by)k(\bx , \by) d\by \leq B(\bx), \\
              &  k(\bx , \by) \leq \hat d + \epsilon, \ \mbox{for all} \ \by.
\end{array}\right.
\ee
$\hfill \Box$

The following Theorem~\ref{thm.Thm10} provides a solution to the above optimization problem (\ref{equ.RefinedUB2}).

\begin{theorem} \label{thm.Thm10}
Given $\bx$, we define the following region
\be \label{equ.RefinedUB4}
A_\bx(\gamma) \dff \Big\{\by: \frac{\hat q(\by|\bx)}{\hat q(\by)} > \gamma\Big\} \
\mbox{and} \ \ \bar A_\bx(\gamma) \dff \Big\{\by: \frac{\hat q(\by|\bx)}{\hat q(\by)} = \gamma\Big\}.
\ee
Assume a threshold $\gamma_\bx$ for which the following is satisfied:
\be \label{equ.RefinedUB5}
&&\gamma_\bx = \sup\{\gamma | (\hat d + \epsilon)\int_{A_\bx(\gamma)} \hat q(\by) d\by \leq B(\bx)\},\nonumber \\
\mbox{and} \ \ &&(\hat d + \epsilon)\int_{A_\bx(\gamma_\bx)} \hat q(\by) d\by + l_\bx\int_{\bar A_\bx(\gamma_\bx)} \hat q(\by) d\by = B(\bx),
\ee
where $0 \leq l_\bx \leq d + \epsilon$. Then, the upper bound $h^U(\bx)$ [c.f.(\ref{equ.RefinedUB2})] is given as follows
\be \label{equ.RefinedUB6}
h^U(\bx) = (d + \epsilon)\int_{A_\bx(\gamma_\bx)} \hat q(\by|\bx) d\by + l_\bx\int_{\bar A_\bx(\gamma_\bx)} \hat q(\by|\bx) d\by.
\ee

\begin{proof}
Similar to $A_\bx(\gamma)$ and $\bar A_\bx(\gamma)$ given in (\ref{equ.RefinedUB4}), we define
\be \label{equ.RefinedUB7}
A^c_\bx(\gamma) \dff \Big\{\by: \frac{q(\by|\bx)}{q(\by)} < \gamma\Big\} = \X \setminus \Big(A_\bx(\gamma) \cup \bar A_\bx(\gamma)\Big).
\ee
Then, since $\tilde h(\bx) \leq \tilde B(\bx)$, we have from (\ref{equ.UpperB8}) and (\ref{equ.RefinedUB2}) that
\be \label{equ.RefinedUB9}
&&\tilde h(\bx) - \tilde B(\bx) \nonumber \\
&=&\int_{A_\bx(\gamma_\bx)} \hat q(\by)\Big(l(\bx , \by) - (d + \epsilon)\Big) d\by
+ \int_{\bar A_\bx(\gamma_\bx)} \hat q(\by)\Big(l(\bx , \by) - l_\bx\Big)d\by \nonumber \\
&& \ \ + \int_{A^c_\bx(\gamma_\bx)} \hat q(\by)l(\bx , \by) d\by \leq 0.
\ee
Then, due to the definition of $A(\bx , \gamma_\bx)$, $\bar A(\bx , \gamma_\bx)$, and $A^c(\bx , \gamma_\bx)$,
we have that
\be \label{equ.RefinedUB11}
&&h(\bx) - h^U(\bx) \nonumber \\
&=&\int_{A_\bx(\gamma_\bx)} \hat q(\by|\bx)\Big(l(\bx , \by) - (d + \epsilon)\Big) d\by
+ \int_{\bar A_\bx(\gamma_\bx)} \hat q(\by|\bx)\Big(l(\bx , \by) - l_\bx\Big)d\by \nonumber \\
&& \ \ + \int_{A^c_\bx(\gamma_\bx)} \hat q(\by|\bx)l(\bx , \by) d\by \nonumber \\
&\leq& \int_{A_\bx(\gamma_\bx)} \gamma\hat q(\by)\Big(l(\bx , \by) - (d + \epsilon)\Big) d\by
+ \int_{\bar A_\bx(\gamma_\bx)} \gamma\hat q(\by)\Big(l(\bx , \by) - l_\bx\Big)d\by \nonumber \\
&& \ \ + \int_{A^c_\bx(\gamma_\bx)} \gamma\hat q(\by)l(\bx , \by) d\by \nonumber \\
&=& \gamma \cdot \Big(\tilde h(\bx) - \tilde B(\bx)\Big) \leq 0.
\ee
Therefore we have that $h(\bx) \leq h^U(\bx)$.
\end{proof}
\end{theorem}

According to Theorems~\ref{thm.Thm60} and~\ref{thm.Thm10}, we can bound the distortion gap $\tilde D^n_Q - D(R_0)$
as follows
\be \label{equ.RefinedUB12}
\tilde D^n_Q - D(R_0) \leq \int_{\B^n_\delta} p(\bx) h^U(\bx) d\bx.
\ee
We have the following result for $\tilde D^n_Q - D(R_0)$, which shows that the proposed upper bound is tighter than that provided in~\cite{DiscreteSources}.

\begin{theorem} \label{thm.Thm101}
We have
\be \label{equ.RefinedUB120}
\tilde D^n_Q - D(R_0) &\leq& \frac{d + \delta}{(e - \eta)^{1 - \beta}} \Big(2^{-(1 - \beta)R} \int \Big(\int \hat q_0(y) \hat q_0(x|y)^{1 / \beta} dy\Big)^\beta dx\Big)^n.
\ee
for sufficiently large quantization block length $n$.
Then, for any $\delta > 0$ the distortion gap $\tilde D^n_Q - D(R_0) < \delta$ for sufficiently large codeword block length $n$.
\end{theorem}
$\hfill \Box$

Note that in~\cite{DiscreteSources} an upper bound is given as follows,
\be \label{equ.RefinedUB121}
\tilde D^n_Q - D(R_0) &\leq& d_m \Big(2^{-(1 - \beta)R} \int \Big(\int \hat q_0(y) \hat q_0(x|y)^{1 / \beta} dy\Big)^\beta dx\Big)^n,
\ee
for bounded source where $d_m = \max_{(\bx,\by)}d(\bx,\by)$.
Then, for bounded source we can prove that $l(\bx,\by) \leq d_m$ and thus the following upper bound
\be \label{equ.RefinedUB122}
\tilde D^n_Q - D(R_0) &\leq& \frac{d_m}{(e - \eta)^{1 - \beta}} \Big(2^{-(1 - \beta)R} \int \Big(\int \hat q_0(y) \hat q_0(x|y)^{1 / \beta} dy\Big)^\beta dx\Big)^n,
\ee
which is tighter than that given in (\ref{equ.RefinedUB121}).

\emph{Remark 3 (Symmetric Cases):}
We consider a special case where the input alphabet is symmetric with respect to the output alphabet.
More specifically, we consider the input and output alphabets for which the following two conditions are satisfied,
\begin{enumerate}
  \item For any $\bx \in \X^n$, the expectation $\mathbb{E}\Big(d(\bx, \by)\Big)$ is a constant, not a function of $\bx$.
  \item For any $\bx \in \X^n$, for any $\gamma$, the probability $\mathbb{P}\Big(A_\bx(\gamma)\Big)$, $\mathbb{P}\Big(\bar A_\bx(\gamma)\Big)$,
  and $\mathbb{P}\Big(A^c_\bx(\gamma)\Big)$ under the distribution $\hat q_0(\by)$ for $\by$ is not a function of $\bx$.
\end{enumerate}
In this case, from the proof of Theorem~\ref{thm.Thm101} it is seen that for all $\bx$, the upper bound $h^U(\bx)$ are the same;
and thus we only need to compute the bound $h^U(\bx)$ for only one $\bx$, as the upper bound for $\tilde D^n_Q$.
An example of this special case is the binary symmetric source.

\subsection{Summary}
From the Algorithmic point of view, the upper bound based on the ordered statistics is easier to compute.
It only involves the optimal quantization conditional probability function for the current quantization rate $R$,
and thus does not need to consider another reference rate as the upper bound based on reference rate.
On the other hand, the upper bound based on ordered statistics depends on the selected reference rate,
and a good upper bound is the minimum among upper bounds for many selected reference rates,
which also significantly increases the computational complexity.

The computational complexity for these bounds depends on the type of sources under consideration.
For the binary symmetric source, since the source is symmetric over all source alphabets $\bx$, we can derive one $h^U(\bx)$ for one $\bx$ as the upper bound.
For the binary non-symmetric source, note that the source is symmetric over all source alphabet $\bx$ of the same weight,
we can sum up the $h^U(\bx)$ for the $\bx$ of all weights from $1,2$ to $n$.

\section{Binary Symmetric Sources} \label{sec.BinarySymSource}
Consider quantizing length-$n$ binary symmetric source sequences, with $p(\bx) = 2^{-n}$ for each sequence $\bx$.
Assume that the quantization alphabet size $Q = 2^{nR}$.
The optimal reconstruction pdf from the rate-distortion theory is given by,
\be \label{equ.BSS1}
\hat q(y|x) = \left\{
\begin{array}{ccc}
q, &&\mbox{if} \ y \neq x,\\
1 - q, &&\mbox{if} \ y = x,
\end{array}\right.
\ee
where $0 < q < 1/2$ and $1 - H(q) = R$.
The corresponding optimal distortion is $D^*=D(R)=q$.

\subsection{Lower Bound} \label{subsec.BinaryLB}
We apply the lower bounds obtained from Corollary~3 and Theorem~\ref{thm.Thm54} to binary symmetric sources.
Note that for binary uniform source, $\theta(\by_j,\epsilon)$ in (\ref{equ.Symmetric3}) is only a function of $\epsilon$.
Moreover, given the quantization codeword $\by_j$, we have that
\be \label{equ.BSSL1}
\log_2\frac{\hat q(\bx|\by_j)}{p(\bx)} = n + H(\bx, \by_j) \log_2 q + \Big(n - H(\bx, \by_j)\Big)\log_2 (1 - q),
\ee
where $H(\bx, \by_j)$ is the Hamming distance between $\bx$ and $\by_j$.
We have the following lower bound on the quantization distortion.
The main idea is to find a distance $D$ where the probability within distance $D$ to any quantization codeword $\by$ is $\frac{1}{Q}$.

\begin{theorem} \label{thm.Thm12}
For length-$n$ binary uniform source and size-$Q$ ($Q = 2^{nR}$) quantization codebook, we have that the distortion
\be \label{equ.BSSL2}
D^n(q_Q) \geq Q 2^{-n}\cdot \Big[\sum^{D - 1}_{j = 0} {n \choose j} \frac{j}{n} + \alpha {n \choose D}\Big],
\ee
where the distance $D$ and the fraction $\alpha$ are specified as follows
\be \label{equ.BSSL3}
&& D = \max\{d| \sum^{d - 1}_{j = 0} {n \choose j} \leq 2^{n(1 - R)}\}, \nonumber \\
\mbox{and} \ && \sum^{D - 1}_{j = 0} {n \choose j} + \alpha {n \choose D} = 2^{n(1 - R)}, \ \ 0 \leq \alpha < 1.
\ee

\begin{proof}
According to Theorem~\ref{thm.Thm54}, we have the following lower bound
\be \label{equ.BSSL4}
D^n(q_Q) \geq D^* + \frac{\hat \lambda}{n}\Big[nR - Q \cdot \theta\Big(\frac{1}{Q}\Big)\Big],
\ee
where $D^* = q$ and from simple calculation using Theorem~\ref{thm.Thm2}
\be \label{equ.BSSL5}
\hat \lambda = \frac{1}{\log_2\frac{1 - q}{q}}.
\ee

Note that, since $\log_2\frac{\hat q(\bx|\by_j)}{p(\bx)}$ decreases with $H(\bx, \by_j)$,
according to Theorem~\ref{thm.Thm54} we have that the optimal solution is given by,
\be \label{equ.BSSL50}
\theta\Big(\frac{1}{Q}\Big) = \sum_{\bx: H(\bx, \by_j) < D} p(\bx)\log_2\frac{\hat q(\bx|\by_j)}{p(\bx)}
+ \alpha \sum_{\bx: H(\bx, \by_j) = D} p(\bx)\log_2\frac{\hat q(\bx|\by_j)}{p(\bx)},
\ee
where the parameter $0 \leq \alpha < 1$ and distance threshold $D$ is specified by,
\be \label{equ.BSSL51}
\frac{1}{Q} = \sum_{\bx: H(\bx, \by_j) < D} p(\bx) + \alpha \sum_{\bx: H(\bx, \by_j) = D} p(\bx)
= 2^{-n}\sum^{D - 1}_{j = 0} {n \choose j} + \alpha 2^{-n} {n \choose D},
\ee
which is equivalent to (\ref{equ.BSSL3}).

Substituting (\ref{equ.BSSL1}), (\ref{equ.BSSL5}), and (\ref{equ.BSSL51}) into (\ref{equ.BSSL4}),
we have that
\be \label{equ.BSSL6}
D(\hat q) + \frac{\hat \lambda}{n}\Big[nR - Q \cdot \theta\Big(\frac{1}{Q}\Big)\Big]
= 2^{-n} Q \cdot \Big[\sum^{D - 1}_{j = 0} {n \choose j} \frac{j}{n} + \alpha \frac{D}{n}\Big],
\ee
and thus from (\ref{equ.BSSL4}) we prove (\ref{equ.BSSL2}).
\end{proof}
\end{theorem}

\subsection{Upper Bound} \label{subsec.BinaryUB}
\subsubsection{Upper Bound Based on Ordered Statistics}
We have the following results based on Theorem~\ref{thm.Thm602}.

\begin{theorem} \label{thm.Thm121}
For any $\epsilon > 0$, we define the threshold $t_\epsilon$ as follows,
\be \label{equ.BSSU1}
\sum^{t_\epsilon - 1}_{j = 0} {n \choose j} 2^{-n} < \frac{\ln \frac{1}{\epsilon}}{Q - 1}
\leq \sum^{t_\epsilon}_{j = 0} {n \choose j} 2^{-n}.
\ee
Then, an upper bound for the average distortion $\bar D^n_Q$ is given as follows,
\be \label{equ.BSSU2}
\bar D^n_Q \leq \frac{(1 - \epsilon) t_\epsilon}{n} + \frac{\epsilon}{2}.
\ee

\begin{proof}
The proof is similar to that of Theorem~\ref{thm.Thm602}, except that there exists no fixed quantization codeword $\by_b$
with finite expected distortion.
It is easily verified that the optimal marginal $\hat q (y) = 2^{-n}$ for all $\by$.

For the binary symmetric source, the average distortion is simply the distortion for each $\bx$,  i.e.,
\be \label{equ.BSSU21}
\bar D^n_Q = \sum_{\{\by_j\}^Q_{j = 1}} \prod^Q_{j = 1} \hat q(\by_j) d(\bx , \Y_Q) = \mathbb{E}\Big(d(\bx , \Y_Q)\Big).
\ee
Define $\bar F_k = \mathbb{P}\Big(d(\bx , \by) \geq \frac{k}{n}\Big) = 1 - \sum^{k-1}_{j = 0} {n \choose j} 2^{-n}$, for $k = 1,2,...$. Then
\be \label{equ.BSSU22}
\bar D^n_Q = \frac{1}{n}\sum^n_{k = 1} \mathbb{P}\Big(d(\bx , \Y_Q) \geq \frac{k}{n}\Big) = \frac{1}{n}\sum^n_{k = 1} \bar F^Q_k.
\ee

Given $t_\epsilon$ specified by (\ref{equ.BSSU1}), we have
\be \label{equ.BSSU23}
1 - \epsilon^{\frac{1}{Q - 1}} \leq \frac{\ln \frac{1}{\epsilon}}{Q - 1} \leq \sum^{t_\epsilon}_{j = 0} {n \choose j} 2^{-n},
\ee
and thus
\be \label{equ.BSSU24}
\bar F_{t_\epsilon + 1} = 1 - \sum^{t}_{j = 0} {n \choose j} 2^{-n} \leq \epsilon^{\frac{1}{Q - 1}},
\ee
and thus $\bar F^{Q - 1}_{j} \leq \bar F^{Q - 1}_{t_\epsilon + 1} = \epsilon$ for $j \geq t_\epsilon + 1$.
Then, we have
\be \label{equ.BSSU25}
\bar D_Q &=& \frac{1}{n}\sum^n_{k = 1} \bar F^Q_k = \frac{1}{n}\sum^{t_\epsilon}_{k = 1} \bar F^Q_k + \frac{1}{n}\sum^n_{k = t + 1} \bar F^Q_k \nonumber \\
&\leq& \frac{1}{n}\sum^{t_\epsilon}_{k = 1} \bar F_k + \frac{\epsilon}{n} \sum^n_{k = t_\epsilon + 1} \bar F_k
= \frac{1 - \epsilon}{n}\sum^{t_\epsilon}_{k = 1} \bar F_k + \frac{\epsilon}{n} \sum^n_{k = 1} \bar F_k \nonumber \\
&\leq& \frac{1 - \epsilon}{n}\sum^{t_\epsilon}_{k = 1} 1 + \frac{\epsilon}{n} \sum^n_{k = 1} \bar F_k
= \frac{(1 - \epsilon) t_\epsilon}{n} + \frac{\epsilon}{2}.
\ee
\end{proof}
\end{theorem}

\subsubsection{Upper Bound Based on Reference Rate}
Note that for the length-$n$ binary uniform source, the symmetric condition specified in Section~\ref{subsec.RefinedUB}
is satisfied, such that the upper bound $h^U(\bx)$ in (\ref{equ.RefinedUB2}) does not depend on $\bx$.
We have the following results for the upper bound on the quantization distortion.

\begin{theorem} \label{thm.Thm122}
Assume size-$Q$ random codebook $\Y_Q = \{\by_j\}^Q_{j = 1}$, where each codeword $\by_j$ is independently and identically distributed
under a uniform distribution $\hat q(\by_j)$.
\begin{itemize}
  \item For all $\bx$ and $\by$,
  \be \label{equ.BSSU30}
  l(\bx , \by) \dff \sum_{\Y_{Q}} p(\Y_{Q}) \Big(d(\bx , \Y_{Q}) -  d(\bx,\by)\Big) \Phi(\bx , \by , \Y_{Q}) \leq \frac{1}{2}.
  \ee
  \item For any $\bx$, we have
  \be \label{equ.BSSU31}
  \tilde h(\bx) = \sum_{\by_0} \hat q(\by_0) l(\bx , \by_0) \leq \frac{1}{2Q}\Big(\frac{Q - 1}{Q}\Big)^{Q - 1} \dff \tilde B.
  \ee
  \item Consider a crossover probability $p$ for which the entropy $H(p) = R_0 < R$. Then an upper bound for the average quantization distortion
  is given by,
  \be \label{equ.BSSU32}
  \bar D^n_Q - p \leq \frac{1}{2} \sum^{D - 1}_{j = 0} {n \choose j} p^j(1 - p)^{n - j} + l {n \choose D} p^D(1 - p)^{n - D},
  \ee
  where the distance threshold $D$ and the parameter $0 \leq l < \frac{1}{2}$ is specified as follows,
  \be \label{equ.BSSU33}
  && D = \max\{d|\frac{1}{2}\sum^{d - 1}_{j = 0} {n \choose j} \leq 2^{n} \tilde B\}, \nonumber \\
  \mbox{and} \ &&\frac{1}{2}\sum^{D - 1}_{j = 0} {n \choose j} + l {n \choose D} = 2^{n} \tilde B.
  \ee
\end{itemize}

\begin{proof}
We sequentially prove the above three statements.
First, for all $\bx$ and $\by$, we have the following
\be \label{equ.BSSU40}
l(\bx , \by) &\dff& \sum_{\Y_{Q}} p(\Y_{Q}) \Big(d(\bx , \Y_{Q}) -  d(\bx,\by)\Big) \Phi(\bx , \by , \Y_{Q}) \nonumber \\
&\leq& \sum_{\Y_{Q}} p(\Y_{Q}) d(\bx , \Y_{Q}) \Phi(\bx , \by , \Y_{Q}) \leq \sum_{\Y_{Q}} p(\Y_{Q}) d(\bx , \Y_{Q}) \nonumber \\
&\leq& \sum_{\Y_{Q}} p(\Y_{Q}) d(\bx , \by_1) = \sum_{\by_1} p(\Y_{Q}) d(\bx , \by_1) = \frac{1}{2}.
\ee

Second, according to (\ref{equ.UpperB180}) we have
\be \label{equ.BSSU41}
\tilde h(\bx) &=& \frac{\mathbb{E}\Big(d^{(2)}\Big) - \mathbb{E}\Big(d^{(1)}\Big)}{Q + 1}
\leq \frac{1}{Q + 1} \frac{Q + 1}{Q} \Big(\frac{Q - 1}{Q}\Big)^{Q - 1} \mathbb{E}\Big(d^1\Big) \nonumber \\
&\leq& \frac{1}{2Q} \Big(\frac{Q - 1}{Q}\Big)^{Q - 1},
\ee
where $d^{(1)}$ and $d^{(2)}$ are the smallest and second smallest ordered statistics among the $Q + 1$ independent and identically distributed
random variables $d^j$ for $1 \leq j \leq Q + 1$, with the distribution that $\mathbb{P}\Big(d^j = \frac{k}{n}\Big) = 2^{-n}{n \choose k}$.

Finally, we select the reference channel transfer function $\hat q_0(y|x) = p$ for $y \neq x$ and $\hat q_0(y|x) = 1 - p$ otherwise.
Note that the expected distortion $D(\hat q_0) = p$.
Then (\ref{equ.BSSU32}) directly follows Theorem~\ref{thm.Thm10}, (\ref{equ.RefinedUB6}) and (\ref{equ.RefinedUB12}).
\end{proof}
\end{theorem}

\subsection{Numerical Evaluations} \label{subsec.BinaryNumerical}


\begin{figure}[htb]
\centering
\includegraphics[width = 0.8\textwidth]{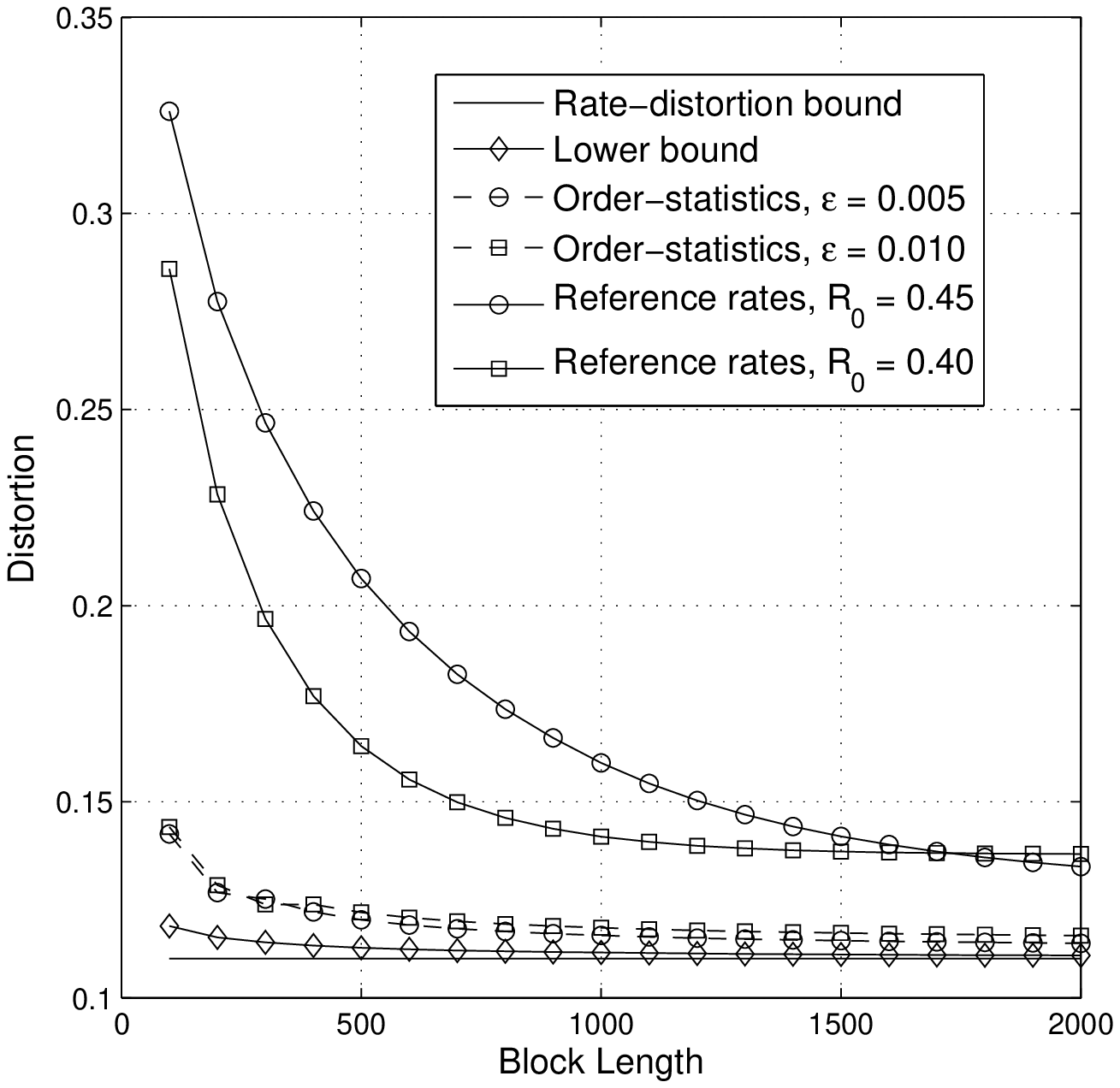}
\caption{\small Distortion for the Binary Symmetric Source with Quantization Rate $1/2$.}
\label{fig.BSC1}
\end{figure}


We consider binary symmetric source with the quantization rate $R = 1/2$.
The rate distortion theory shows that the lower bound for the distortion for all codeword length is $D^* = 0.109$.
All the numerical computations involved are performed in the log domain, e.g., $2^{1000}$ is represented by $\ln 2^{1000} = 1000\ln 2$.
For finite quantization block length, we plot distortion lower bound and upper bounds, as well as the asymptotic distortion $0.109$, in Fig.~\ref{fig.BSC1}.
Note that the upper bound based on ordered statistics are plotted for $\epsilon = 0.005$ and $0.010$,
and the upper bound based on the reference rate are plotted for $R_0 = 0.40$ and $0.45$.
It is seen that the upper bound based on ordered statistics becomes tighter for small $\epsilon$;
and for upper bound based on the reference rate, smaller $R_0$ causes faster attenuation from the beginning but larger converged values,
and larger $R_0$ causes slower attenuation from the beginning but smaller converged values.


\section{Binary Non-symmetric Sources}
Consider the length-$n$ independent and identically distributed binary source with non-uniform distribution,
with the probability $p$ for bit one and $1-p$ for bit zero,
where the probability for a length $n$ sequence with $k$ bits one and $n-k$ bits zero is $p^{k}(1-p)^{n-k}$.
Without loss of generality, assume $p \leq 0.5$.
Assume that the quantization alphabet size $Q = 2^{nR}$.
The optimal quantization conditional probability function from the rate-distortion theory is given as follows,
\be \label{equ.BNS1}
\hat q(x|y) = \left\{
\begin{array}{ccc}
D &&\mbox{for} \ y \neq x;\\
1 - D &&\mbox{for} \ y = x;
\end{array}\right.
\ee
for the distortion $D$, $0 < D < p$.
Then, the probability of $y$ is given by,
\be \label{equ.BNS2}
\mathbb{P}(y = 1) = \frac{p-D}{1-2D}, \ \ \mbox{and} \ \ \mathbb{P}(y = 0) = 1 - \mathbb{P}(y = 1).
\ee
The rate-distortion function is given by,
\be \label{equ.BNS3}
R(D) = H(p) - H(D)
\ee
for $0 \leq D \leq p$ and $R(D) = 0$ otherwise.

\subsection{Lower Bound}
According to \emph{Corollary 3}, for any quantization using a codebook $\Y_Q = \{\by_1,\by_2,...,\by_Q\}$
and quantization region $\R_j$ for $\by_j$ for $1 \leq j \leq Q$, the distortion is given by
\be \label{equ.BNS4}
D^n(q_Q) = D^* + \frac{\hat \lambda}{n}\Big(nR - \sum^Q_{j = 1} \sum_{\bx \in \R_j} p(\bx) \log\frac{\hat q(\bx|\by_j)}{p(\bx)}\Big),
\ee
where $\hat \lambda = \log_2((1-D)/D)$. Note that
\be \label{equ.BNS5}
\sum^Q_{j = 1} \sum_{\bx \in \R_j} p(\bx) \log\frac{\hat q(\bx|\by_j)}{p(\bx)}
&=& \sum^Q_{j = 1} \sum_{\bx \in \R_j} p(\bx) \log\hat q(\bx|\by_j) + \sum_{\bx} p(\bx) \log_2 \frac{1}{p(\bx)} \nonumber \\
&=& \sum^Q_{j = 1} \sum_{\bx \in \R_j} p(\bx) \log\hat q(\bx|\by_j) + n H(p).
\ee
We are interested in an upper bound for $\sum^Q_{j = 1} \sum_{\bx \in \R_j} p(\bx) \log\hat q(\bx|\by_j)$.
The following result provides an upper bound for the sum of the product of two arrays.

\begin{theorem} \label{thm.BNST1}
Assume two arrays $\{a_i\}^N_{i = 1}$ and $\{b_i\}^N_{i = 1}$ satisfy $a_1 \geq a_2 ... \geq a_N \geq 0$ and $b_1 \geq b_2 ... \geq b_N$.
Then for any permutation $j_1$, $j_2$, ..., $j_N$ of $1$, $2$, ..., $N$, we have
\be \label{equ.BNS6}
\sum^N_{i = 1} a_i b_{j_i} \leq \sum^N_{i = 1} a_i b_i.
\ee
\end{theorem}

According to Theorem~\ref{thm.BNST1}, we can obtain an upper bound for $\sum^Q_{j = 1} \sum_{\bx \in \R_j} p(\bx) \log\hat q(\bx|\by_j)$ via ranking the two arrays,
$\{p(\bx), \bx \in \X^n \}$ and $\bigcup^Q_{j=1}\{\log\hat q(\bx|\by_j), \bx \in \R_j\}$.
Although the latter depends on the selection of the quantization regions $\R_j$, we can further provide an upper bound on $\bigcup^Q_{j=1}\{\log\hat q(\bx|\by_j), \bx \in \R_j\}$
that is independent of $\R_j$.

More specifically, for Hamming distance $H(\bx,\by_j) = i$ we have,
\be \label{equ.BNS11}
\log\hat q(\bx|\by_j) = i \log D + (n-i) \log(1-D) \dff d_{i}.
\ee
which is independent of $\by_j$. Then, we can grab the largest $Q = 2^{nR}$ values of $\log\hat q(\bx|\by_j)$ via finding the distance threshold $D_T$ as follows,
\be \label{equ.BNS12}
&&D_T = \max\{D|Q\sum^{D-1}_{j=0}{n \choose j} \leq 2^n\}, \nonumber \\
\mbox{and $0 \leq K < {n \choose D}$ such that} \ &&Q\sum^{D-1}_{j=0}{n \choose j} + K  = 2^n.
\ee
Let $\{b_i\}^{2^n}_{i=1}$ denote the array consisting of $Q \cdot {n \choose j}$ elements of $d_j$ for $1 \leq j \leq D-1$
and $K$ elements of $d_D$, in descending order;
and let $\{a_i\}^{2^n}_{i=1}$ denote the $2^n$ elements of $p(\bx)$ for all $\bx \in \X^n$, also in descending order.
Then we have the following result.

\begin{theorem} \label{thm.BNSU2}
We have
\be \label{equ.BNS13}
\sum^Q_{j = 1} \sum_{\bx \in \R_j} p(\bx) \log\hat q(\bx|\by_j) \leq \sum^{2^{n}}_{i=1} a_i b_i.
\ee
Furthermore, we have
\be \label{equ.BNS14}
\sum^{2^{n}}_{i=1} a_i b_i + n H(p) \leq nR,
\ee
such that the lower bound obtained from (\ref{equ.BNS13}) is tighter than the infinite-length distortion $D^*$.

\begin{proof}
We rank $\{\log\hat q(\bx|\by_j), \bx \in \R_j, 1 \leq j \leq Q\}$ in descending order, denoted as  $c_1 \geq c_2 ... \geq c_{2^n}$.
Since $b_i$ is the largest $2^{n}$ values of $\{\log\hat q(\bx|\by_j)\}$, we have that $c_i \leq b_i$ for $1 \leq i \leq 2^n$, and thus
\be \label{equ.BNS15}
\sum^Q_{j = 1} \sum_{\bx \in \R_j} p(\bx) \log\hat q(\bx|\by_j) \leq \sum^{2^{n}}_{i=1} a_i c_i \leq \sum^{2^{n}}_{i=1} a_i b_i.
\ee
To prove (\ref{equ.BNS14}), let $\bar B = \sum^{2^{n}}_{i=1} 2^{b_i} \leq Q = 2^{nR}$,
and $\tilde b_i = 2^{b_i} / \bar B$ such that $\sum^{2^{n}}_{i=1} \tilde b_i = 1$.
We have that
\be \label{equ.BNS16}
\sum^{2^{n}}_{i=1} a_i b_i + n H(p) &=& \sum^{2^{n}}_{i=1} a_i \log \frac{\tilde b_i}{a_i} + \sum^{2^{n}}_{i=1} a_i \log \bar B \nonumber \\
&\leq& \frac{1}{\ln 2} \sum^{2^{n}}_{i=1} a_i \Big(\frac{\tilde b_i}{a_i} - 1\Big) + \sum^{2^{n}}_{i=1} a_i \log \bar B \nonumber \\
&=& \sum^{2^{n}}_{i=1} a_i \log \bar B = \log \bar B \leq nR.
\ee
\end{proof}
\end{theorem}

\emph{\textbf{Remark:}} We discuss the computational issues for (\ref{equ.BNS13}).
Note that there are $n+1$ and $D_T+1$ different values for $a_i$ and $b_i$. Then, for $\sum^{2^{n}}_{i=1} a_i b_i$ all in the descending order,
actually there are at most $n+D_T+1$ different values of $a_ib_i$.
Computing $\sum^{2^{n}}_{i=1} a_i b_i$ is to compute the sum of the product of the $n+D_T+1$ different values and their frequencies.
Here we also compute the sum and product operations in the logarithm domain.

\subsection{Upper Bound}
We consider the following mean distortion over the codebook
\be \label{thm.BNSU1}
\bar D_Q &=& \sum_{\Y_Q} p(\Y_Q) \sum_{\bx} p(\bx) d(\bx, \Y_Q) = \sum_{\bx} p(\bx) \sum_{\Y_Q} p(\Y_Q) d(\bx, \Y_Q) \nonumber \\
&\dff& \sum_{\bx} p(\bx) h_{Q}(\bx).
\ee
Note that, due to the symmetricity of $\bx$ for a given weight of $\bx$, $h_{Q}(\bx)$ only depends on the weight of $\bx$.
We derive an upper bound for each weight of $\bx$.
We let $z \dff \mathbb{P}(y = 1) = \frac{p-D}{1-2D}$.

\subsubsection{Upper Bound based on Ordered Statistics}
Note that for $\bx$ with weight $w$, we can split it into two parts, $w$ bits one and $n-w$ bits zero.
The following result shows the probability $\mathbb{P}\Big(d(\bx, \by) = d\Big)$ for $\bx$ with weight $w$.
The proof numerates all combinations of the different numbers of bits $i$ and $j$
among the $w$ bits one and $n-w$ bits zero of $\bx$, respectively.

\begin{theorem} \label{thm.BNSU3}
For $\bx$ with weight $w$, we have the following probability
\be \label{equ.BNS17}
\mathbb{P}\Big(d(\bx, \by) = d\Big) = \sum_{i+j=d, 0 \leq i \leq w, 0 \leq j \leq n-w} {w \choose i} z^{w-i}(1-z)^i {n-w \choose j} z^{j}(1-z)^{n-w-j}.
\ee
\end{theorem}

\

Similar to Theorem~\ref{thm.Thm121}, we have the following result on the upper bound based on ordered statistics.
\begin{theorem} \label{thm.BNSU4}
For any $\epsilon > 0$, we define the threshold $t_\bx$ as follows,
\be \label{equ.BNSU18}
\sum^{t_\bx - 1}_{j = 0} \mathbb{P}\Big(d(\bx, \by) = j\Big) < \frac{\ln \frac{1}{\epsilon}}{Q - 1}
\leq \sum^{t_\bx}_{j = 0} \mathbb{P}\Big(d(\bx, \by) = j\Big).
\ee
\begin{enumerate}
  \item For weight-$w$ sequence $\bx$, we have the following upper bound
\be \label{equ.BNSU19}
\sum_{\Y_Q} p(\Y_Q) d(\bx, \Y_Q) \leq \frac{(1 - \epsilon) t_\bx}{n} + \frac{\epsilon}{2} \dff D_w.
\ee
  \item
Then the upper bound for the distortion $\bar D_Q$ is given as follows,
\be \label{equ.BNSU20}
\bar D_Q \leq \sum^n_{w=0} {n \choose w} p^w(1-p)^{n-w} D_w.
\ee
\end{enumerate}
\end{theorem}

\

\subsubsection{Upper Bound based on Reference Rates}
Similarly to Theorem~\ref{thm.Thm122}, we consider a larger distortion $D_0 > D$
such that the optimal transfer function $\hat q(x|y) = D_0$ for $y \neq x$
and the corresponding probability $\mathbb{P}(y = 1) = \frac{p-D_0}{1-2D_0} \dff z_0$.
We have the following result for the upper bound based on the reference rates for binary non-symmetric source.

\begin{theorem} \label{thm.BNSU5}
We consider a random codebook where each bit of codeword $\by$ satisfies the i.i.d. distribution $\mathbb{P}(y = 1)$.
\begin{enumerate}
  \item For weight $w$ sequence $\bx$, we have
  \be \label{equ.BSSU30}
  l(\bx , \by) &\dff& \sum_{\Y_{Q}} p(\Y_{Q}) \Big(d(\bx , \Y_{Q}) -  d(\bx,\by)\Big) \Phi(\bx , \by , \Y_{Q}) \nonumber \\
  &\leq& z_0(1-\frac{w}{n})+(1-z_0)\frac{w}{n} \dff u_w.
  \ee
  \item For weight $w$ sequence $\bx$, we have
  \be \label{equ.BSSU31}
  \tilde h(\bx) = \sum_{\by_0} \hat q(\by_0) l(\bx , \by_0) \leq u_w \frac{1}{Q}\Big(\frac{Q - 1}{Q}\Big)^{Q - 1} \dff \tilde B_w.
  \ee
  \item For the weight $w$ sequence $\bx$, we consider the distance threshold $d_\bx$ and a fraction $0 \leq l < 1$ such that
\be \label{equ.BNSU32}
&&\sum^{d_\bx - 1}_{j = 0} \mathbb{P}\Big(d(\bx, \by) = j\Big) \leq \frac{1}{Q}\Big(\frac{Q - 1}{Q}\Big)^{Q - 1} < \sum^{d_\bx}_{j = 0} \mathbb{P}\Big(d(\bx, \by) = j\Big); \nonumber \\
\mbox{and} \ &&\sum^{d_\bx - 1}_{j = 0} \mathbb{P}\Big(d(\bx, \by) = j\Big) + l \cdot \mathbb{P}\Big(d(\bx, \by) = d_\bx\Big) = \frac{1}{Q}\Big(\frac{Q - 1}{Q}\Big)^{Q - 1}.
\ee
The distortion with respect to $\bx$ is given as follows,
\be \label{equ.BNSU33}
\tilde h(\bx) &\leq& \sum^{d_\bx - 1}_{j = 0} \mathbb{P}\Big(d(\bx, \by) = j\Big) \frac{D^j_0(1-D_0)^{n-j}}{p^w(1-p)^{n-w}}
+ l \cdot \mathbb{P}\Big(d(\bx, \by) = d_\bx\Big)\frac{D^{d_\bx}_0(1-D_0)^{n-d_\bx}}{p^w(1-p)^{n-w}} \nonumber \\
&\dff& \tilde h^U_w,
\ee
where the probability $\mathbb{P}\Big(d(\bx, \by) = j\Big)$ is given by (\ref{equ.BNS17}).
Then, we have,
\be \label{equ.BNSU34}
\bar D_Q \leq D_0 + \sum^n_{w=0} {n \choose w} p^w(1-p)^{n-w} \tilde h^U_w.
\ee
\end{enumerate}

\begin{proof}
\begin{enumerate}
\item We have the following
\be \label{equ.BSSU35}
l(\bx , \by) &=& \sum_{\Y_{Q}} p(\Y_{Q}) \Big(d(\bx , \Y_{Q}) -  d(\bx,\by)\Big) \Phi(\bx , \by , \Y_{Q}) \nonumber \\
&\leq& \sum_{\Y_{Q}} p(\Y_{Q}) d(\bx , \Y_{Q}) \Phi(\bx , \by , \Y_{Q})
\leq \sum_{\Y_{Q}} p(\Y_{Q}) d(\bx , \Y_{Q}) \nonumber \\
&\leq& \sum_{\Y_{Q}} p(\Y_{Q}) d(\bx , \by_1) = \sum_{\by_1} p(\by_1) d(\bx , \by_1) = \mathbb{E}\Big( d(\bx , \by_1)\big) \nonumber \\
&\leq& z_0(1-\frac{w}{n})+(1-z_0)\frac{w}{n} \dff u_w.
\ee
\item The proof is similar to that of Theorems~\ref{thm.Thm7} and~\ref{thm.Thm8} and thus omitted here.
\item The proof is follows the steps in Theorem~\ref{thm.Thm10}. We need to note that
\be \label{equ.BSSU35}
p(\by|\bx) = p(\by)\frac{p(\bx|\by)}{p(\bx)} = \mathbb{P}\Big(d(\bx, \by) = j\Big) \frac{D^j_0(1-D_0)^{n-j}}{p^w(1-p)^{n-w}}.
\ee
\end{enumerate}
\end{proof}
\end{theorem}


\subsection{Numerical Results}

\begin{figure}[htb]
\centering
\includegraphics[width = 0.8\textwidth]{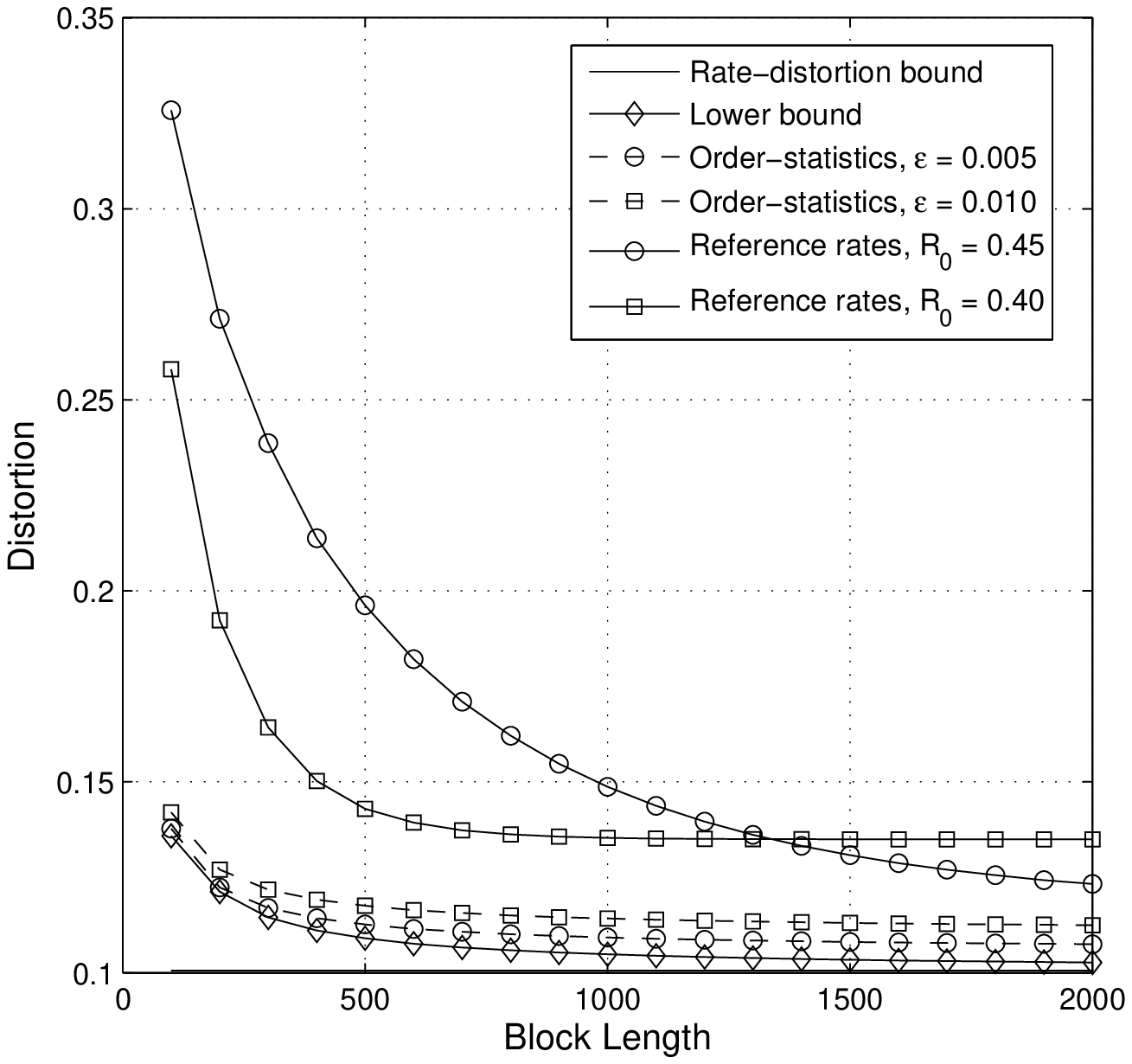}
\caption{\small Distortion for the Binary Non-symmetric Source with Distribution $\mathbb{P}(x = 1) = 0.4$ and Quantization Rate $1/2$.}
\label{fig.BSC1}
\end{figure}

We consider binary source with the probability that $\mathbb{P}(x = 1) = 0.4$, and the quantization rate $R = 1/2$.
The rate distortion theory shows that the lower bound for the distortion for all codeword length is $D^* = 0.101$.
We show the lower bound, the upper bound from ordered statistics for parameters $\epsilon = 0.005$ and $0.010$,
and the upper bound from reference rates for parameters $R_0 = 0.45$ and $0.40$.
Again, it is seen that the upper bound based on ordered statistics becomes tighter for small $\epsilon$;
and for upper bound based on the reference rate, smaller $R_0$ causes faster attenuation from the beginning but larger converged values,
and larger $R_0$ causes slower attenuation from the beginning but smaller converged values.

\section{Gaussian Sources} \label{sec.GaussianSource}
We consider $n$-dimension Gaussian source with the following probability density function
\be \label{equ.GaussianPDF}
p(\bx) = \frac{1}{(2\pi\sigma^2)^{\frac{n}{2}}}e^{-\frac{\|\bx\|^2}{2\sigma^2}}.
\ee
The distortion measured by the norm-$2$ distortion $d(\bx , \by) = \frac{\|\bx - \by\|^2}{n}$.
We employ a size-$Q$ codebook, where $Q = 2^{nR}$.
From the rate-distortion theory, as the dimension $n$ approaches infinity,
the quantization distortion approaches to $D$, which satisfies
\be \label{equ.GaussianPDF2}
\frac{1}{2}\log_2 \frac{\sigma^2}{D} = R;
\ee
and the asymptotic random reconstruction function $\hat q(\bx|\by)$ and $\hat q(\by)$
is given as follows,
\be \label{equ.GaussianPDF3}
\hat q(\bx|\by) = \frac{1}{(2\pi D)^{\frac{n}{2}}}e^{-\frac{\|\bx - \by\|^2}{2D}}, \ \mbox{and} \ \
\hat q(\by) = \frac{1}{(2\pi (\sigma^2 - D))^{\frac{n}{2}}}e^{-\frac{\|\by\|^2}{2(\sigma^2 - D)}}.
\ee

In this Section we derive lower and upper bounds for the optimal quantization distortion
using a size-$Q$ codebook.
More specifically, we consider the following two cases for the codebook $\Y_Q$,
\begin{itemize}
  \item bounded codebook: all codewords are constrained within the ball $\|\by\| \leq R_m$, i.e., $\|\by_j\| \leq R_m$ for all $1 \leq j \leq Q$;
  \item unbounded codebook: all codewords can be chosen from the entire $n$-dimensional real space $\mathbb{R}^n$.
\end{itemize}

\subsection{Lower Bound for the Quantization Distortion} \label{subsec.GaussianLB}
From (\ref{equ.Duality11}), we have that
\be \label{equ.GaussianPDF4}
G(\bx) = \min_{1 \leq j \leq Q} \Big\{\frac{\|\bx - \by_j\|^2}{2D} - \frac{\|\bx\|^2}{2\sigma^2}\Big\}
= \min_{1 \leq j \leq Q} G_j(\bx),
\ee
where $G_j(\bx) = \frac{\|\bx - \by_j\|^2}{2D} - \frac{\|\bx\|^2}{2\sigma^2}$.
We derive a lower bound on the distortion gap $\Delta D^n(q_Q)$ based on Theorem~\ref{thm.Thm40}
in Section~\ref{subsec.Duality}, which provides the following lower bound
\be \label{equ.GaussianPDF5}
\Delta D^n(q_Q) \geq \int^{+\infty}_{0} \mathbb{P}\Big(G(\bx) - 1 + e^{-G(\bx)} \geq \mu\Big) d\mu
\dff \Delta \tilde D^n.
\ee

To efficiently derive a lower bound for $\Delta \tilde D$ over all possible codebooks $\Y_Q$, i.e., $\min_{\Y_Q} \Delta \tilde D$,
we add a special codeword $\by_0 \dff {\bf 0}$ into the codebook.
Let $\tilde \Y_Q \dff \Y_Q \cup \{\by_0\}$ be the new codebook, and
\be \label{equ.GaussianPDF6}
\tilde G(\bx) \dff \min_{0 \leq j \leq Q} \Big\{\frac{\|\bx - \by_j\|^2}{2D} - \frac{\|\bx\|^2}{2\sigma^2}\Big\} \leq G(\bx).
\ee
Moreover, we define function $E(\mu)$ as the inverse of the function $t - 1 + e^{-t}$ as follows,
\be \label{equ.GaussianPDF7}
E(\mu) = t \ \Leftrightarrow \mu = t - 1 + e^{-t} \ \mbox{and} \ t \geq 0.
\ee
Then, for $t \geq E(\mu)$, we have $t - 1 + e^{-t} \geq \mu$.
We have the following result for a lower bound for $\Delta \tilde D$.

\begin{theorem} \label{thm.Thm130}
We have the following the lower bound,
\be \label{equ.GaussianLB0}
\Delta \tilde D \geq \int^{+\infty}_{0} \mathbb{P}\Big(\tilde G(\bx) \geq E(\mu)\Big) d\mu.
\ee

\begin{proof}
Note that, due to (\ref{equ.GaussianPDF6}), for $\tilde G(\bx) \geq E(\mu)$ we have $G(\bx) \geq \tilde G(\bx) \geq E(\mu)$,
and thus $G(\bx) - 1 + e^{-G(\bx)} \geq \mu$. Therefore
\be \label{equ.GaussianLB11}
\mathbb{P}\Big(G(\bx) - 1 + e^{-G(\bx)} \geq \mu\Big) \geq \mathbb{P}\Big(\tilde G(\bx) \geq E(\mu)\Big).
\ee
Via integrating (\ref{equ.GaussianLB11}) for $\mu$ from $0$ to $+\infty$, we can prove (\ref{equ.GaussianLB0}).
\end{proof}
\end{theorem}

Based on Theorem~\ref{thm.Thm130}, in the following we evaluate the lower bound on $\Delta \tilde D$
via evaluating an upper bound on $\mathbb{P}\Big(\tilde G(\bx) \leq E(\mu)\Big) = 1 - \mathbb{P}\Big(\tilde G(\bx) \geq E(\mu)\Big)$.

\subsubsection{Modified Union Bound for $\mathbb{P}\Big(\tilde G(\bx) \leq t\Big)$}
Here we again let $t = E(\mu)$.
Let $\C_j(t) \dff \{\bx: G_j(\bx) \leq t\}$ for $0 \leq j \leq Q$.
Via simple calculation, we have
\be \label{equ.GaussianLB3}
\bx \in \C_j(t) \Leftrightarrow
\|\bx - \frac{\sigma^2}{\sigma^2 - D}\by_j\|^2 \leq \frac{ \sigma^2 D \|\by_j\|^2}{(\sigma^2 - D)^2} + \frac{2D\sigma^2t}{\sigma^2 - D}
\dff R(t , \|\by_j\|), \ 0 \leq j \leq Q;
\ee
and in particular
\be \label{equ.GaussianLB4}
\bx \in \C_0(t) \Leftrightarrow \|\bx\|^2 \leq \frac{2D\sigma^2t}{\sigma^2 - D} = R(t , 0).
\ee
According to (\ref{equ.GaussianLB3}), for each $0 \leq j \leq Q$, $\C_j(t)$ is a ball with center $\frac{\sigma^2}{\sigma^2 - D}\by_j$
and radius $\sqrt{R(t , \|\by_j\|^2)}$.

We have the following modified union bound for $\mathbb{P}\Big(\tilde G(\bx) \leq t\Big)$.

\begin{theorem} \label{thm.Thm132}
We have the following upper bound for $\mathbb{P}\Big(\tilde G(\bx) \leq t\Big)$,
\be \label{equ.GaussianLB8}
\mathbb{P}\Big(\tilde G(\bx) \leq t\Big) \leq \mathbb{P}\Big(\C_0(t)\Big) + \sum^Q_{j = 1} \mathbb{P}\Big(\C_j(t) \setminus \C_0(t)\Big).
\ee
\begin{proof}
Note that
\be \label{equ.GaussianLB9}
G(\bx) = \min_{0 \leq j \leq Q} G_j(\bx) \leq t \Leftrightarrow \bx \in \cup^Q_{j = 0} \C_j(t);
\ee
and $\cup^Q_{j = 0} \C_j(t) = \C_0(t) \cup \cup^Q_{j = 1} \Big(\C_j(t) \setminus \C_0(t)\Big)$.
Then, we have
\be \label{equ.GaussianLB10}
\mathbb{P}\Big(G(\bx) \leq t\Big) = \mathbb{P}\Big(\cup^Q_{j = 0} \C_j(t)\Big)
\leq \mathbb{P}\Big(\C_0(t)\Big) + \sum^Q_{j = 1} \mathbb{P}\Big(\C_j(t) \setminus \C_0(t)\Big).
\ee
\end{proof}
\end{theorem}

Note that $\C_j(t) \setminus \C_0(t)$ denotes the space in the ball $\C_j(t)$ but not in $\C_0(t)$.
Due to the sphere symmetric property of Gaussian distribution, $\mathbb{P}\Big(\C_j(t) \setminus \C_0(t)\Big)$
is only a function of $\|\by_j\|$, and $t$.
Based on the computational methods in Section~\ref{subsec.InterTwoBalls}, we have the following result on
the probability $\mathbb{P}\Big(\C_j(t) \setminus \C_0(t)\Big)$.

\begin{theorem} \label{thm.Thm134}
Let
\be \label{equ.GaussianLB160}
r_{min} &=& \max\Big\{\sqrt{R(t , 0)}, \frac{\sigma^2}{\sigma^2 - D}\|\by_j\| - \sqrt{R(t , \|\by_j\|)}\Big\}, \nonumber\\
r_{max} &=& \frac{\sigma^2}{\sigma^2 - D}\|\by_j\| + \sqrt{R(t , \|\by_j\|)}.
\ee
The probability $\mathbb{P}\Big(\C_j(t) \setminus \C_0(t)\Big)$ is given as follows,
\be \label{equ.GaussianLB161}
\mathbb{P}\Big(\C_j(t) \setminus \C_0(t)\Big) = \int^{r_{max}}_{r_{min}} \frac{1}{(2\pi \sigma^2)^{\frac{n}{2}}}
e^{-\frac{r^2}{2\sigma^2}} r^{n - 1} \Omega_n(\theta_j(r)) dr,
\ee
where the semiangle $\theta_j(r)$ is given as follows,
\be \label{equ.GaussianLB162}
\theta_j(r) = \cos^{-1}\frac{\Big(\frac{\sigma^2}{\sigma^2 - D}\Big)^2\|\by_j\|^2 + r^2 - R(t , \|\by_j\|)}
{2\frac{\sigma^2}{\sigma^2 - D}\|\by_j\|r}.
\ee

\begin{proof}
Consider the intersection of $\C_j(t) \setminus \C_0(t)$ with a radius-$r$ sphere centered at the origin,
which is essentially a radius-$r$ sphere cut out by a cone.
The cone can be described by a triangular with the lengths $\Big(\frac{\sigma^2}{\sigma^2 - D}\|\by_j\|, r, \sqrt{R(\|\by_j\|^2 , t)}\Big)$
of the three edges; and the semiangle is the angle between the edges of lengths $\frac{\sigma^2}{\sigma^2 - D}\|\by_j\|$ and $r$.
Then, from the cosine formula, we have that
\be \label{equ.GaussianLB163}
\Big(\frac{\sigma^2}{\sigma^2 - D}\Big)^2\|\by_j\|^2 + r^2 - 2\frac{\sigma^2}{\sigma^2 - D}\|\by_j\|r \cos\theta_j(r) = R_j(t);
\ee
and thus (\ref{equ.GaussianLB162}) follows (\ref{equ.GaussianLB163});
and the sphere area is given by $r^{n - 1} \Omega_n(\theta_j(r))$.

Next we consider such radius-$r$ sphere that can have non-empty intersection with $\C_j(t) \setminus \C_0(t)$.
Note that, if $\sqrt{R(t) , 0} < \frac{\sigma^2}{\sigma^2 - D}\|\by_j\| - \sqrt{R(t , \|\by_j\|)}$, then $\C_j(t) \cap \C_0(t) = \emptyset$
and thus $\C_j(t) \setminus \C_0(t) = \C_j(t)$.
The range of $r$ is from $\frac{\sigma^2}{\sigma^2 - D}\|\by_j\| - \sqrt{R(t , \|\by_j\|)}$ to $r_{max}$ [c.f. (\ref{equ.GaussianLB160})].
Otherwise if $\sqrt{R(t,0)} \geq \frac{\sigma^2}{\sigma^2 - D}\|\by_j\| - \sqrt{R(t , \|\by_j\|)}$, the range of $r$
is from $\sqrt{R(t,0)}$ to $r_{max}$ [c.f. (\ref{equ.GaussianLB160})].
Thus the range of $r$ is from $r_{min}$ to $r_{max}$ as specified by (\ref{equ.GaussianLB160}).

Finally, via integrating the following probability density of Gaussian distribution on a radius-$r$ sphere,
\be \label{equ.GaussianLB164}
\frac{1}{(2\pi \sigma^2)^{\frac{n}{2}}} e^{-\frac{r^2}{2\sigma^2}},
\ee
over the intersection with $\C_j(t) \cap \C_0(t)$ with the semiangle $\Omega_n(\theta_j(r))$,
we have the expression of the probability $\mathbb{P}\Big(\C_j(t) \setminus \C_0(t)\Big)$
as in (\ref{equ.GaussianLB161}).
\end{proof}
\end{theorem}
$\hfill \Box$

Note that the upper bounds obtained from Theorems~\ref{thm.Thm132} and~\ref{thm.Thm134} may exceed $1$.
In the following we propose another bound upper bound for bounded codeword constraint, i.e., $\|\by_j\| \leq R_m$ for all $1 \leq j \leq Q$.
More specifically, for each $1 \leq j \leq Q$, we consider a codeword $\by_j$ lying on the boundary $\|\by_j\| = R_m$ with a radius $\sqrt{R(t , R_m)}$,
denoted as $\tilde \C_j(t)$; and for consistency let $\tilde \C_0(t) = \C_0(t)$.
We compute the an upper bound for \emph{the volume} of $\cup^Q_{j = 0} \tilde \C_j(t)$, denoted as $\V_n(\cup^Q_{j = 0} \tilde \C_j(t))$.
The upper bound for the volume $\V_n\Big(\cup^Q_{j = 0} \tilde \C_j(t)\Big)$ can be proved to be an upper bound
for the volume $\V_n\Big(\cup^Q_{j = 0} \C_j(t)\Big)$.
The probability $\mathbb{P}\Big(\cup^Q_{j = 0} \tilde \C_j(t)\Big)$ is upper bounded by
the probability of the ball with the center at zero with the same volume.
The following results show that, although the exact value of $\V_n\Big(\cup^Q_{j = 0} \tilde \C_j(t)\Big)$ is difficult to compute,
we are able to derive an upper bound for $\V_n\Big(\cup^Q_{j = 0} \tilde \C_j(t)\Big)$ and
the associated upper bound for the probability $\mathbb{P}\Big(\cup^Q_{j = 0} \tilde \C_j(t)\Big)$.

\begin{theorem} \label{thm.Thm135}
For bounded codewords $\|\by_j\| \leq R_m$ for $1 \leq j \leq Q$,
an upper for the volume $\V_n\Big(\cup^Q_{j = 0} \C_j(t)\Big)$ is given as follows,
\be \label{equ.GaussianLB170}
\V_n\Big(\cup^Q_{j = 0} \C_j(t)\Big) \leq \V_n\Big(\tilde \C_0(t)\Big) + \sum^Q_{j = 1} \V_n\Big(\tilde \C_j(t) \setminus \C_0(t)\Big) \dff \tilde V_n.
\ee
Define the radius $r_n$ as follows,
\be \label{equ.GaussianLB171}
r_n \dff \Big(\frac{\tilde V_n}{V_n}\Big)^{\frac{1}{n}},
\ee
where $V_n$ is the volume of a unit ball in a dimension-$n$ space; and define another radius
\be \label{equ.GaussianLB172}
\tilde r_n \dff \frac{\sigma^2R_m}{\sigma^2 - D} + \sqrt{R(t , R_m)}.
\ee
Let $r_E = \min\{r_n , \tilde r_n\}$, an upper bound for the probability $\mathbb{P}\Big(\cup^Q_{j = 0} \C_j(t)\Big)$ is given as follows,
\be \label{equ.GaussianLB173}
\mathbb{P}\Big(\cup^Q_{j = 0} \C_j(t)\Big) \leq \Upsilon_n\Big(\frac{r^2_E}{\sigma^2}\Big) \dff \Gamma_n(t).
\ee
\end{theorem}
$\hfill \Box$

The following result provides an upper bound on $\V_n\Big(\tilde \C_j(t) \setminus \C_0(t)\Big)$.
The proof is similar to that of Theorem~\ref{thm.Thm134}, and thus omitted here.

\begin{theorem} \label{thm.Thm136}
Given $\|\tilde \by_j\| = R_m$ and the ball $\tilde \C_j(t)$, given by
\be \label{equ.GaussianLB180}
\tilde \C_j(t) = \Big\{\bx: \|\bx - \frac{\sigma^2}{\sigma^2 - D}\tilde \by_j\| \leq \sqrt{R(t , R_m)} \Big\}.
\ee
Letting
\be \label{equ.GaussianLB181}
\tilde r_{min} &=& \max\Big\{\sqrt{R(t , 0)}, \frac{\sigma^2}{\sigma^2 - D}R_m - \sqrt{R(t , R_m)}\Big\}, \nonumber\\
\tilde r_{max} &=& \frac{\sigma^2}{\sigma^2 - D}R_m + \sqrt{R(t , R_m)},
\ee
we have
\be \label{equ.GaussianLB182}
\V_n\Big(\tilde \C_j(t) \setminus \C_0(t)\Big) = \int^{\tilde r_{max}}_{\tilde r_{min}} r^{n - 1} \Omega_n(\tilde \theta_j(r)) dr,
\ee
where the semiangle
\be \label{equ.GaussianLB183}
\tilde \theta_j(r) = \cos^{-1}\frac{\Big(\frac{\sigma^2}{\sigma^2 - D}\Big)^2R_m^2 + r^2 - R(t , R_m)}
{2\frac{\sigma^2}{\sigma^2 - D}R_m r}.
\ee
\end{theorem}
$\hfill \Box$

Note that the above upper bound is for the bounded quantization codewords $\|\by_j\| \leq R_m$.
For unbounded quantization codewords, we simply set $\Gamma_n(t) = 1$.
Moreover, the probability $\mathbb{P}\Big(\C_0(t)\Big)$ is a function of $t$, denoted as $K_0(t)$;
and the probability $\mathbb{P}\Big(\C_0(t) \setminus \C_j(t)\Big)$ is a function of $\|\by_j\|$ and $t$, denoted as $K(t , \|\by_j\|)$.
Thus we can write the upper bound in Theorems~\ref{thm.Thm132} and~\ref{thm.Thm134} as follows,
\be \label{equ.GaussianLB19}
\mathbb{P}\Big(\cup^Q_{j = 0} \C_j(t)\Big) \leq K_0(t) +  \sum^Q_{j = 1} K(t , \|\by_j\|),
\ee
where $K_0(t) = \mathbb{P}\Big(\C_0(t)\Big)$.
Via combining the results in Theorems~\ref{thm.Thm132} to~\ref{thm.Thm136},
we have the following result on an upper bound for $\mathbb{P}\Big(\tilde G(\bx) \leq t\Big) = \mathbb{P}\Big(\cup^Q_{j = 0} \C_j(t)\Big)$.

\begin{theorem} \label{thm.Thm137}
An upper bound on $\mathbb{P}\Big(G(\bx) \leq t\Big)$ is given as follows,
\be \label{equ.GaussianLB180}
\mathbb{P}\Big(G(\bx) \leq t\Big) = \mathbb{P}\Big(\cup^Q_{j = 0} \C_j(t)\Big) \leq \min\Big\{K_0(t) + \sum^Q_{j = 1} K(\|\by_j\|,t), \Gamma(t) \Big\}.
\ee
\end{theorem}
$\hfill \Box$

\subsubsection{A Lower Bound for $\Delta \tilde D$}
Via letting $t = E(\mu)$, from (\ref{equ.GaussianLB180}) we have
\be \label{equ.GaussianLB190}
\mathbb{P}\Big(\tilde G(\bx) \leq E(\mu)\Big)
&\leq& \min\Big\{K_0(E(\mu)) + \sum^Q_{j = 1} K(\|\by_j\|,E(\mu)), \Gamma(E(\mu)) \Big\} \nonumber \\
&\dff& \min\Big\{\tilde K_0(\mu) + \sum^Q_{j = 1} \tilde K(\|\by_j\|,\mu), \tilde \Gamma(\mu) \Big\}.
\ee
Then, we have the following lower bound,
\be \label{equ.GaussianLB191}
\Delta \tilde D \geq \int^{+\infty}_{0}
\Big(1 - \min\Big\{\tilde K_0(\mu) + \sum^Q_{j = 1} \tilde K(\|\by_j\|,\mu), \tilde \Gamma(\mu) \Big\}\Big) d\mu
\dff \Delta \hat D.
\ee

Since the codewords $\{\by_j\}_{1 \leq j \leq Q}$, can be arbitrarily selected, we need to minimize $\hat D$ subject to all possible
codewords $\{\by_j\}_{1 \leq j \leq Q}$, i.e., to obtain a lower bound of the following,
\be \label{equ.GaussianLB192}
\min_{\{\by_j\}_{1 \leq j \leq Q}} \Delta \hat D.
\ee
However, directly solving (\ref{equ.GaussianLB192}) incurs prohibitive computational complexity.
The following result provides a further lower bound of $\Delta \hat D$, which is significantly more computational feasible.

It is observed that,
\be \label{equ.GaussianLB193}
\tilde K_0(\mu) + \sum^Q_{j = 1} \tilde K(\|\by_j\|,\mu) < \tilde \Gamma(\mu),
\ee
for small $\mu$ and the vice versa for large $\mu$.
Based on this observation, we set up a parameter $\mu_0 \geq 0$, and let
\be \label{equ.GaussianLB194}
\Delta \hat D(\mu_0) \dff \int^{\mu_0}_{0} \Big(1 - \tilde K_0(\mu) - \sum^Q_{j = 1} \tilde K(\|\by_j\|,\mu)\Big) d\mu
+ \int^{+\infty}_{\mu_0} \Big(1 - \tilde \Gamma(\mu)\Big) d\mu.
\ee

It is easily seen that, for any $\{\by_j\}^Q_{j = 1}$, we have that $\Delta \hat D(\mu_0) \leq \Delta \hat D$.
However, since in (\ref{equ.GaussianLB194}) all $\by_j$ for $1 \leq j \leq Q$ are independent,
the lower bound $\min_{\{\by_j\}_{1 \leq j \leq Q}}\Delta \hat D(\mu_0)$
can be solved via solving,
\be \label{equ.GaussianLB195}
\max_{\|\by_j\|} \int^{\mu_0}_{0} \tilde K(\|\by_j\|,\mu) d\mu.
\ee
Note that the above optimization problems are the same for all $1 \leq j \leq Q$,
and we can solve (\ref{equ.GaussianLB195}) for all $\by_j$.
The following Theorem~\ref{thm.Thm138} formalizes the above arguments.

\begin{theorem} \label{thm.Thm138}
For any $\mu_0 \geq 0$, we have that
\be \label{equ.GaussianLB195}
\min_{\{\by_j\}_{1 \leq j \leq Q}}\Delta \hat D(\mu_0) \leq \min_{\{\by_j\}_{1 \leq j \leq Q}}\Delta \hat D.
\ee
\begin{itemize}
  \item For bounded codebook constraints $\|\by_j\| \leq R_m$ for $1 \leq j \leq Q$, we have
\be \label{equ.GaussianLB196}
\min_{\{\by_j\}_{1 \leq j \leq Q}}\Delta \hat D(\mu_0)
&=& \min_{0 \leq r \leq R_m} \int^{\mu_0}_{0} \Big(1 - \tilde K_0(\mu) - Q \tilde K( r ,\mu)\Big) d\mu
+ \int^{+\infty}_{\mu_0} \Big(1 - \tilde \Gamma(\mu)\Big) d\mu \nonumber \\
&\dff& \min_{0 \leq r \leq R_m} \Delta \hat D(\mu_0, r).
\ee
Thus, we have
\be \label{equ.GaussianLB197}
\sup_{\mu_0} \min_{0 \leq r \leq R_m} \Delta \hat D(\mu_0, r) \leq \min_{\{\by_j\}_{1 \leq j \leq Q}}\Delta \hat D.
\ee
  \item For unbounded codebook, then $\tilde \Gamma(\mu) = 1$ for all $\mu \geq 0$, we have
\be \label{equ.GaussianLB204}
\min_{\{\by_j\}_{1 \leq j \leq Q}}\Delta \hat D(\mu_0)
&=& \inf_{r \geq 0} \int^{\mu_0}_{0} \Big(1 - \tilde K_0(\mu) - Q \tilde K( r ,\mu)\Big) d\mu \nonumber \\
&\dff& \inf_{r \geq 0} \Delta \hat D(\mu_0, r).
\ee
Thus we have
\be \label{equ.GaussianLB205}
\sup_{\mu_0 \geq 0} \inf_{r \geq 0} \Delta \hat D(\mu_0, r) \leq \min_{\{\by_j\}_{1 \leq j \leq Q}}\Delta \hat D.
\ee
\end{itemize}

\begin{proof}
Note that for any $\{\by_j\}_{1 \leq j \leq Q}$ we have $\Delta \hat D(\mu_0) \leq \Delta \hat D$.
Then, minimizing the left side among all $\{\by_j\}_{1 \leq j \leq Q}$ we have
\be \label{equ.GaussianLB200}
\min_{\{\by_j\}_{1 \leq j \leq Q}}\Delta \hat D(\mu_0) \leq \Delta \hat D;
\ee
and via minimizing the right side among all $\{\by_j\}_{1 \leq j \leq Q}$ we have
\be \label{equ.GaussianLB201}
\min_{\{\by_j\}_{1 \leq j \leq Q}}\Delta \hat D(\mu_0) \leq \min_{\{\by_j\}_{1 \leq j \leq Q}}\Delta \hat D,
\ee
and thus prove (\ref{equ.GaussianLB195}).

Then, according to (\ref{equ.GaussianLB194}), we have that
\be \label{equ.GaussianLB202}
\min_{\{\by_j\}_{1 \leq j \leq Q}} \Delta \hat D(\mu_0)
&=& \int^{\mu_0}_{0} \Big(1 - \tilde K_0(\mu)\Big) d\mu - \sum^Q_{j = 1} \max_{\by_j} \int^{\mu_0}_{0} \tilde K(\|\by_j\|,\mu) d\mu
+ \int^{+\infty}_{\mu_0} \Big(1 - \tilde \Gamma(\mu)\Big) d\mu \nonumber \\
&=& \int^{\mu_0}_{0} \Big(1 - \tilde K_0(\mu)\Big) d\mu - Q \max_{0 \leq r \leq R_m} \int^{\mu_0}_{0} \tilde K(r,\mu) d\mu
+ \int^{+\infty}_{\mu_0} \Big(1 - \tilde \Gamma(\mu)\Big) d\mu \nonumber \\
&=& \min_{0 \leq r \leq R_m} \int^{\mu_0}_{0} \Big(1 - \tilde K_0(\mu)- Q \tilde K(r,\mu)\Big) d\mu
+ \int^{+\infty}_{\mu_0} \Big(1 - \tilde \Gamma(\mu)\Big) d\mu \nonumber \\
&=& \min_{0 \leq r \leq R_m} \Delta \hat D(\mu_0, r),
\ee
and thus prove (\ref{equ.GaussianLB196}).

Finally, since
\be \label{equ.GaussianLB203}
\min_{0 \leq r \leq R_m} \Delta \hat D(\mu_0, r) \leq \min_{\{\by_j\}_{1 \leq j \leq Q}}\Delta \hat D,
\ee
for all $\mu_0 \geq 0$, we can select the supreme of the left side, and thus prove (\ref{equ.GaussianLB197}).
\end{proof}
\end{theorem}

\emph{Remark 3:}
It is easily seen that for $\mu_0 = 0$, $\Delta \hat D(\mu_0, r) \geq 0$ for all $r \geq 0$. Thus we have
\be \label{equ.GaussianLB206}
\sup_{\mu_0 \geq 0} \inf_{r \geq 0} \Delta \hat D(\mu_0, r) \geq \inf_{r \geq 0} \Delta \hat D(0, r) \geq 0.
\ee
Therefore, the lower bound $\sup_{\mu_0 \geq 0} \inf_{r \geq 0} \Delta \hat D(\mu_0, r)$ is well-defined.

\subsection{Upper Bound} \label{subsec.GaussianUB}
Due to the high computational complexity of the upper bound based on the reference rate, we only provide an upper
bound based on ordered statistics.

\subsubsection{Upper Bound for Unbounded Sources}
We derive an upper for Gaussian sources based on Theorem~\ref{thm.Thm602}.
Let $\by_b = {\bf 0}$ and $\B^n_\delta = \{\bx: \|\bx\|^2 \leq (1 + \delta)\sigma^2\}$,
such that for all $\bx \in \B^n_\delta$ we have that $d(\bx , \by_b) \leq 1 + \delta$.
In the following we evaluate each term involved in Theorem~\ref{thm.Thm602}
and specify the upper bound for Gaussian sources.

We evaluate the probability $\mathbb{P}\Big(d(\bx,\by) \leq t_\bx\Big)$ based on non-central chi-squared distribution.
Note that due to (\ref{equ.GaussianPDF3}), we have that
\be \label{equ.GaussianLB211}
P\Big(\|\by - \bx\|^2 \leq \beta^2\Big) = \Upsilon_n\Big(\frac{\beta^2}{\sigma^2 - D}, \frac{\|\bx\|^2}{\sigma^2 - D}\Big),
\ee
where $\Upsilon_(\cdot,\cdot)$ is the non-central chi-squared specified in (\ref{equ.ChiSquared5}).
Since given $\bx$, $P\Big(\|\by - \bx\|^2 \leq \beta^2\Big)$ is strictly increasing with $\beta^2$, we can
define the following inverse function,
\be \label{equ.GaussianLB212}
\frac{\beta^2}{\sigma^2 - D} = \Theta_n\Big(\frac{\|\bx\|^2}{\sigma^2 - D} , P_0\Big),
\ee
if $P_0 = P\Big(\|\by - \bx\|^2\Big)$.
Moreover, let $\upsilon_n(x)$ be the probability density function be the order-$n$ Chi-squared distribution given as follows,
\be \label{equ.ChiSquared2}
\upsilon_n(x) &=& \frac{1}{2^{n/2}\Gamma(n/2)} x^{n/2 - 1} e^{-x/2},
\ee
which is the pdf of the squared sum of $n$ independently unit Gaussian distributed variables.
We integrate in the $n$-dimensional space according to that squared sum and have the following result.

\begin{theorem} \label{thm.Thm140}
For any $\delta, \epsilon > 0$, we have the following upper bound
\be \label{equ.GaussianLB214}
\Bar D_Q &\leq& \int^{n(\sigma^2 + \delta)}_{0} \frac{1}{\sigma^2} \upsilon_n\Big(\frac{x}{\sigma^2}\Big)
\min\Big\{\frac{x}{n}, \frac{(\sigma^2 - D)}{n} \Theta_n\Big(\frac{x}{\sigma^2 - D} , \frac{\ln\frac{1}{\epsilon}}{Q - 2}\Big)\Big\} dx \nonumber \\
&& \ \ \ +  \int^{+\infty}_{n(\sigma^2 + \delta)} \frac{x}{\sigma^2} \upsilon_n\Big(\frac{x}{\sigma^2}\Big) dx + \epsilon (2\sigma^2 - D).
\ee

\begin{proof}
We analyze each term involved in Theorem~\ref{thm.Thm602}.
First, we have that
\be \label{equ.GaussianLB215}
\int_{\bar \B^n_\delta} p(\bx) d(\bx , \by_b)d\bx &=& \int_{\|\bx\|^2 \geq n(\sigma^2 + \delta)} \|\bx\|^2p(\bx) d\bx \nonumber \\
&=& \int^{+\infty}_{n(\sigma^2 + \delta)} \frac{x}{\sigma^2} \upsilon_n\Big(\frac{x}{\sigma^2}\Big) dx.
\ee

Then, according to Theorem~\ref{thm.Thm602}, we have the following
\be \label{equ.GaussianLB216}
&&\epsilon \int \int p(\bx) \hat q(\by) d(\bx , \by) d\bx d\by \nonumber \\
&=& \epsilon \int \int p(\bx) \hat q(\by) \frac{\|\bx - \by\|^2}{n} d\bx d\by \nonumber \\
&=& \epsilon \int \int p(\bx) \hat q(\by) \frac{\|\bx \|^2}{n} d\bx d\by
+ \epsilon \int \int p(\bx) \hat q(\by) \frac{\|\by\|^2}{n} d\bx d\by
- \epsilon \int \int p(\bx) \hat q(\by) \frac{2\bx^T\by}{n} d\bx d\by \nonumber \\
&=& \epsilon \int p(\bx) \frac{\|\bx \|^2}{n} d\bx + \epsilon \int \hat q(\by) \frac{\|\by\|^2}{n} d\by \nonumber \\
&=& \epsilon \sigma^2 + \epsilon (\sigma^2 - D) = \epsilon (2\sigma^2 - D).
\ee

Finally, given $\bx$, we have that the radius $t_\bx$ is given as follows,
\be \label{equ.GaussianLB217}
P\Big(\frac{\|\by - \bx\|^2}{n} \leq t_\bx\Big) = \Upsilon_n\Big(\frac{\|\bx\|^2}{\sigma^2 - D}, \frac{n t_\bx}{\sigma^2 - D}\Big)
= 1 - \epsilon^{\frac{1}{Q - 2}} \leq \frac{\ln \frac{1}{\epsilon}}{Q - 2},
\ee
and thus we have that
\be \label{equ.GaussianLB218}
t_\bx \leq \frac{(\sigma^2 - D)}{n} \Theta_n\Big(\frac{\|\bx\|^2}{\sigma^2 - D} , \frac{\ln\frac{1}{\epsilon}}{Q - 2}\Big).
\ee
Then, we have the following,
\be \label{equ.GaussianLB219}
\min\Big\{d(\bx, {\bf 0}) , t_\bx\Big\} \leq
\min\Big\{\frac{\|\bx\|^2}{n}, \frac{(\sigma^2 - D)}{n} \Theta_n\Big(\frac{\|\bx\|^2}{\sigma^2 - D} , \frac{\ln\frac{1}{\epsilon}}{Q - 2}\Big)\Big\},
\ee
and thus
\be \label{equ.GaussianLB2110}
&&\int_{\B^n_\delta} p(\bx) \min\Big\{d(\bx , {\bf 0}), t_\bx\Big\} d\bx \nonumber \\
&\leq& \int_{\|\bx\|^2 \leq n(\sigma^2 + \delta) } p(\bx)
\min\Big\{\frac{\|\bx\|^2}{n}, \frac{(\sigma^2 - D)}{n} \Theta_n\Big(\frac{\|\bx\|^2}{\sigma^2 - D} , \frac{\ln\frac{1}{\epsilon}}{Q - 2}\Big)\Big\} d\bx \nonumber \\
&=& \int^{n(\sigma^2 + \delta)}_{0} \frac{1}{\sigma^2} \upsilon_n\Big(\frac{x}{\sigma^2}\Big)
\min\Big\{\frac{x}{n}, \frac{(\sigma^2 - D)}{n} \Theta_n\Big(\frac{x}{\sigma^2 - D} , \frac{\ln\frac{1}{\epsilon}}{Q - 2}\Big)\Big\} dx.
\ee

Then, via combining (\ref{equ.GaussianLB215}), (\ref{equ.GaussianLB216}), and (\ref{equ.GaussianLB2110}),
we can prove (\ref{equ.GaussianLB214}).
\end{proof}
\end{theorem}

\subsubsection{Upper Bound for Bounded Sources Codewords}
We consider the upper bound for the bounded source codewords $\|\by_j\| \leq R_m$ for $1 \leq j \leq Q$.
Assume that the codewords satisfy the following distribution
\be \label{equ.GaussianLB220}
\tilde q(\by) = \frac{\hat q(\by)}{\int_{|\by| \leq R_m} \hat q(\by) d\by} \dff \frac{\hat q(\by)}{C_m},
\ee
for $\|\by\| \leq R_m$ and $\tilde q(\by) = 0$ otherwise.
We then analyze the three terms involved in Theorem~\ref{thm.Thm602} as follows.

First, the term $\int_{\bar \B^n_\delta} p(\bx) d(\bx,\by_b) d\bx$ is the same as that given in Theorem~\ref{thm.Thm140},
given by
\be \label{equ.GaussianLB221}
\int_{\bar \B^n_\delta} p(\bx) d(\bx,\by_b) d\bx = \int^{+\infty}_{n(\sigma^2 + \delta)} \frac{x}{\sigma^2} \upsilon_n\Big(\frac{x}{\sigma^2}\Big)dx.
\ee

Second, the term $\epsilon\int_{\bx}\int_{\by}p(\bx)\tilde q(\by)d(\bx , \by)d\by d\bx$ can be given as follows,
\be \label{equ.GaussianLB222}
\epsilon\int\int p(\bx)\tilde q(\by)d(\bx , \by)d\by d\bx
&=& \epsilon\int p(\bx)\frac{\|\bx\|^2}{n} d\bx + \epsilon\int \tilde q(\by)\frac{\|\by\|^2}{n} d\by \nonumber \\
&=& \epsilon \sigma^2 + \frac{\epsilon}{C_m} \int^{R^2_m}_{0} \frac{x}{\sigma^2} \upsilon_n\Big(\frac{x}{\sigma^2}\Big) dx.
\ee

Finally, for the term $\int_{\bx \in \B^n_\epsilon} p(\bx)\min\Big\{\frac{\|\bx\|^2}{n}, t_\bx \Big\} d\bx$,
the key point is to evaluate the $t_\bx$.
This is equivalent to evaluating the probability $\mathbb{P}\Big(\|\by - \bx\|^2 \leq t^2\Big)$,
which is shown in the following Theorem~\ref{thm.Thm141}.

\begin{theorem} \label{thm.Thm141}
The probability $\mathbb{P}\Big(\|\by - \bx\|^2 \leq t^2\Big)$ can be expressed as follows,
\begin{itemize}
  \item If $R_m + t \leq \|\bx\|$, we have that $\mathbb{P}\Big(\|\by - \bx\|^2 \leq t^2\Big) = 0$;
  \item otherwise, letting $\mathbb{P}\Big(\|\by - \bx\|^2 \leq t^2\Big) = P_1 + P_2$, where
\begin{itemize}
  \item $P_1 = \frac{1}{C_m}\Upsilon_n\Big(\frac{(t - \|\bx\|)^2}{\sigma^2 - D}\Big)$ for $\|\bx\| \leq t$ and $P_1 = 0$ for $\|\bx\| > t$;
  \item Let $r_{min} = \Big|\|\bx\| - t\Big|$, and $r_{max} = \min\{R_m, \|\bx\| + t\}$. We have that the probability $P_2$ can be given as follows,
\be \label{equ.GaussianLB223}
P_2 = \int^{r_{max}}_{r_{min}} \frac{1}{(2\pi(\sigma^2 - D))^{\frac{n}{2}}}e^{-\frac{r^2}{2(\sigma^2 - D)}} r^{n-1} \Omega_n(\theta(r)) dr,
\ee
where $\theta(r)$ can be specified as follows,
\be \label{equ.GaussianLB224}
\cos \theta(r) = \frac{r^2 + \|\bx\|^2 - t^2}{2r\|\bx\|}.
\ee
\end{itemize}
\end{itemize}
\begin{proof}
We prove using the computational methods in Section~\ref{subsec.InterTwoBalls}.
If $R_m + t \leq \|\bx\|$, then $\{\by: \|\by - \bx\|^2 \leq t^2\} \bigcap \{\by: \|\by\| \leq R_m\}$
does not have positive measure.
Therefore $\mathbb{P}\Big(\|\by - \bx\|^2 \leq t^2\Big) = 0$.

Otherwise, we consider the probability of the intersection $\U = \{\by: \|\by - \bx\| \leq t\} \cap \{\by: \|\by\| \leq R_m\}$.
If $\|\bx\| \leq t$, it contains a radius-$r$ ball for $r \leq t - \|\bx\|$.
Note that the probability of the radius-$(t - \|\bx\|)$ ball is
\be \label{equ.GaussianLB225}
P_1 &=& \int_{\|\by\|^2 \leq (t - \|\bx\|)^2} \tilde q(\by) d\by = \frac{1}{C_m}\int_{\|\by\|^2 \leq (t - \|\bx\|)^2} \hat q(\by) d\by \nonumber \\
&=& \frac{1}{C_m}\Upsilon_n\Big(\frac{(t - \|\bx\|)^2}{\sigma^2 - D}\Big).
\ee
Then, we consider the radius-$r$ sphere of $\|\by\|$ which is not entirely contained in $\U$, and integrate according to the radius of $\|\by\|$.
Via computation, it can be seen that the range of such $r$ is given by $r_{min} \leq r \leq r_{max}$.
For a radius $r$ sphere, $r_{min} \leq r \leq r_{max}$, let $\theta(r)$ be the semi-angle of its cone
contained in $\U$, which can be specified by (\ref{equ.GaussianLB224}).
We integrate this cone in the range $r_{min} \leq r \leq r_{max}$, and thus obtained
the probability $P_2$ given in (\ref{equ.GaussianLB223}).
\end{proof}
\end{theorem}

\subsection{Numerical Results}
\begin{figure}[htb]
\centering
\includegraphics[width = 0.8\textwidth]{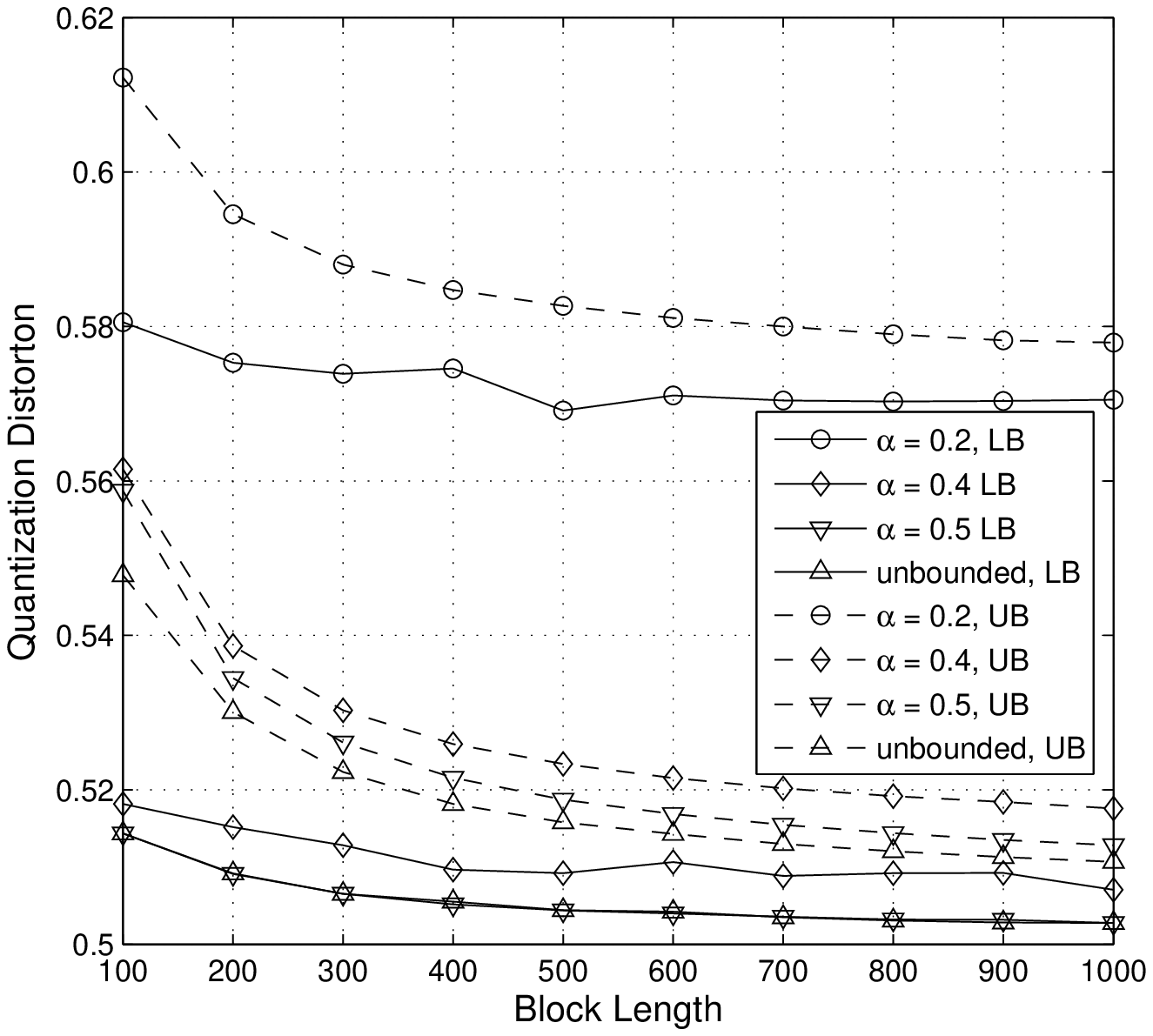}
\caption{\small Distortion for the unit Gaussian Source with Quantization Rate $1/2$.}
\label{fig.GS1}
\end{figure}

Assume identically independent distributed unit Gaussian source with variance $\sigma^2 = 1.0$ per dimension.
We show the upper and lower bounds for the optimal quantization for the quantization rate $R = 1/2$.
Again, according to the rate-distortion theory, the asymptotic lower bound for the quantization distortion
is given by $D^n(\hat q) = 0.5$.
We set the parameters $\epsilon = 0.005$ and $\delta = 0.50$.
We consider bounded and unbounded codebook constraints, and plot the upper and lower bounds
with respect to source sequence lengths $100 \leq n \leq 1000$, along with the infinite-length distortion $0.50$.
For bounded codewords, we plot in Fig.~\ref{fig.GS1} the upper and lower bounds for the bound $|\by_j|^2 \leq \alpha n$,
for $\alpha = 0.20$, $0.40$, and $0.50$.

It is seeing that the upper and lower bounds decrease with for larger $\alpha$.
Moreover, the upper and lower bounds for unbounded codebook ($\alpha = +\infty$) are very close to those for bounded codebook with $\alpha = 0.5$.
This can be explained by the typical set of the quantization codebook distribution $\by_j$ around the typical set, which is given by
\be \label{equ.TypicalSet}
\A^\zeta = \Big\{\by \Big| 0.5 - \zeta < \frac{|\by|^2}{n} < 0.5 + \zeta \Big\}.
\ee
The region $|\by_j|^2 \leq \alpha n$ for $\alpha = 0.5$ ``almost'' covers the typical set,
and thus further increasing the value of $\alpha$ will not bring significant decrease of the quantization distortion.


\section{Conclusions} \label{sec.Conclusions}
We have proposed the upper and lower bounds of the optimal quantization
for identically and independently distributed source in the finite-block length regime.
The lower bound can be proved to be larger than the asymptotical distortion of the rate-distortion theory.
The upper bounds can be proved to approach the asymptotical distortion of the rate-distortion theory.
We have also applied the upper and lower bounds to binary symmetric source, binary non-symmetric source, and Gaussian source.
Numerical results show reasonable gap between the upper and lower bounds.

One important open question is the one-curve approximation of the quantization distortion.
For the finite-block length regime of the block error probability for channel codes,
the one-curve Gaussian approximation is first proposed in~\cite{GaussianApproximationCC} and then refined in~\cite{FiniteLength},
based on the Gaussian approximation of the Neyman-Pearson detection.
For the rate-distortion counterparts, a possibly feasible way to solving this question is
from the lower bound given by \emph{Corollary~3}, which remains to be an open question for further research.

\section{Appendix - Background Knowledge} \label{sec.Appendix}

\subsection{Chi-squared and Non-centralized Chi-squared Distributions}
\subsubsection{Chi-squared Distribution}
Consider $n$ independent Gaussian random variables $Z_j \sim \calN(0,\sigma^2_j)$ for $1 \leq j \leq n$,
and
\be \label{equ.ChiSquared1}
Q = \sum^n_{j = 1} \Big(\frac{Z_j}{\sigma_j}\Big)^2.
\ee
Then, $Q$ satisfies the order-$n$ Chi-squared distribution, denoted as $\chi^2(n)$.

The probability density function and cumulative distribution function for the order-$n$ Chi-squared distribution,
denoted as $\upsilon_n(x)$ and $\Upsilon_n(x)$, respectively, are given as follows,
\be \label{equ.ChiSquared2}
\upsilon_n(x) &=& \frac{1}{2^{n/2}\Gamma(n/2)} x^{n/2 - 1} e^{-x/2}, \ \mbox{and} \ \Upsilon_n(x) = \int^x_{0} \upsilon_n(t) dt. 
\ee

\subsubsection{Non-centered Chi-squared Distribution}
Consider $n$ independent Gaussian random variables $Z_j \sim \calN(\mu_j,\sigma^2_j)$ for $1 \leq j \leq n$.
Let
\be \label{equ.ChiSquared3}
Q = \sum^n_{j = 1} \Big(\frac{Z_j}{\sigma_j}\Big)^2, \ \mbox{and} \ \lambda = \sum^n_{j = 1} \Big(\frac{\mu_j}{\sigma_j}\Big)^2.
\ee
Then, $Q$ satisfies the order-$n$ noncentral Chi-squared distribution with the noncentrality parameter $\lambda$.
The probability density function and cumulative distribution function, denoted as $\upsilon_n(x,\lambda)$ and $\Upsilon_n(x,\lambda)$,
respectively, are given as follows,
\be \label{equ.ChiSquared5}
\upsilon_n(x,\lambda) = \sum^{+\infty}_{i = 0} \frac{e^{-\lambda/2}(\lambda/2)^i}{i!} \upsilon_{n + 2i}(x), \ \mbox{and} \
\Upsilon_n(x,\lambda) = \int^x_{0} \upsilon_n(t,\lambda) dt.
\ee

Another equivalent definition for non-centered Chi-squared distribution is given as follows.
Consider $n$ independent Gaussian random variables $Z_j \sim \calN(0,\sigma^2_j)$ for $1 \leq j \leq n$,
and fixed $y_j$ for $1 \leq j \leq n$. Let
\be \label{equ.ChiSquared6}
Q = \sum^n_{j = 1} \Big(\frac{Z_j - y_j}{\sigma_j}\Big)^2.
\ee
Then, $Q$ satisfies order-$n$ noncentral Chi-squared distribution with noncentrality parameter
\be \label{equ.ChiSquared7}
\lambda = \sum^n_{j = 1} \Big(\frac{y_j}{\sigma_j}\Big)^2.
\ee

\subsection{Computational Methods for the Intersection of Two Balls} \label{subsec.InterTwoBalls}
We introduce a methods for computing the volume and probability measure for the intersection of two balls.
More specifically, we consider the following two balls
\be \label{equ.TwoBalls1}
\C_0 = \{\bx: \|\bx\|^2 \leq r^2_0\} \ \mbox{and} \ \C_1 = \{\bx: \|\bx - \bx_1\|^2 \leq r^2_1\},
\ee
as well as the following Gaussian distribution,
\be \label{equ.TwoBalls2}
p(\bx) = \frac{1}{(2\pi\sigma^2)^{\frac{n}{2}}}e^{-\frac{\|\bx\|^2}{2\sigma^2}}.
\ee
For a set $S \subseteq \mathbb{R}^n$, let $\V_n(S)$ be the volume of $S$,
and $\mathbb{P}(S)$ be the probability measure of $S$ under the distribution $p(\bx)$.
In the following we present a computational method for the volume and probability of
$\C_1 \setminus \C_0$ and $\C_1 \cap \C_0$.

The area of a unit sphere in $\R^n$, denoted as $A_n$, is given as follows
\be \label{equ.GaussianLB101}
A_n = \frac{2\pi^{n/2}}{\Gamma(n/2)};
\ee
and the volume of a unit ball, denoted as $V_n$, is given by
\be \label{equ.GaussianLB102}
V_n = \frac{A_n}{n} = \frac{\pi^{n/2}}{\Gamma(1 + n/2)}.
\ee

\begin{figure}[htbp]
\center{ \epsfxsize=6.0in \centerline{\epsffile{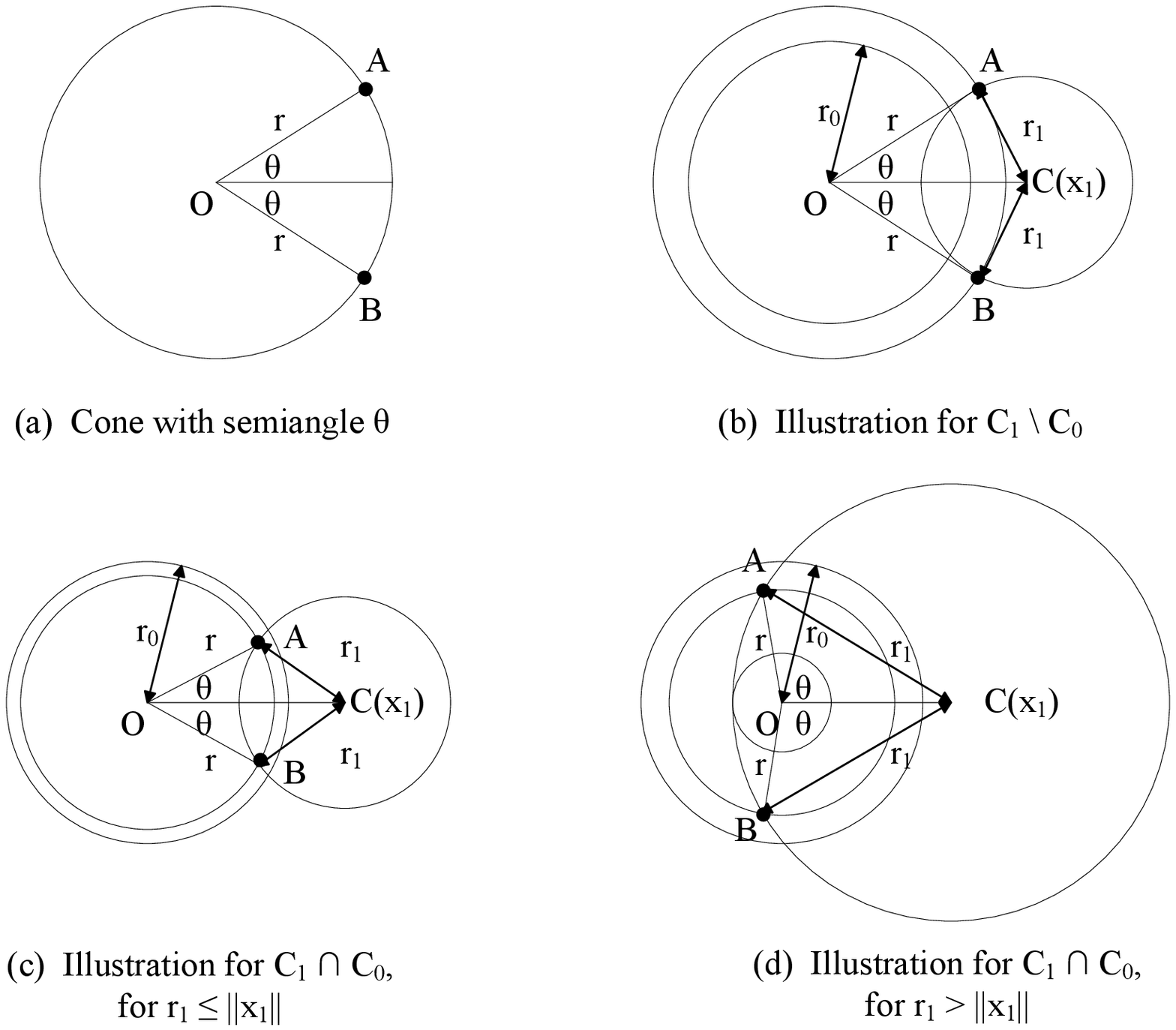}}
\caption{\small The illustration of a cone with semiangle $\theta$, the parts $\C_1 \setminus \C_0$, and $\C_1 \cap \C_0$.}
\label{fig.Intersection}}
\end{figure}

\subsubsection{Sphere Area of a Cone}
Consider the radius-$r$ sphere in $\R^n$ cut out by a cone with semiangle $\theta$,
as shown in Fig.~\ref{fig.Intersection} (a).
The area of that cone, denoted as $\Omega_n(\theta)$, is given by
\be \label{equ.GaussianLB100}
\Omega_n(\theta) = \frac{2\pi^{(n - 1)/2}}{\Gamma((n - 1)/2)} \int^{\theta}_{0} (\sin\phi)^{n - 2} d\phi,
\ee
where $\Gamma(\cdot)$ is gamma function.

\subsubsection{Volume and Probability of $\C_1 \setminus \C_0$}
The volume of $\C_1 \setminus \C_0$ can be computed via integrating the intersection of a radius-$r$ sphere with $\C_1 \setminus \C_0$.
Note that the intersection is not empty for $r_0 \leq r \leq \|\bx_1\| + r_1$.
The semiangle of the cone cut out, denoted as $\theta(r)$, can be specified by triangle with the lengths of three edges, $r$, $\|\bx_1\|$, and $r_1$,
where $\theta(r)$ is between the two edges of lengths $r$ and $\|\bx_1\|$.
Thus, we have
\be \label{equ.TwoBalls3}
\theta(r) = \cos^{-1} \frac{\|\bx_1\|^2 + r^2 - r^2_1}{2\|\bx_1\|r};
\ee
and the area of the sphere cut out is given by $r^{n-1}\Omega_n(\theta(r))$.

The volume $\V_n\Big(\C_1 \setminus \C_0\Big)$ is computed via integrating the area of radius-$r$ sphere
with semi-angle $\theta(r)$ in the range $r_0 \leq r \leq \|\bx_1\| + r_1$, given by
\be \label{equ.TwoBalls4}
\V_n\Big(\C_1 \setminus \C_0\Big) = \int^{\|\bx_1\| + r_1}_{r_0} r^{n - 1} \Omega_n(\theta(r)) dr;
\ee
and the probability $\mathbb{P}_n\Big(\C_1 \setminus \C_0\Big)$ is computed via integrating the area weighted by the probability density $p(\bx)$ for $\|\bx\| = r$,
given by
\be \label{equ.TwoBalls5}
\mathbb{P}\Big(\C_1 \setminus \C_0\Big) = \int^{\|\bx_1\| + r_1}_{r_0} \frac{1}{(2\pi \sigma^2)^{\frac{n}{2}}} e^{-\frac{r^2}{2\sigma^2}}
r^{n - 1} \Omega_n(\theta(r)) dr.
\ee

\subsubsection{Volume and Probability of $\C_1 \cap \C_0$ for $r_1 \leq \|\bx_1\|$}
For $r_1 \leq \|\bx_1\|$, the center of $\C_0$ is not contained in the interior of $\C_1$, as shown in Fig.~\ref{fig.Intersection} (c).
The volume is computed via integrating the area $r^{n-1}\Omega_n(\theta(r))$ among $\|\bx_1\| - r_1 \leq r \leq r_0$,
where $\theta(r)$ is specified by (\ref{equ.TwoBalls3}), given by
\be \label{equ.TwoBalls6}
\V_n\Big(\C_1 \cap \C_0\Big) = \int^{r_0}_{\|\bx_1\| - r_1} r^{n - 1} \Omega_n(\theta(r)) dr;
\ee
and the probability  $\mathbb{P}\Big(\C_1 \setminus \C_0\Big)$ is computed via integrating the sphere area weighted by the probability density, given by
\be \label{equ.TwoBalls7}
\mathbb{P}\Big(\C_1 \cap \C_0\Big) =  \int^{r_0}_{\|\bx_1\| - r_1}
\frac{1}{(2\pi \sigma^2)^{\frac{n}{2}}} e^{-\frac{r^2}{2\sigma^2}} r^{n - 1} \Omega_n(\theta(r)) dr.
\ee

\subsubsection{Volume and Probability of $\C_1 \cap \C_0$ for $r_1 \geq \|\bx_1\|$}
For $\|\bx_1\| \geq r_1$, $\C_1 \cap \C_0$ contains a ball of radius $\|\bx_1\| - r_1$.
The volume of the rest part can be computed via integrating $r^{n-1}\Omega_n(\theta(r))$ among $r_1 - \|\bx_1\| \leq r \leq r_0$,
where $\theta(r)$ is specified by (\ref{equ.TwoBalls3}).
Then, we have
\be \label{equ.TwoBalls8}
\V_n\Big(\C_1 \cap \C_0\Big) = \Big(r_1 - \|\bx_1\|\Big)^n V_n + \int^{r_0}_{r_1 - \|\bx_1\|} r^{n - 1} \Omega_n(\theta(r)) dr;
\ee
and $\mathbb{P}\Big(\C_1 \setminus \C_0\Big)$ can be computed by the sum probability of the two parts, given by
\be \label{equ.TwoBalls9}
\mathbb{P}\Big(\C_1 \cap \C_0\Big) = \Upsilon_n\Big((\frac{r_1 - \|\bx_1\|}{\sigma})^2\Big)
+ \int^{r_0}_{r_1 - \|\bx_1\|} \frac{1}{(2\pi \sigma^2)^{\frac{n}{2}}} e^{-\frac{r^2}{2\sigma^2}} r^{n - 1} \Omega_n(\theta(r)) dr.
\ee

\section{Appendix - Proof of Theorems} \label{sec.Appendix2}
\subsection{Proof of Theorem~\ref{thm.Thm603}}
Let $s_\bx = \min\{d(\bx,\by_b), t_\bx\}$. Then, from (\ref{equ.Achieve238}) and Theorem~\ref{thm.Thm602} we have that
\be \label{equ.Achieve290}
\bar F(\bx , s_\bx) \geq \epsilon^{\frac{1}{Q - 2}},
\ee
and that the probability
\be \label{equ.Achieve291}
\mathbb{P}\Big(d(\bx , \by) < s_\bx\Big) \leq 1 - \epsilon^{\frac{1}{Q - 2}}.
\ee
Note that, due to the convexity of the function $\epsilon^x$ in terms of $x$, we have the following
\be \label{equ.Achieve292}
\mathbb{P}\Big(d(\bx , \by) < s_\bx\Big) &\leq& 1 - \epsilon^{\frac{1}{Q - 2}} = \epsilon^{0} - \epsilon^{\frac{1}{Q - 2}} \nonumber \\
&\leq& \frac{1}{Q - 2} \frac{\partial \epsilon^x}{\partial x}\Big|_{x = 0} = \frac{1}{Q - 2} \ln \frac{1}{\epsilon}.
\ee

To prove (\ref{equ.Achieve276}), we have the following
\be \label{equ.Achieve293}
\Delta \tilde D^n_Q &\dff& \int_{\B^n_\epsilon} p(\bx) s_\bx d\bx
- \int_{\B^n_\epsilon} \int p(\bx) \hat q_0(\by|\bx) d(\bx,\by) d\by d\bx \nonumber \\
&\leq& \int_{\B^n_\epsilon} p(\bx) s_\bx d\bx - \int_{\B^n_\epsilon} \int p(\bx) \hat q_0(\by|\bx) d(\bx,\by) d\by d\bx \nonumber \\
&=& \int_{\B^n_\epsilon} p(\bx) d\bx \int \Big(s_{\bx} - d(\bx , \by)\Big) \hat q_0(\by|\bx) d\by \nonumber \\
&\leq& \int_{\B^n_\epsilon} p(\bx) d\bx \int \Big(s_{\bx} - d(\bx , \by)\Big) \hat q_0(\by|\bx) \mathbf{1}_{s_{\bx} > d(\bx , \by)} d\by.
\ee
Since for $\bx \in \B^n_\epsilon$, $s_\bx = \min\{d(\bx,\by_b), t_\bx\} \leq d(\bx,\by_b) \leq d + \epsilon$, we have
\be \label{equ.Achieve294}
\Big(s_{\bx} - d(\bx , \by)\Big) \cdot \mathbf{1}_{s_{\bx} > d(\bx , \by)} \leq (\hat d + \epsilon) \cdot \mathbf{1}_{s_{\bx} > d(\bx , \by)},
\ee
Based on this, we have the following to further bound $\Delta \tilde D_Q$ based on (\ref{equ.Achieve293}),
\be \label{equ.Achieve294}
\Delta \tilde D^n_Q &\leq& \int_{\B^n_\epsilon} p(\bx) \int (\hat d + \epsilon) \hat q_0(\by|\bx) \mathbf{1}_{s_{\bx} > d(\bx , \by)} d\by d\bx \nonumber \\
&\leq& (\hat d + \epsilon) \int \int p(\bx)\hat q_0(\by|\bx) \mathbf{1}_{s_{\bx} > d(\bx , \by)} d\by d\bx \nonumber \\
&=& (\hat d + \epsilon) \int \int \hat q_0(\by)\hat q_0(\bx|\by) \mathbf{1}_{s_{\bx} > d(\bx , \by)} d\by d\bx.
\ee

Using Holder inequality, we can further derive the upper bound for any $0 < \beta < 1$, as follows,
\be \label{equ.Achieve295}
\Delta \tilde D^n_Q &\leq& (\hat d + \epsilon) \int \int \hat q_0(\by)\hat q_0(\bx|\by) \mathbf{1}_{s_{\bx} > d(\bx , \by)} d\by d\bx \nonumber \\
&\leq& (\hat d + \epsilon) \int \Big(\int \hat q(\by) \hat q_0(\bx|\by)^{1 / \beta} d\by\Big)^\beta
\Big(\int \hat q(\by)\mathbf{1}_{s_{\bx} > d(\bx , \by)}  d\by\Big)^{1 - \beta} d\bx \ \mbox{[Holder]} \nonumber \\
&\leq& (\hat d + \epsilon) \int \Big(\int \hat q_0(\by) \hat q_0(\bx|\by)^{1 / \beta} d\by\Big)^\beta
\Big(\frac{1}{Q - 2}\ln \frac{1}{\epsilon}\Big)^{1 - \beta} d\bx \nonumber \\
&\leq& (\hat d + \epsilon) \ln \frac{1}{\epsilon} \int \Big(\int \hat q_0(\by) \hat q_0(\bx|\by)^{1 / \beta} d\by\Big)^\beta
\Big(\frac{1}{Q - 2}\Big)^{1 - \beta} d\bx \nonumber \\
&=& (\hat d + \epsilon) \ln \frac{1}{\epsilon} \cdot \frac{2^{n(1 - R)}}{2^{n(1 - R)} - 2}
\Big(2^{-(1 - \beta)R}\int \Big(\int \hat q_0(y) \hat q_0(x|y)^{1 / \beta} dy\Big)^\beta dx \Big)^n.
\ee

Note that since~\cite{DiscreteSources},
\be \label{equ.Achieve296}
\lim_{\beta \uparrow 1} \frac{\log_2 \int \Big(\int \hat q_0(y) \hat q_0(x|y)^{1 / \beta} dy\Big)^\beta dx}{1 - \beta} = I(\hat q_0) = R_0,
\ee
we have that there exists a $\beta_0$ sufficient close to $1$, such that
\be \label{equ.Achieve297}
\frac{\log_2 \int \Big(\int \hat q_0(y) \hat q_0(x|y)^{1 / \beta_0} dy\Big)^{\beta_0} dx}{1 - \beta_0} < \frac{R + R_0}{2},
\ee
such that
\be \label{equ.Achieve298}
2^{-(1 - \beta_0)R}\int \Big(\int \hat q_0(y) \hat q_0(x|y)^{1 / \beta_0} dy\Big)^{\beta_0} dx < 2^{-(1 - \beta_0)(R - R_0)},
\ee
and thus
\be \label{equ.Achieve299}
\Delta \tilde D^n_Q \leq (\hat d + \epsilon) \ln \frac{1}{\epsilon} \cdot 2^{-(1 - \beta_0)(R - R_0)n}.
\ee

Finally, since $\tilde D_0 < \infty$ [c.f. (\ref{equ.Achieve275})], we have that
\be \label{equ.Achieve2910}
\tilde D^n_Q - \int \int p(\bx) \hat q_0(\by|\bx) d(\bx,\by) d\bx d\by
\leq (\hat d + \epsilon) \ln \frac{1}{\epsilon} \cdot 2^{-(1 - \beta_0)(R - R_0)n} + \epsilon \tilde D_0.
\ee
Then, for sufficiently small $\zeta > 0$, we select a sufficient small $\epsilon$, the second term $\epsilon \tilde D_0 < \zeta / 2$;
and given the selected $\epsilon$, for sufficient large $n$ have that the first terms is also smaller than $\zeta / 2$,
and thus we have that
\be \label{equ.Achieve2911}
\tilde D^n_Q < \int \int p(\bx) \hat q_0(\by|\bx) d(\bx,\by) d\bx d\by  + \zeta.
\ee

\subsection{Proof of Theorem~\ref{thm.Thm101}}
For any $\bx \in \B^n_\delta$, we let $\Psi(\bx , \by) = d + \delta$ for $\by \in A_\bx(\gamma_\bx)$,
$\Psi(\bx , \by) = l_\bx$ for $\by \in \bar A_\bx(\gamma_\bx)$, and $\Psi(\bx , \by) = 0$ for $\by \in A^c_\bx(\gamma_\bx)$;
and thus we have that
\be \label{equ.RefinedUB13}
\int \hat q_0(\by) \Psi(\bx , \by) d\by = \tilde B(\bx).
\ee

We can write the right side of (\ref{equ.RefinedUB12}) as follows
\be \label{equ.RefinedUB14}
\int_{\B^n_\delta} p(\bx) h^U(\bx) d\bx &=& \int_{\B^n_\delta} \int p(\bx) \hat q_0(\by|\bx) \Psi(\bx , \by) d\bx d\by \nonumber \\
&=& \int_{\B^n_\delta} \int \hat q_0(\by) \hat q_0(\bx|\by) \Psi(\bx , \by) d\bx d\by.
\ee
For any $0 < \beta < 1$, also from Holder inequality we have the following
\be \label{equ.RefinedUB15}
\tilde D^n_Q - D(R_0)
&\leq&\int_{\B^n_\delta} \int \hat q_0(\bx) \hat q_0(\bx|\by) \Psi(\bx , \by) d\bx d\by \nonumber \\
&\leq& \int_{\B^n_\delta} \Big(\int \hat q_0(\by) \hat q_0(\bx|\by)^{1 / \beta} d\by\Big)^\beta
\Big(\int \hat q_0(\by) \Psi(\bx , \by)^{1 / (1 - \beta)} d\by\Big)^{1 - \beta}d\bx \nonumber \\
&\leq& \int_{\B^n_\delta} \Big(\int \hat q_0(\by) \hat q_0(\bx|\by)^{1 / \beta} d\by\Big)^\beta
\Big(\int \hat q_0(\by) \Psi(\bx , \by) (d + \delta)^{\beta / (1 - \beta)} d\by\Big)^{1 - \beta}d\bx \nonumber \\
&=& (d + \delta)^{\beta} \int_{\B^n_\delta} \Big(\int \hat q_0(\by) \hat q_0(\bx|\by)^{1 / \beta} d\by\Big)^\beta
\Big(\int \hat q_0(\by) \Psi(\bx , \by) d\by\Big)^{1 - \beta}d\bx \nonumber \\
&\leq& (d + \delta)^{\beta} \int_{\B^n_\delta} \Big(\int \hat q_0(\by) \hat q_0(\bx|\by)^{1 / \beta} d\by\Big)^\beta
\Big(\frac{\mathbb{E}\Big(\rho^{0}_\bx\Big)}{Q(e - \eta)}\Big)^{1 - \beta}d\bx \nonumber \\
&\leq& (d + \delta)^{\beta} \Big(\frac{d + \delta}{e - \eta}\Big)^{1 - \beta}
 \frac{1}{Q^{1 - \beta}} \int_{\B^n_\delta} \Big(\int \hat q_0(\by) \hat q_0(\bx|\by)^{1 / \beta} d\by\Big)^\beta d\bx \nonumber \\
&\leq& \frac{d + \delta}{(e - \eta)^{1 - \beta}}
\frac{1}{Q^{1 - \beta}} \int \Big(\int \hat q_0(\by) \hat q_0(\bx|\by)^{1 / \beta} d\by\Big)^\beta d\bx \nonumber \\
&\leq& l_m^{\beta} \Big(\frac{\rho_m}{e - \eta}\Big)^{1 - \beta} \Big(2^{-(1 - \beta)R} \int \Big(\int \hat q_0(y) \hat q_0(x|y)^{1 / \beta} dy\Big)^\beta dx\Big)^n.
\ee
Then, similar to the arguments in the proof of Theorem~\ref{thm.Thm603}, we have that for $\beta$ sufficient close to $1$,
\be \label{equ.RefinedUB16}
2^{-(1 - \beta)R} \int \Big(\int \hat q_0(y) \hat q_0(x|y)^{1 / \beta} dy\Big)^\beta dx < 1,
\ee
and thus according to (\ref{equ.RefinedUB15}), the upper bound of $\tilde D_Q - D(\hat q)$
exponentially attenuates with the codeword block length $n$.

\subsection{Proof of Theorem~\ref{thm.Thm135}}
Note that
\be \label{equ.GaussianLB174}
\V_n\Big(\cup^Q_{j = 0} \C_j(t)\Big) \leq \V_n\Big(\C_0(t)\Big) + \sum^Q_{j = 1} \V_n\Big(\C_j(t) \setminus \C_0(t)\Big).
\ee
Then we need to prove that
\be \label{equ.GaussianLB175}
\V_n\Big(\C_j(t) \setminus \C_0(t)\Big) \leq \V_n\Big(\tilde \C_j(t) \setminus \C_0(t)\Big).
\ee
Note that, for $\C_j(t)$ the center and radius are $\by_j$ and $\sqrt{R(t , \|\by_j\|)}$, respectively;
and for $\tilde \C_j(t)$ they are $\tilde \by_j$ and $\sqrt{R(t , R_m)}$, respectively.
We define another ball with center $\by_j$ and radius $\sqrt{R(t , R_m)}$, denoted as $\hat \C_j(t)$.
Then, we can prove
\be \label{equ.GaussianLB176}
\V_n\Big(\C_j(t) \setminus \C_0(t)\Big) \leq \V_n\Big(\hat \C_j(t) \setminus \C_0(t)\Big) \leq \V_n\Big(\tilde \C_j(t) \setminus \C_0(t)\Big).
\ee
The first inequality follows that fixing the center the volume $\V_n\Big(\hat \C_j(t) \setminus \C_0(t)\Big)$ increases via
increasing radius from $\sqrt{R(t,\|\by_j\|)}$ to $\sqrt{R(t,R_m)}$;
and the second inequality follows that fixing the radius the volume $\V_n\Big(\tilde \C_j(t) \setminus \C_0(t)\Big)$
increases via increasing the distance between the centers of the two balls.
Then, from (\ref{equ.GaussianLB174}) and (\ref{equ.GaussianLB176}) we have
\be \label{equ.GaussianLB177}
\V_n\Big(\cup^Q_{j = 0} \C_j(t)\Big) \leq \V_n\Big(\C_0(t)\Big) + \sum^Q_{j = 1} \V_n\Big(\tilde \C_j(t) \setminus \C_0(t)\Big).
\ee

Via letting $\bar V_n = \V_n\Big(\cup^Q_{j = 0} \C_j(t)\Big)$, then according to $\bar V_n \leq \tilde V_n$ [c.f. (\ref{equ.GaussianLB170})],
we have
\be \label{equ.GaussianLB178}
\mathbb{P}\Big(\cup^Q_{j = 0} \C_j(t)\Big) \leq \max_{S, \V_n(S) \leq \bar V_n}\mathbb{P}\Big(S\Big)
\leq \max_{S, \V_n(S) \leq \tilde V_n}\mathbb{P}\Big(S\Big).
\ee
It is well known that, the given the volume $\V_n(S)$, the set $S$ that maximizes $\mathbb{P}\Big(S\Big)$
is the ball centered at the origin.
Note that the radius of such ball is given by
\be \label{equ.GaussianLB179}
r_n = \Big(\frac{\tilde V_n}{V_n}\Big)^{\frac{1}{n}};
\ee
and the probability mass is given by the chi-square cdf $\Upsilon_n(r^2_n)$.
Thus we have
\be \label{equ.GaussianLB1710}
\mathbb{P}\Big(\cup^Q_{j = 0} \C_j(t)\Big) \leq \max_{S, \V_n(S) \leq \tilde V_n}\mathbb{P}\Big(S\Big) = \Upsilon_n(r^2_n).
\ee

Also, since for all $\bx \in \cup^Q_{j = 0} \C_j(t)$, there exist a $j$, $0 \leq j \leq Q$,
such that $\|\bx - \frac{\sigma^2}{\sigma^2 - D}\by_j\| \leq \sqrt{R(t,\|\by_j\|)}$,
and thus
\be \label{equ.GaussianLB1711}
\|\bx\| \leq \frac{\sigma^2}{\sigma^2 - D}\|\by_j\| + \sqrt{R(t,\|\by_j\|)} \dff \tilde r_n,
\ee
and thus the probability
\be \label{equ.GaussianLB1712}
\mathbb{P}\Big(\cup^Q_{j = 0} \C_j(t)\Big) \leq \mathbb{P}\Big(\|\bx\| \leq \tilde r_n\Big) = \Upsilon_n(\tilde r^2_n).
\ee

Therefore, from (\ref{equ.GaussianLB1710}) and (\ref{equ.GaussianLB1712}), we have that
\be \label{equ.GaussianLB1713}
\mathbb{P}\Big(\cup^Q_{j = 0} \C_j(t)\Big) \leq \min\Big\{\Upsilon_n(r^2_n), \Upsilon_n(\tilde r^2_n)\Big\}
= \Upsilon_n(\tilde r^2_E),
\ee
where $r_E = \min\{r_n , \tilde r_n\}$.

\small{\baselineskip = 10pt
\bibliographystyle{./IEEEtran}
\bibliography{./gong_11}
}

\newpage

\end{document}

%% file: pream_bm.tex
\newtheorem{prop}{Proposition}

\newtheorem{cor}{Corollary}

\newtheorem{lm}{Lemma}

\newtheorem{thm}{Theorem}

\newcommand{\be}{\begin{eqnarray}}
\newcommand{\ee}{\end{eqnarray}}
\newcommand{\benn}{\begin{eqnarray*}}
\newcommand{\eenn}{\end{eqnarray*}}
\def\IR{\rm I \kern-0.20em R}
\newcommand{\utwi}[1]{\mbox{\boldmath $ #1$}}

\newcommand{\bthm}{\begin{thm}}
\newcommand{\ethm}{\end{thm}}

\newcommand{\bcor}{\begin{cor}}
\newcommand{\ecor}{\end{cor}}
\newcommand{\bprop}{\begin{prop}}
\newcommand{\eprop}{\end{prop}}
\newcommand{\blm}{\begin{lm}}
\newcommand{\elm}{\end{lm}}
\newcommand{\beq}{\begin{equation}}
\newcommand{\eeq}{\end{equation}}
\newcommand{\ber}{\begin{eqnarray}}
\newcommand{\eer}{\end{eqnarray}}

\newcommand{\bproof}{\begin{proof}}
\newcommand{\eproof}{\end{proof}}

\newcommand{\argmax}{\mathop{\mbox{\rm arg\,max}}}

\newcommand{\bit}{\begin{itemize}}
\newcommand{\eit}{\end{itemize}}
\newcommand{\ben}{\begin{enumerate}}
\newcommand{\een}{\end{enumerate}}
\newcommand{\bdesc}{\begin{description}}
\newcommand{\edesc}{\end{description}}
\newcommand{\beqarrn}{\begin{eqnarray*}}
\newcommand{\eeqarrn}{\end{eqnarray*}}
\newcommand{\bproofof}{\begin{proofof}}
\newcommand{\eproofof}{\end{proofof}}
\newenvironment{rem}{\begin{trivlist}\item[]{\bf
Remark:}\hspace{4mm}}{\end{trivlist}}
\newcommand{\brem}{\begin{rem}}
\newcommand{\erem}{\end{rem}}
\newenvironment{rems}{\begin{trivlist}\item[]{\bf
Remarks}\begin{itemize}}{\end{itemize}\end{trivlist}}
\newcommand{\brems}{\begin{rems}}
\newcommand{\erems}{\end{rems}}
\newtheorem{fact}{Fact}
\newcommand{\bfact}{\begin{fact}}
\newcommand{\efact}{\end{fact}}
\newtheorem{examp}{Example}
\newcommand{\bexamp}{\begin{examp}\rm}
\newcommand{\eexamp}{\end{examp}}
\newtheorem{defn}{Definition}
\newcommand{\bdefn}{\begin{defn}\rm}
\newcommand{\edefn}{\end{defn}}

\newtheorem{alg}{Algorithm}
\newcommand{\balg}{\begin{alg}}
\newcommand{\ealg}{\end{alg}}

\newtheorem{prob}{Problem}
\newcommand{\bprob}{\begin{prob}}
\newcommand{\eprob}{\end{prob}}

\newcommand{\bvtm}{\begin{verbatim}}
\newcommand{\bfig}{\begin{figure}}
\newcommand{\efig}{\end{figure}}
\newcommand{\bcen}{\begin{center}}
\newcommand{\ecen}{\end{center}}

\long\def\comment#1{}

\def \n2{{N_0 \over 2}}

\def \h5{\hspace{0.5in}}

\newcommand{\bx}{{\utwi{x}}}
\newcommand{\by}{{\utwi{y}}}

\newcommand{\dff}{\stackrel{\triangle}{=}}

%% file: RateDistortion_0609.bbl
\begin{thebibliography}{10}
\providecommand{\url}[1]{#1}
\def\UrlFont{\rmfamily}
\providecommand{\newblock}{\relax}
\providecommand{\bibinfo}[2]{#2}
\providecommand\BIBentrySTDinterwordspacing{\spaceskip=0pt\relax}
\providecommand\BIBentryALTinterwordstretchfactor{4}
\providecommand\BIBentryALTinterwordspacing{\spaceskip=\fontdimen2\font plus
\BIBentryALTinterwordstretchfactor\fontdimen3\font minus
  \fontdimen4\font\relax}
\providecommand\BIBforeignlanguage[2]{{%
\expandafter\ifx\csname l@#1\endcsname\relax
\typeout{** WARNING: IEEEtran.bst: No hyphenation pattern has been}%
\typeout{** loaded for the language `#1'. Using the pattern for}%
\typeout{** the default language instead.}%
\else
\language=\csname l@#1\endcsname
\fi
#2}}

\bibitem{ElementsInfoTheory}
T.~M. Cover and J.~A. Thomas, \emph{Elements of Information Theory},
  2nd~ed.\hskip 1em plus 0.5em minus 0.4em\relax Wiley Interscience, 2006.

\bibitem{ExtremeProperty}
R.~Ahlswede, ``Extremal properties of rate-distortion functions,'' \emph{IEEE
  Trans. Info. Theory}, vol.~36, no.~1, pp. 166--171, Jan. 1990.

\bibitem{EstimateRD}
M.~T. Harrison and I.~Kontoyiannis, ``Estimation of the rate distortion
  function,'' \emph{IEEE Trans. Info. Theory}, vol.~54, no.~8, pp. 3757--3762,
  Aug. 2008.

\bibitem{DSCRD}
D.~Krithivasan and S.~S. Pradhan, ``Distributed source coding using abelian
  group codes: A new achievable rate-distortion region,'' \emph{IEEE Trans.
  Info. Theory}, vol.~57, no.~3, pp. 1495--1519, Mar. 2011.

\bibitem{RDFeedforward}
R.~Venkataramanan and S.~S. Pradhan, ``Source coding with feed-forward:
  Rate-distortion theorems and error exponents for a general source,''
  \emph{IEEE Trans. Info. Theory}, vol.~53, no.~6, pp. 2154--2179, Jun. 2007.

\bibitem{RDMemory}
Y.-Q. Zhang, R.~L. Pickholtz, and M.~H. Leow, ``Rate-distortion bound for a
  class of non-gaussian sources with memory,'' \emph{IEEE Trans. Info. Theory},
  vol.~39, no.~5, pp. 1697--1701, Sept. 2003.

\bibitem{RDQuotient}
A.~Buzo, F.~Kuhlmann, and C.~Rivera, ``Rate-distortion bounds for
  quotient-based distortions with application to itakura-saito distortion
  measures,'' \emph{IEEE Trans. Info. Theory}, vol. IT-32, no.~2, pp. 141--147,
  Mar. 1986.

\bibitem{ListExponent}
N.~Merhav, ``On list size exponents in rate-distortion coding,'' \emph{IEEE
  Trans. Info. Theory}, vol.~43, no.~2, pp. 765--769, Feb. 1997.

\bibitem{DiscreteSources}
J.~K. Omura, ``A coding theorem for discrete-time sources,'' \emph{IEEE Trans.
  Info. Theory}, vol. IT-19, no.~4, pp. 490--498, Jul. 1973.

\bibitem{TrellisCoding}
I.~Hen and N.~Merhav, ``On the error exponent of trellis source coding,''
  \emph{IEEE Trans. Info. Theory}, vol.~51, no.~11, pp. 3734--3741, Nov. 2005.

\bibitem{TradeoffExponents}
T.~Weissman and N.~Merhav, ``Tradeoffs between the excess-code-length exponent
  and the excess-distortion exponent in lossy source coding,'' \emph{IEEE
  Trans. Info. Theory}, vol.~48, no.~2, pp. 396--415, Feb. 2002.

\bibitem{RDNoisySource}
T.~Weissman, ``Universally attainable error exponents for rate-distortion
  coding of noisy sources,'' \emph{IEEE Trans. Info. Theory}, vol.~50, no.~6,
  pp. 1229--1246, Jun. 2004.

\bibitem{ConvexOptimization}
S.~Boyd and L.~Vandenberghe, \emph{Convex Optimization}.\hskip 1em plus 0.5em
  minus 0.4em\relax Cambridge University Press, 2004.

\bibitem{GeneralRD}
T.~BerGer, ``Rate distortion theory for sources with abstract alphabets and
  memory,'' \emph{Informaiton and Control}, vol.~13, pp. 254--273, 1968.

\bibitem{GaussianApproximationCC}
V.~Strassen, ``Asymptotische abschatzungen in shannons informationstheorie,''
  in \emph{Trans. Third Prague Conf. Information Theory}, Prague, 1962, pp.
  689--723.

\bibitem{FiniteLength}
Y.~Polyanskiy, H.~V. Poor, and S.~Verdu, ``Channel coding rate in the finite
  blocklength regime,'' \emph{IEEE Trans. Info. Theory}, vol.~56, no.~5, pp.
  2307--2359, May 2010.

\end{thebibliography}
